\documentclass[prl,twocolumn,showpacs,preprintnumbers,amsmath,amssymb,superscriptaddress]{revtex4-1}
\usepackage{amssymb}
\usepackage{bm}
\usepackage{amsmath}
\usepackage[dvips]{graphicx}
\usepackage{color}
\usepackage{comment}
\usepackage{wrapfig}
\usepackage{hyperref}
\usepackage{textcomp}

\begin{document}

\title{Coulomb-interaction-induced Majorana edge modes in nanowires}
\author{Tommy Li}
\affiliation{Center for Quantum Devices, Niels Bohr Institute,
	University of Copenhagen, DK-2100 Copenhagen, Denmark}
\author{Michele Burrello}
\affiliation{Center for Quantum Devices, Niels Bohr Institute, University of Copenhagen, DK-2100 Copenhagen, Denmark}
\affiliation{Niels Bohr International Academy, Niels Bohr Institute, University of Copenhagen, DK-2100 Copenhagen, Denmark}
\author{Karsten Flensberg}
\affiliation{Center for Quantum Devices, Niels Bohr Institute, University of Copenhagen, DK-2100 Copenhagen, Denmark}
\begin{abstract}
We show that Majorana edge modes appear in a strongly correlated phase of semiconducting nanowires with discrete rotational symmetry in the cross section. These modes exist in the absence of spin-orbit coupling, magnetic fields and superconductivity. They appear purely due to the combination of the three-dimensional Coulomb interaction and orbital physics, which generates a fermionic condensate  exhibiting a topological ground state degeneracy in a  sector of the spectrum which is gapped to continuum modes. The gap can be comparable in magnitude to the topological superconducting gap in other solid-state candidate systems for Majorana edge modes, and  may similarly be probed via tunnel spectroscopy.
\end{abstract}
\maketitle

The quantum engineering of Majorana modes is one of the pinnacles of the study of topological phases of matter. In condensed matter systems, these modes are fermionic quasiparticles equal to their own charge conjugates, and thus a counterpart of the Majorana fermions proposed as elementary particles. Majorana modes were first predicted in two-dimensional $p$-wave superfluids \cite{Volovik1989} and superconductors \cite{ReadGreen}, but their peculiar non-local behaviour was fully understood only after they were demonstrated to exist as zero-energy modes in one-dimensional (1D) topological superconductors \cite{Kitaev2001}. Similarly to certain fractional quantum Hall excitations \cite{MooreRead,NayakWilczek}, Majorana modes possess non-Abelian braiding statistics \cite{Ivanov2001}, which makes them an extremely attractive potential platform for fault-tolerant topological quantum computation \cite{NayakReview}.


Following suggestions that a $p$-wave superconducting phase could be engineered in semiconductor nanowires using a combination of proximitization to a conventional $s$-wave superconductor \cite{Lutchyn2010,Oreg2010}, the native spin-orbit interaction in the semiconductor and external magnetic fields, several experiments have shown  evidence for the formation of Majorana zero modes above a critical magnetic field \cite{Das2012,Deng2012,Mourik2012,Finck2013,Deng2016,Nichele2017,Gul2018,Zhang2018}. However, the realization and control of topological superconducting states in such systems remains challenging and it is desirable to envision additional routes for the creation of Majorana modes, aimed at simplifying the experimental setups as much as possible.


\begin{figure}
\includegraphics[width = 0.46\textwidth]{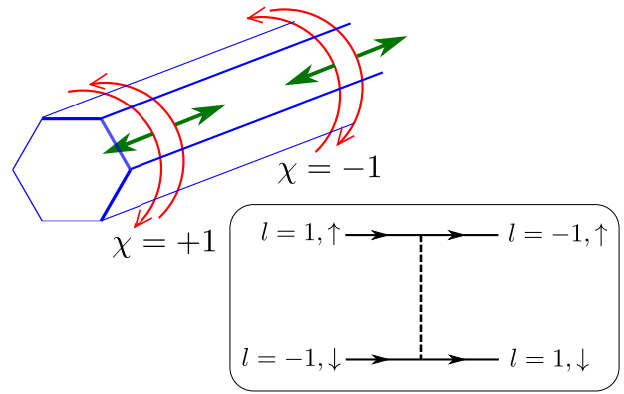}
\caption{A hexagonal wire, with electron orbital trajectories and spins (bold arrows) corresponding to positive and negative helicities ($\chi$). \emph{Inset:} the angular momentum and spin structure of the central interaction driving the topological physics.}
\label{fig:system}
\end{figure}

In this paper we propose that semiconducting nanowires can host Majorana zero modes in the absence of spin-orbit interaction, magnetic fields, and even superconductivity, thus eliminating some of the challenges present in conventional approaches to their realization. Our mechanism relies simply on the combination of the orbital physics of the nanowire and the three-dimensional Coulomb interaction. We show that in nanowires with rotational symmetry, interactions involving finite angular momentum transfer between electrons (Fig. ~\ref{fig:system}) create a strongly correlated phase with gaps appearing in the collective modes. In our system the electrons are fractionalized and the elementary excitations each carry only a part of the original electronic degrees of freedom. A topogical phase emerges within the sector associated with the product of the spin and orbital momentum, or \emph{helicity} $\chi = ls$, establishing the presence of Majorana edge modes. The situation in which we find the destruction of Luttinger liquid behaviour due to the formation of gaps in some sectors of the excitation spectrum was first demonstrated to exist by Luther and Emery in quantum wires with attractive backscattering \cite{LutherEmery}, and these phases have also appeared in studies of coupled spinless chains \cite{Cheng2011,Kraus2013,Lang2015,Iemini2015,Guther2017,Cornfeld2015},  ultracold atoms with artificial spin-orbit coupling \cite{Iemini2017}, and nanowires with commensurate Fermi and spin-orbit momenta \cite{OregSelaStern2014},  while interacting topological 1D models without violation of fermion number conservation have also been explored in coupled superconducting nanowires \cite{Sau2011} and 1D superfluid chains \cite{Ortiz2014}.


Our system has phenomenologically similar characteristics to $p$-wave superconductors, namely a gap to fermionic quasiparticle excitations which are visible in tunneling transport and a ground state degeneracy due to the edge modes. However, our system exhibits $Z_2$ rather than continuous symmetry breaking and  the topological sector consists of condensate of emergent fractional fermions which are both spinless and neutral (as opposed to conventional superconductivity), and time-reversal symmetry is not lifted as is the case for $p$-wave superconductors.

\emph{Model Hamiltonian}. We model the interacting nanowire by the three-dimensional Hamiltonian which includes the Coulomb interaction,
\begin{gather}
H = \sum_s{ \int{ \psi^\dagger_s(\bm{r})\left[ \frac{\bm{p}^2}{2m^*} + V(x,y) \right] \psi_s(\bm{r}) d^3 \bm{r}} }\nonumber \\
+ \frac{e^2}{2\epsilon_r}\sum_{s,s'}{ \int{\frac{ \psi^\dagger_{s'}(\bm{r}') \psi^\dagger_s(\bm{r}) \psi_s(\bm{r}) \psi_{s'}(\bm{r}')}{ | \bm{r} - \bm{r}'| } d^3 \bm{r} d^3 \bm{r}' } } \ \ , 
\label{Hamiltonian}
\end{gather}
where $\bm{r} = (x,y,z), \bm{r}'$ are coordinate vectors, $\bm{p}$ is the momentum operator and $V(x,y)$ is the transverse potential, which confines electrons in the cross-section. Assuming an $N$-fold rotational symmetry of the potential, the single electron states $|n,l, k, s\rangle$ possess a principal quantum number $n$ indexing radial excitations, conserved momentum $k$ along the wire axis, spin $s = \uparrow, \downarrow$ and angular momentum $l = 0, 1, \dots N-1$ associated with the phase $e^{2\pi i l/N}$ acquired under $2\pi/N$ rotation. For $l \neq 0$, there is a double degeneracy between states with equal and opposite values of $l$, which are related by time reversal. Assuming that the radial excitations exist at much higher energy than the angular excitations in the cross-section of the wire, the lowest three 1D bands consist of states with $n = 0$, $l = 0, -1, 1$.

The Coulomb interaction connects initial and final two-electron states  $|i \rangle = |k_1, l_1, s_1 \rangle \otimes | k_2, l_2, s_2 \rangle$, $|f \rangle = |k_3, l_3, s_3 \rangle \otimes | k_4, l_4, s_4 \rangle$ and discrete rotational symmetry implies $l_1 + l_2 = l_3 + l_4$ mod $N$. For a square wire, for example, interactions with $l_1 = l_2 = 1$, $l_3 = l_4 = -1$ are allowed and may significantly affect the physical characteristics of the system. For $N>4$, however, the interactions are well-approximated by those in a cylindrical geometry. This is the case we consider, which includes some of the most common experimental setups with $N = 6$. We may assume further that surface effects confine the wavefunction to the surface of the cylinder. The matrix elements of the interaction term $U$ (i.e. the second line in Eq. (\ref{Hamiltonian})) assuming the Coulomb interaction is long ranged are given by
\begin{gather}
\langle f | U | i \rangle  = 
\frac{2 e^2}{\epsilon_r} I_{l_3 - l_1}(qR) K_{l_3 - l_1}(qR) \ \ , 
\label{Bessel}
\end{gather}
where $I_j$ and $K_j$ are the modified Bessel functions and $q = k_3 - k_1$ is the exchanged momentum. For the case of forward scattering ($q = 0$) with non-zero angular momentum transfer ($l_3 \neq l_1$), the interactions become $e^2/(\epsilon_r |l_3 - l_1|)$. For $l_3 = l_1$, the expression (\ref{Bessel}) diverges logarithmically as $q \rightarrow 0$. Assuming the existence of a screening plane at distance $L \gg R$, these interactions are screened to the values $(2 e^2/\epsilon_r) \ln ( 2L/R)$ which is $\approx 6.0 e^2/\epsilon_r - 9.2 e^2/\epsilon_r$ for $L/R =  10 - 50$.

\emph{Renormalization Group analysis.} Concentrating on scattering processes close to the Fermi level (i.e. at momenta $k \approx \pm k_{l}$ equal to the Fermi momenta in the $l$ bands), we find a total of 12 independent couplings, with six forward scattering interactions, which we denote by $\Gamma_{l_1 l_2 l_3l_4}$ with momentum transfer $k_{l_3} - k_{l_1}$, and six backscattering interactions $\hat{\Gamma}_{l_1l_2l_3l_4}$ with momentum transfer $k_{l_1} + k_{l_3}$. In order to deal with the large number of competing interactions, we have calculated the running of the couplings under the renormalization group (RG), which reduces the number of interactions that become relevant at energy scales much lower than the Fermi energy.

The RG equations were obtained to second order in the couplings (listed in the Supplemental Material) and solved numerically. We find that below a critical density $\epsilon_F < E^*$ where $\epsilon_F$ is the Fermi energy as measured from the edge of the $l =\pm 1$ bands, the density of states in the upper ($l = \pm 1$) bands is sufficiently larger than that in the lower ($l = 0$) band that the interaction between upper and lower bands become irrelevant. The value of the critical density depends on weakly on the long-range interactions $U_{l_1 l_1 l_3 l_3}(q = 0) = (2e^2/\epsilon_r)\ln ( 2L/R)$; we have $E^* \approx 0.45 E_L - 0.60 E_L $ for $L/R = 10-50$ where $E_L = 1/2m^* R^2$ is the splitting between the upper and lower $l$ bands. In the following we will consider only densities corresponding to $\epsilon_F < E^*$. In this case all interactions involving the upper bands flow to strong coupling.

\emph{Bosonisation.} Having observed from the RG solutions that  only interactions involving the upper bands flow to strong coupling, we will hereafter  focus on their description in terms of eight chiral modes labeled by $l = \pm 1$ and $s = \pm 1$ with right or left chirality $\mu = \pm 1$ corresponding to the sign of the velocity. We introduce a bosonized description \cite{Giamarchi2003} of the fermionic modes by expressing the fermionic fields as
\begin{gather}
\psi_{ls, \mu}(x) =
\sum_{ \mu k - k_l <\Lambda/v_1} {
e^{i k x} c_{kls} 
}
 \nonumber \\
=\sqrt{ \Lambda/v_1}\exp\{\frac{ i}{2}( \varphi^\mu_\rho(x) + s \varphi^\mu_\sigma(x) + l \varphi^\mu_l(x) + ls \varphi^\mu_\chi(x)) \} \ \ ,
\end{gather}
where $v_1$ is the Fermi velocity in the $l = \pm 1$ bands and $\Lambda$ is the running energy scale. (For simplicity of presentation we omit the required Klein factors). The chiral bosonic fields $\partial_x \varphi^\mu_i$ represent the chiral densities of charge ($i = \rho$), orbital momentum ($i = l$), spin ($i = \sigma$), and helicity ($i = \chi$), with the latter being  the difference between densities of electrons with angular momentum parallel to spin ($ls = 1$) and those with angular momentum antiparallel to spin $(ls = -1)$. We also introduce the pairs of dual fields $\phi_i = \varphi^R_i - \varphi^L_i $ and $\theta_i =  \varphi^R_i + \varphi^L_i$ satisfying $\left[ \partial_x \theta_j(x'),  \phi_i(x) \right] = 4 \pi  i \delta_{ij} \delta(x - x')$ which define the low-energy fluctuations of densities and currents respectively within each sector $i= \rho, l, \sigma, \chi$ of the excitation spectrum. The Hamiltonian density of the system is given by
\begin{align}
\mathcal{H} = 
&\sum_{i = \rho, l, \sigma, \chi}{\frac{ v_i}{8\pi K_i} ( \partial_x \phi_i)^2 + \frac{K_i v_i}{8\pi} ( \partial_x \theta_i)^2 } \nonumber \\
+ 4 &\sum_{i,j = l, \sigma, \chi}{ G_{ij} \cos \phi_i \cos\phi_j}
+ 4 \sum_{i = l, \sigma}{ \widetilde{G}_i \cos\theta_\chi\cos\phi_i} \ .
\label{Hamilbos}
\end{align}
Here we have introduced the Luttinger parameters $K_i$ and velocities $v_i$ for each sector. The explicit relation between the running couplings $\Gamma_{l_1, l_2, l_3, l_4}, \hat{\Gamma}_{l_1, l_2, l_3, l_4}$ and the parameters $K_i, v_i, G_{ij}, \widetilde{G}_i$ of the sine-Gordon theory is provided in the Supplemental Material.

Under the RG flow, $K_\rho$ is fixed, $K_l, K_\sigma \rightarrow 0$ and $K_\chi \rightarrow \infty$. This implies that the interaction terms in Eq. (\ref{Hamilbos}) open a gap in the spin, orbital momentum and helicity sectors, while the charge sector remains gapless. This scenario is similar to the one studied in \cite{Iemini2017} in the context of ultracold atoms with spin-orbit coupling. Concerning the spin and orbital sectors, only the fields $\phi_\sigma$ and $\phi_l$ are involved in the interaction terms. Hence these fields fluctuate around one of their semiclassical energy minima, giving rise to vacuum expectation values $\langle \cos  \phi_\sigma \rangle \sim \langle \cos  \phi_l \rangle \sim \pm 1$, since $G_{\sigma l} < 0$ \cite{supp}. The helicity sector presents instead both $\phi_\chi$ and $\theta_\chi$ interactions. Due to the flow of the Luttinger parameter $K_\chi \rightarrow \infty$, the $\theta_\chi$ terms dominate, thus $\langle \cos \theta_\chi \rangle \sim \mp 1$ (since $\widetilde{G}_\sigma, \widetilde{G}_l > 0$ \cite{supp}) and hereafter we will neglect the $\phi_\chi$ interactions for the sake of simplicity.

In the strong-coupling limit of the theory, terms involving products of cosines may be decomposed via mean-field theory, $\cos \phi_i \cos \phi_j \rightarrow \langle \cos \phi_i\rangle \cos\phi_j + \langle \cos \phi_j\rangle \cos \phi_i$, \emph{etc.}, which allows  the Hamiltonian to be separated into four sectors, $\mathcal{H} = \mathcal{H}_\rho + \mathcal{H}_l + \mathcal{H}_\sigma + \mathcal{H}_\chi$. $\mathcal{H}_\rho$ describes a free bosonic theory, while $\mathcal{H}_l, \mathcal{H}_\sigma, \mathcal{H}_\chi$ describe gapped  excitations the corresponding combinations of the electronic densities.

In the gapped sectors, we observe the existence of fermionic quasiparticle excitations after introduction of  operators which  act as local raising and lowering operators $\psi_{i,\mu} \sim e^{i\varphi^\mu_i}$ \cite{vonDelft1998}
where $i = l, \sigma, \chi$.
The Hamiltonian densities describing the new spinless fermionic fields then consist of both quadratic and quartic terms, with the latter representing self-interaction of the $\psi_{i,\mu}$ fermions. Since the theory is gapped, self-interactions may be treated in mean-field theory, which leads to the effective quadratic Hamiltonian density
\begin{align}
& \ \ \ \ \ \mathcal{H} = \nonumber \\
&-i v_l \left[ \psi^\dagger_{l,R} \partial_x \psi_{l,R} - \psi^\dagger_{l,L} \partial_x \psi_{l,L} \right] + \Delta_l \left[ \psi^\dagger_{l,R} \psi_{l,L} + h.c. \right] \nonumber \\
-
&i v_\sigma  \left[ \psi^\dagger_{\sigma,R} \partial_x \psi_{\sigma,R} - \psi^\dagger_{\sigma,L} \partial_x \psi_{\sigma,L} \right] + \Delta_\sigma \left[ \psi^\dagger_{\sigma,R} \psi_{\sigma,L} + h.c. \right] \nonumber \\
-&i v_\chi \left[ \psi^\dagger_{\chi,R} \partial_x \psi_{\chi,R} - \psi^\dagger_{\chi,L} \partial_x \psi_{\chi,L} \right] + \Delta_\chi\left[ \psi^\dagger_{\chi,R} \psi^\dagger_{\chi,L} + h.c. \right] .
\label{Hamilref}
\end{align}
The gaps $\Delta_l, \Delta_\sigma, \Delta_\chi$ must be determined from the self-consistency equations involving the  coefficients $G_{ij}, \widetilde{G}_i$ of the couplings, which are listed in the Supplemental Material. These are dependent on the running couplings, however the gaps may be calculated using any set of couplings along the RG trajectory: in the mean-field equations, the running of the energy scale is compensated by the logarithmic enhancement of the couplings, and the system enters the strong-coupling regime when $\Lambda \approx \Delta_l \approx \Delta_\sigma \approx \Delta_\chi$. We therefore find that the resulting gaps are independent of the choice of $\Lambda$.

For the $l$ and $\sigma$ sectors, the refermionized Hamiltonians contain \emph{excitonic} pairing interactions $\propto \psi^\dagger_{i,R} \psi_{i,L} + h.c. $, while in the $\chi$-sector, the refermionized Hamiltonian contains a superconducting pairing interaction $\propto \psi^\dagger_{\chi,R} \psi^\dagger_{\chi,L} + h.c.$ and is equivalent to the Kitaev model for 1D spinless superconductors. The excitation spectra then become gapped,
\begin{gather}
E_{i} = \pm \sqrt{v_i^2 k^2 + \Delta_i^2} \ \ .
\end{gather}
While we refer to  the $\psi^\dagger_{R, i} \psi_{L, i}$ and $\psi^\dagger_{R, \chi} \psi^\dagger_{L, \chi}$ terms respectively as excitonic and superconducting pairing interactions because they have the form of the pairing interactions occurring in the mean-field descriptions of excitonic and $p$-wave superconductors, it is crucial to note that the order parameters $\Delta_i$ consist of real expectation values $\langle \psi^\dagger_{j, R} \psi_{j, L} + h.c. \rangle$ and $\langle \psi^\dagger_{\chi,R} \psi^\dagger_{\chi,L} + h.c. \rangle$, which do not possess a complex phase and thus there is no spontaneously broken U(1) symmetry, as required by the Mermin-Wagner theorem.

The existence of the gaps in the $\sigma, l, \chi$ sectors thus allow for the description of the system in terms of nearly-free fermionic quasiparticles, with corrections to the free motion being suppressed by the gap. Unlike the original electrons, which each carry charge, spin, angular momentum and helicity simultaneously, the true excitations each carry only one of these quantum numbers. We find that Luttinger liquid behaviour ordinarily expected for interacting quantum wires survives only in the gapless charge excitations. In the remaining sectors, operators pairing oppositely moving quasiparticles develop expectation values in the ground state and we may describe the gapped sectors as condensed states of emergent, fractional fermions. This feature defines the Luther-Emery state \cite{LutherEmery}, and we may also refer to the emergent fermionic excitations as \emph{Luther-Emery fermions.}

In the helicity sector, the superconducting term $\propto \psi^\dagger_{\chi R} \psi^\dagger_{\chi L}$ has a simple physical origin. The action of this product of fermionic operators is to locally change the helicity densities $\partial_x \phi_\chi$ without affecting the spin, charge or orbital angular momentum. This originates from collisions involving angular momentum transfer with structure $\psi^\dagger_{-1 \uparrow} \psi^\dagger_{1\downarrow} \psi_{1\uparrow} \psi_{-1\downarrow}$, in which one pair of  positive helicity ($ l = s$) electrons is converted into one pair of negative helicity ($l = -s$) electrons (see Fig. ~\ref{fig:system}). This interaction connects quantum states in which the number of electrons of a certain helicity is raised by two. It is also simple to see why the superconducting term only occurs within the helicity sector:  two-electron collisions in which either the total spin or the total angular momentum of the pair are changed are forbidden by symmetry. However, it is possible for either spin or angular momentum to be transferred between right and left moving electrons, and when expressed in terms of quasiparticle operators, such interactions take the form $\psi^\dagger_{l, R} \psi_{l, L}$ and $\psi^\dagger_{\sigma, R} \psi_{\sigma, L}$. It should also be noted that terms containing products $\psi^\dagger_{\chi,R} \psi_{\chi,L}$ are also present in the exact Hamiltonian, which correspond to the $\phi_\chi$ interactions in Eq. (\ref{Hamilbos}), however they are fully screened by the condensate, as is consistent with the running of $K_\chi \rightarrow \infty$ under RG. In the mean-field solution of (\ref{Hamilref}) we find $\Delta_l = \Delta_\sigma = \Delta_\chi = \Delta$.

Solution of the mean-field Hamiltonians for the Luther-Emery fermions (\ref{Hamilref}) in the presence of boundary conditions corresponding to full reflection at the edges of the system yields zero-energy modes  $\gamma_{i, 1}$ ($\gamma_{i,2}$) localized at the left (right) of the system and decaying into the bulk with characteristic length $\xi = v_i/\Delta$. These edge modes have also been demonstrated in previous studies of Luther-Emery systems \cite{Ruhman2015,Montorsi2017}. In the helicity sector, the presence of the superconducting term implies that these edge modes are self-conjugate, $\gamma_{\chi,1} = \gamma^\dagger_{\chi,1}$ and $\gamma_{\chi,2} = \gamma^\dagger_{\chi,2}$ and therefore emergent Majorana fermions. In total the system possesses a 32-fold ground state degeneracy arising from the occupations of the four Dirac edge modes in the $l$ and $\sigma$ sectors and the Majorana fermionic parity $i \gamma_{\chi, 1} \gamma_{\chi,2}$.

\begin{figure}
	\includegraphics[width = 0.5\textwidth]{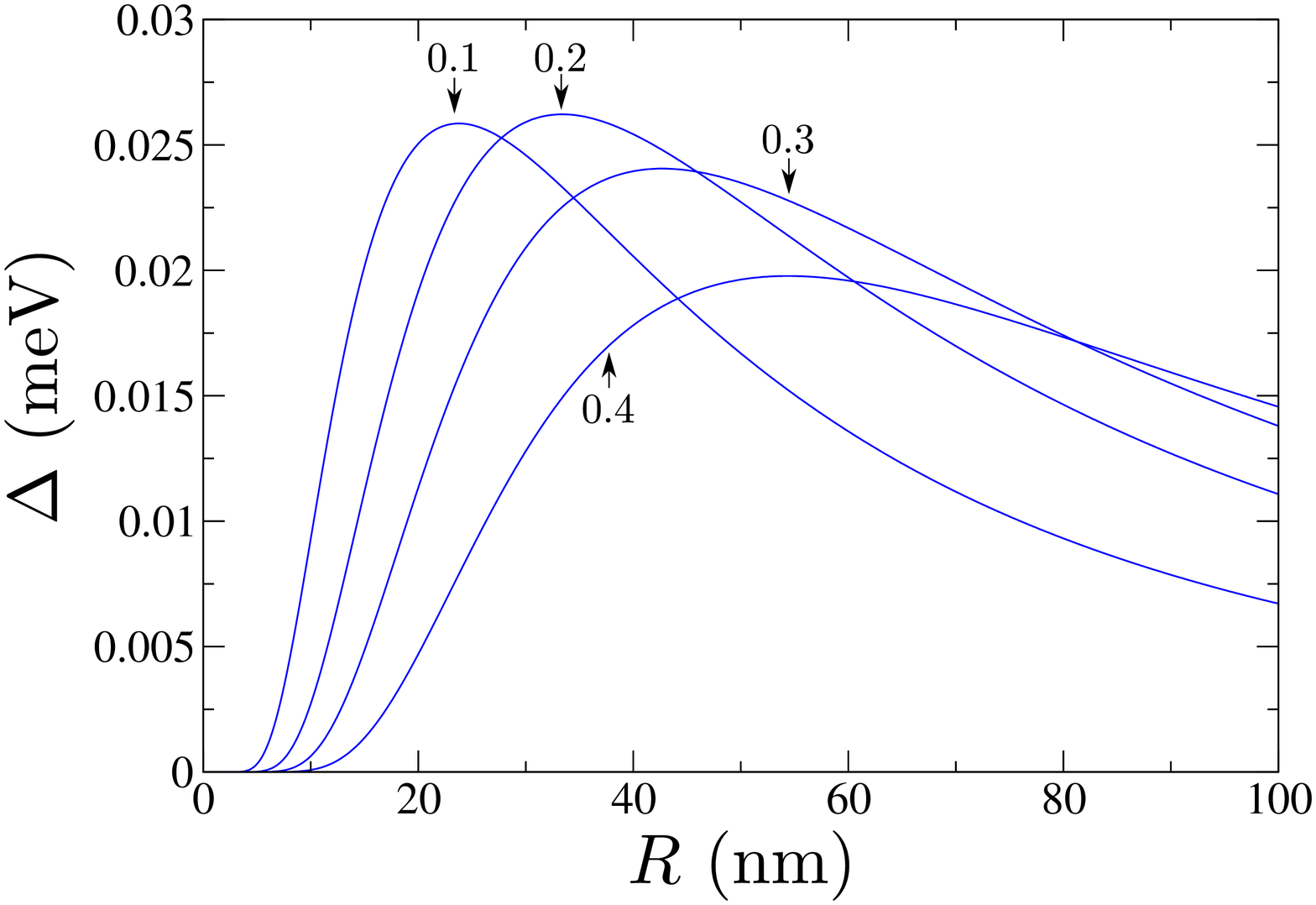}
	\caption{\emph{Solid curves}: the topological gap $\Delta$ (meV) as a function of the wire radius $R$, for various values of the Fermi energy $\epsilon_F$ measured from the bottom of the upper bands, $\epsilon_F/E_L = 0.1, 0.2, 0.3, 0.4$. The material parameters for GaAs are used, $m^* = 0.067m_e$, $\epsilon_r = 12.9$.}
\label{fig:gap}
\end{figure}

From Eq. (\ref{Bessel}), the bare values of the interactions $U_{l_1, l_2, l_3, l_4}(q) \sim e^2/\epsilon_r = g$, and we may crudely estimate the gap to be
\begin{gather}
\Delta  \sim \epsilon_F e^{-\frac{2\pi v_F}{\alpha g}} 
= \epsilon_F e^{ - \sqrt{\frac{ \epsilon_F}{ E_g}}} \ \ , \nonumber \\
E_g =\frac{\alpha^2 }{4\pi^2}\frac{ m^* e^4}{2 \epsilon_r^2} \sim \alpha^2 \frac{m^*/m_e}{4\pi^2 \epsilon_r^2} \cdot 13.6 \text{ eV}
\end{gather}
where $\alpha$ is a numerical factor of the order of unity and $m^*$ is the effective mass in the semiconductor. The gap is enhanced in materials with larger effective mass. For GaAs, $m^* = 0.067m_e$, and the gap is of the order of $\epsilon_F$ when $\epsilon_F \approx E_g \approx  \alpha^2 \cdot 0.14\text{meV}$. The calculated values of the gap as a function of the ratio $\epsilon_F/E_L$ (where $E_L= 1/2m^*R^2$) for different values of $R$ are plotted in Fig. ~\ref{fig:gap}. We find that the maximum value of the gap $\Delta \sim 0.025~\text{meV} \approx 0.3~\text{K}$, so $\alpha \approx 2$.

\begin{figure}
	\includegraphics[width = 0.5\textwidth]{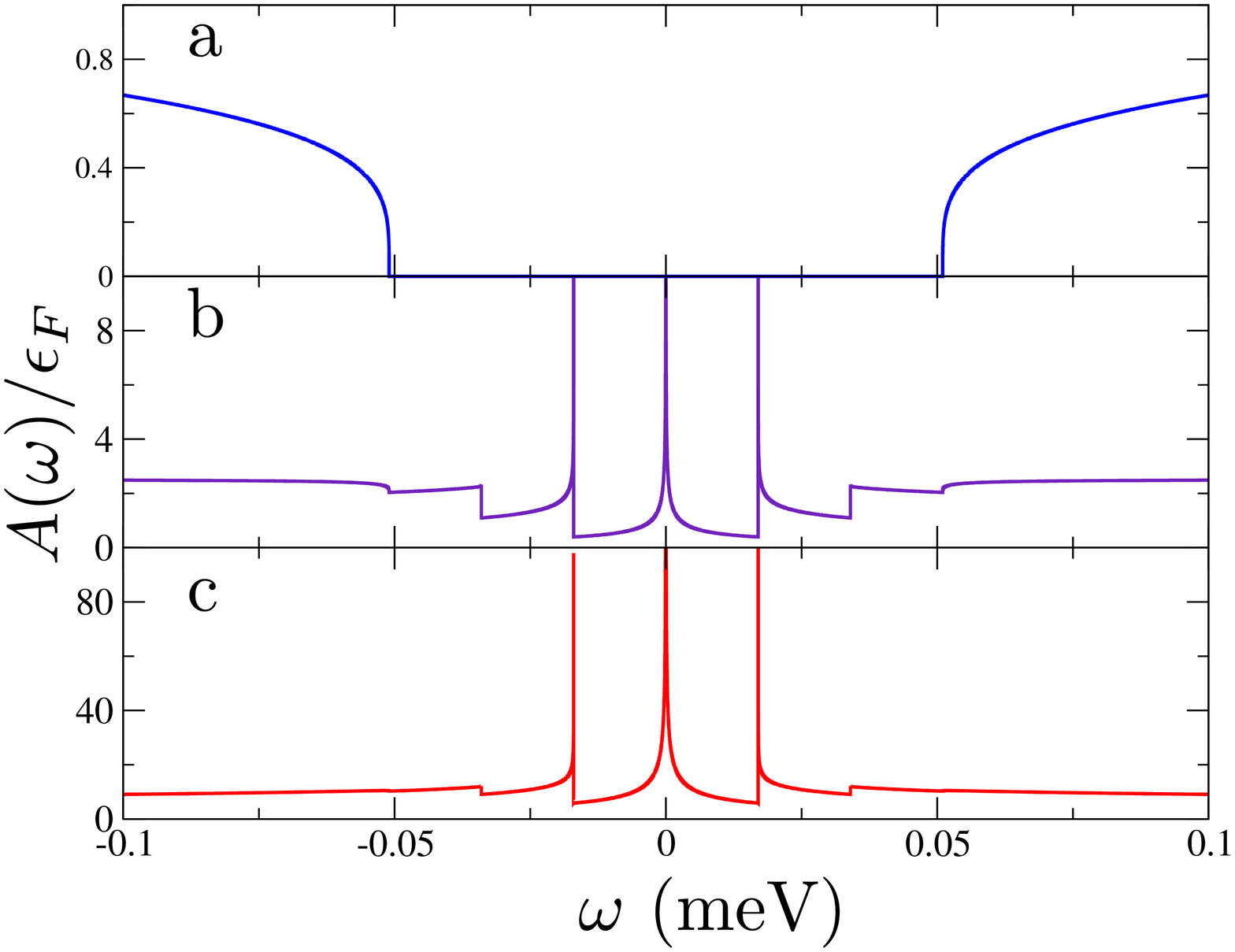}
	\caption{\emph{Color online.} The LDOS $A(x,\omega)$ for varying values (a) $x/\xi = 10$, (b) 1 and (c) 0.1. Parameters are taking for a GaAs nanowire with $R = 25~\text{nm}$ and $\epsilon_F = 0.3E_L$. Note that the scales on the vertical axes increase by an order of magnitude between each panel.}
	\label{fig:ldos}
\end{figure}

The existence of subgap edge states in the   $\sigma$, $l$ and $\xi$ sectors in combination with a gapless total charge mode has profound consequences on the tunneling transport characteristics of the system. The local one-electron density of states $A(x,\omega)$ for the upper bands is a convolution of two-point correlation functions involving the gapless $\rho$ field as well as the $\sigma$, $l$ and $\chi$ fields, which possess both gaps to continuum excitations and subgap states localized at the edges of the system. $A(x,\omega)$ is plotted in Fig.~\ref{fig:ldos} for several values of $R$ and $\epsilon_F/E_C$ and for varying distances $x$ from an edge. The explicit expression for $A(x,\omega)$ is provided in the Supplemental Material. We observe an algebraic peak at zero frequency, $A(x, \omega\approx 0) = F(x) |\omega|^{ - \gamma}$ with $F(x)$ decaying in the bulk of the system due to the localization of the zero-energy modes, in addition to bands existing at energies $|\omega| > \Delta, 2\Delta, 3\Delta$. The bands correspond to incoherent multi-particle  continuum excitations of the Luther-Emery fermions. The zero-frequency peak becomes narrower and eventually disappears at distances larger than the decay length of the edge modes, $x \gg \xi$ with $\xi = v_\chi/\Delta \approx$ 0.9 \textmu m for the parameters corresponding to Fig.~\ref{fig:ldos}. In the bulk of the system, the local density of states is vanishing for frequencies below the gaps, $|\omega| < \Delta$. Thus for certain distances along the wire, the tunneling density of states bears a striking resemblance to the zero-bias peak in Andreev reflection which has been observed several times in superconducting nanowires  \cite{Das2012,Deng2012,Mourik2012,Finck2013,Deng2016,Nichele2017,Gul2018,Zhang2018}, even though in our system superconductivity is absent. We also note that structure of the tunneling density of states is dramatically different to the one expected for a spinful Luttinger liquid, which always exhibits a power-law decay at zero frequency, $A(x,\omega) \propto |\omega|^\gamma$ with $\gamma > 0$.

\emph{Summary}. We have presented a study of an interacting quantum wire with multiple degenerate orbitals occupied. The purely repulsive Coulomb interaction produces a strongly correlated phase which supports both gaps to collective excitations and a high ground state degeneracy arising from zero-energy edge modes. We find a pair of Majorana fermions, as in 1D $p$-wave superconductors, despite the absence of superconductivity or indeed continuous symmetry breaking of any kind. Using the second-order renormalization group with multi-band interaction constants obtained for a cylindrical nanowire, we have calculated the gaps to continuum excitations as well as the local density of states, which we find to exist at experimentally accessible energy scales.

Having focused on the physical properties of the Majorana modes, we have not explored ways in which they might be manipulated  or how their non-Abelian nature might be demonstrated, which is of interest to future work. Since our system appears to be close to experimental realization, it is also an interesting prospect to investigate how  alternative nanowire-based designs might also exhibit Majorana physics as a result of the intrinsic Coulomb interaction rather than relying on the spin-orbit interaction, magnetic fields or superconductivity.


\emph{Acknowledgments.} The authors wish to acknowledge M. Leijnse and Y. Oreg for valuable discussions. M. B. acknowledges support from the Villum foundation. This work was supported by the Danish National Research Foundation.

\bibliography{Majorana3}
\bibliographystyle{apsrev4-1}

\end{document}


\title{Supplemental Material}

\maketitle

\section{2nd order RG equations}

The vertices $\Gamma_{l_1 l_2 l_3 l_4},\hat{\Gamma}_{l_1 l_2 l_3 l_4}$ are defined by matrix elements between two-electron states,
\begin{gather}
\langle f | U | i \rangle =
\begin{cases}
 \Gamma_{l_1 l_2 l_3 l_4} \ \ , \ \   k_1, k_3  > 0 \ , \ \  k_2, k_4 < 0 \ \ , \\
\hat{\Gamma}_{l_1 l_2 l_3 l_4} \ \ , \ \ k_1, k_4 > 0 \ , \ \  k_2, k_3 < 0
\end{cases} \nonumber \\
|i \rangle =| k_1, l_1, s_1 \rangle \otimes |k_2, l_2, s_2 \rangle \ \ , \nonumber \\
 |f \rangle = |k_3, l_3, s_3 \rangle \otimes | k_4, l_4, s_4 \rangle \ \ ,
\end{gather}
where $U$ is the two-electron potential. In the following it will become convenient to introduce notation distinguishing four kinds of interactions,
\begin{align}
&U_{l_1 l_2} = \Gamma_{l_1 l_2 l_1 l_2} \ , \ \ \widetilde{U}_{l_1 l_2} = \hat{\Gamma}_{l_1 l_2 l_2 l_1} \ ,  \nonumber \\
&V_1 = \Gamma_{0,0,1,-1} \ , \ \ \ \ \ \widetilde{V}_1 = \hat{\Gamma}_{0,0,1,-1}  \ ,  \nonumber \\
&V_2 = \Gamma_{1,-1,-1,1} \  , \  \ \ \widetilde{V}_2 = \hat{\Gamma}_{1,-1,1,-1} \  ,
\label{intsUV}
\end{align}
so that $U_{l_1 l_2}$ ($\widetilde{U}_{l_1 l_2}$) correspond to forward (backward) scattering interactions in which angular momentum is conserved for both left and right movers, while $V_l$ ($\widetilde{V}_l$) correspond to forward (backward) scattering interactions in which angular momentum $\pm l$ is transferred between the left and right moving densities.

Denoting
\begin{gather}
\lambda_0 = \frac{1}{2\pi v_0} \ \ , \ \ \lambda_1 = \frac{1}{2\pi v_1} \ \ , \ \ \lambda_{01} = \frac{1}{\pi(v_0 + v_1)}
\end{gather}
with $v_0, v_1$ being the Fermi velocities corresponding to the $l = 0$ and $l = 1$ bands respectively, the flow of the couplings to second order are given by
\begin{align}
& \frac{d U_{11}}{dl} =  \lambda_1 \left[ V_2^2 -\widetilde{U}_{11}^2\right] \ , \nonumber \\
& \frac{dU_{1,-1}}{dl} =  - \lambda_1 \left[ V_2^2 + \widetilde{V}_{2}^2 + \widetilde{U}_{1,-1}^2 \right] \ , \nonumber \\
& \frac{d V_2}{dl} =
2\lambda_1 \left[( U_{11}-U_{1,-1})- \widetilde{V}_2 \widetilde{U}_{1,-1}
\right]
-\lambda_0 \left[ V_1^2 + \widetilde{V}_1^2 \right] \  ,  \nonumber \\
&\frac{d U_{00}}{dl} = - \lambda_0 \widetilde{U}_{00}^2 -2 \lambda_1 \left[ V_1^2 + \widetilde{V}_1^2 \right] \  , \nonumber \\
&\frac{d U_{01}}{dl} = \lambda_{01} \left[  V_1^2 - \widetilde{U}_{01}^2 \right]  \ , \nonumber \\
& \frac{d V_1}{dl} 
=- \lambda_0 \left[ U_{00} V_1 + \widetilde{U}_{00} \widetilde{V}_1 \right]  \nonumber \\
& \ \ \ - \lambda_1 \left[ (U_{1,-1}  + V_2) V_1 + (\widetilde{V}_{2} + \widetilde{U}_{1,-1}) \widetilde{V}_1 \right]  + 2\lambda_{01} U_{01} V_1  \ , \nonumber \\
& \frac{d \widetilde{U}_{11}}{dl} =2 \lambda_1 \left[
\widetilde{V}_2 V_2 -  \widetilde{U}_{11}^2 -  \widetilde{V}_2^2
\right] \ , 
\nonumber \\
& \frac{d\widetilde{V}_2 }{dl} = 2\lambda_1 \left[
(U_{11} - U_{1,-1}) \widetilde{V}_2 + ( \widetilde{U}_{11} - \widetilde{U}_{1,-1}) V_2 - 2\widetilde{U}_{11} \widetilde{V}_2 
\right] \nonumber \\
& \ \ \ \ \ 
 - 2\lambda_0 V_1 \widetilde{V}_1 \,
 \nonumber \\
&\frac{d \hat{U}_{1,-1}}{dl} = 
-2\lambda_1 \left[
V_2 \widetilde{V}_2 
+ \widetilde{U}_{1,-1}^2
\right] - 2 \lambda_0 V_1 \widetilde{V}_1 \ , \nonumber \\
&\frac{ d \widetilde{U}_{00}}{dl} = -2\lambda_0 \widetilde{U}_{00}^2 -4 \lambda_1V_1 \widetilde{V}_1 \ \ , \nonumber \\
&\frac{d \widetilde{U}_{01}}{dl} = 2\lambda_{01} \left[ V_1 \widetilde{V}_1 -\widetilde{U}_{01}^2 - \widetilde{V}_1^2 \right]
\  , \nonumber \\
&\frac{d \widetilde{V}_1}{dl} =- \lambda_0 \left[  U_{00} \widetilde{V}_1 + \widetilde{U}_{00} V_1 \right] \nonumber \\
& \ \ \ \ \ - \lambda_1 \left[
V_1 \widetilde{V}_2 + V_1 \widetilde{U}_{1,-1}  +  \widetilde{V}_1 U_{1,-1} + \widetilde{V}_1 V_2 \right]
 \nonumber \\ 
 & \ \ \ \ \ +2 \lambda_{01} \left[
\widetilde{V}_1 U_{01} + \widetilde{U}_{01} V_1  -2 \widetilde{U}_{01} \hat{V}_1
\right] \ \ .
\label{RGoneloop}
\end{align}

\begin{figure}
	\includegraphics[width = 0.5\textwidth] {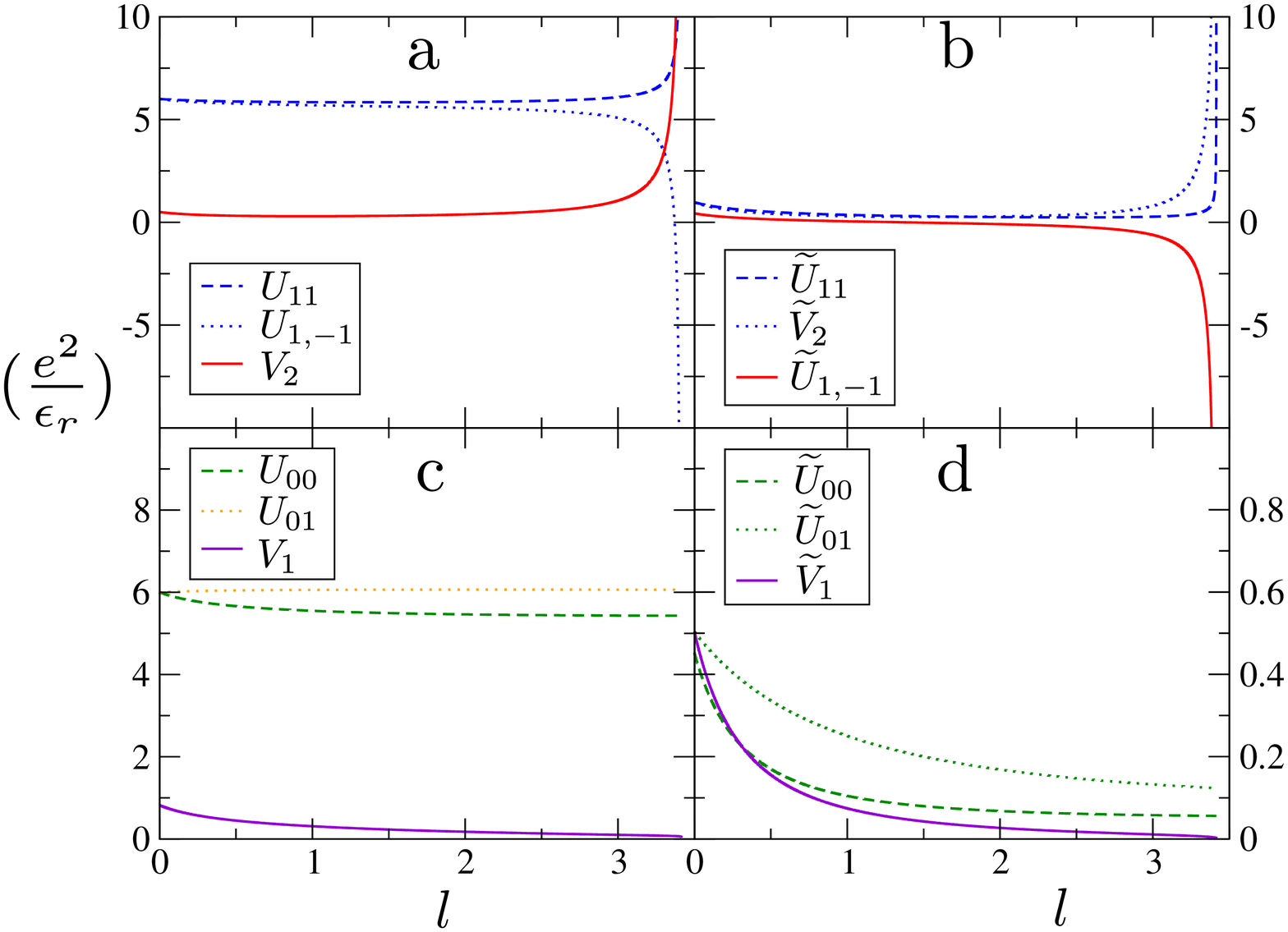}
	\caption{
		 The second-order RG flow of the interactions (\ref{intsUV}), with bare parameters corresponding to a GaAs nanowire with $R = 25~\text{nm}$ and $\epsilon_F = 0.3E_L$. Panels (a,b) show interactions that only involve the $l = \pm 1$ bands, which flow to strong coupling. Panels (a,c) show forward scattering interactions while (b,d) show backscattering interactions. Note that the scales on the axes in panels (a,b) and those in panels (c,d) are different.
	}
	\label{fig:RGflow} 
\end{figure}

The solution of the RG equations (\ref{RGoneloop}) with initial couplings corresponding to a GaAs nanowire ($m^* = 0.067m_e$, $\epsilon_r = 12.9$) with $R = 25~\text{nm}$ and $\epsilon_F = 0.3 E_L$ is plotted in Fig. \ref{fig:RGflow}. The plots show that the six interactions involving the $l = \pm 1$ bands $U_{11}, U_{1,-1}, V_2, \widetilde{U}_{11}, \widetilde{U}_{1,-1}, \widetilde{V}_2$ flow to strong coupling while $U_{00}, U_{01}$ remain marginal. The remaining interactions are irrelevant.

\section{Technical details on bosonization and refermionization}

	Following the notation of the previous section, the  Hamiltonian density of the $l, \sigma, \chi$  fields is
	\begin{gather}
 \frac{1}{4\pi}\sum_{i = \sigma, l, \chi}{ \left[ 
		\frac{v_i}{2K_i} ( \partial_x \phi_i)^2
		+ \frac{ K_i v_i}{2} ( \partial_x \theta_i)^2 \right]
	} \nonumber \\
	+ 4\sum_{i,j = \sigma, l, \chi}{ G_{ij} \cos \phi_i \cos \phi_j} + 4\sum_{i = \sigma, l}{ \widetilde{G}_i \cos \phi_i\cos \theta_\chi}
	\end{gather}
	where
	\begin{align}
& v_\sigma = v_1 + \frac{ V_2^{(0)}}{2\pi} , \ \ v_l =  v_1 - \frac{V_2^{(0)}}{2\pi} , \ \ v_\chi = v_1 - \frac{V_2^{(0)}}{2\pi} , \nonumber \\
& \ \ \ \ \ K_i = \sqrt{ \frac{\widetilde{v}_i - U_i/2\pi}{ \widetilde{v}_i + U_i/2\pi}}  , \nonumber \\
& U_\sigma = - \widetilde{U}_{11} - \widetilde{U}_{1,-1} , \nonumber \\
& U_l = 2 U_{11} - 2 U_{1,-1} - \widetilde{U}_{11} + \widetilde{U}_{1,-1}  , \nonumber \\
&   U_\chi = - \widetilde{U}_{11} + \widetilde{U}_{1,-1} ,  \nonumber \\
& G_{l \sigma} = - \widetilde{V}_2 \ , \ \ G_{\sigma \chi} = - \widetilde{U}_{11} \  , \ \ G_{l \chi} = V_2 - \widetilde{V}_2 , 
\nonumber \\
& \widetilde{G}_l = V_2  , \ \ \widetilde{G}_\sigma = - \widetilde{U}_{1,-1}  \ \ .
\end{align}
	with $V_2^{(0)}$ being the bare (unrenormalized) value of $V_2$.	Following the main text, we perform a mean-field decomposition of terms involving cosines:
	\begin{gather}
	\cos \phi_i \cos \phi_j \rightarrow 
	\langle \cos\phi_i \rangle \cos\phi_j + \langle \cos\phi_j \rangle \cos \phi_i
	\nonumber \\+  ( \cos\phi_i  - \langle \cos \phi_i \rangle)( \cos \phi_j - \langle \cos\phi_j \rangle) \ \ , 
	\label{MFdecomp1}
	\end{gather}
	and
	\begin{gather}
	\cos \theta_\chi \cos\phi_i \rightarrow \langle \cos \theta_\chi \rangle \cos \phi_i + \langle \cos\phi_i \rangle \cos \theta_\chi
\nonumber \\
	+( \cos \theta_\chi  - \langle \cos  \theta_\chi \rangle)( \cos \phi_i - \langle \cos  \phi_i \rangle)
	\label{MFdecomp2} \ \ .
	\end{gather}
	In the limit of strong coupling the final terms in Eqs. (\ref{MFdecomp1},\ref{MFdecomp2}), which describe fluctuations of the fields about their averages, may be neglected. We then obtain separate Hamiltonian densities $\mathcal{H} = \mathcal{H}_\sigma + \mathcal{H}_l + \mathcal{H}_\chi$ where
	\begin{gather}
	\mathcal{H}_i = \frac{v_i K_i}{8\pi} ( \partial_x \theta_i)^2 + \frac{v_i}{8\pi K_i}( \partial_x \phi_i)^2
	\nonumber \\
	+ 4\left[  \sum_{j \neq i}{ G_{ij} \langle \cos\phi_j \rangle } + \widetilde{G}_i \langle \cos \theta_\chi\rangle \right] \cos \phi_i
	\end{gather}
	for $i = \sigma, l$ and
	\begin{gather}
	\mathcal{H}_\chi =\frac{v_\chi K_\chi}{8\pi} ( \partial_x \theta_\chi)^2 + \frac{v_\chi}{8\pi K_\chi}( \partial_x \phi_\chi)^2 
	\nonumber \\
	+ 4
	\sum_{j}{
		\langle \cos \phi_j \rangle 
		\left[ G_{\chi j} \cos\phi_\chi +\widetilde{G}_j \cos  \theta_\chi\right] } \ \ .
	\end{gather}

	As in the main text, we express each bosonic theory $\mathcal{H}_i$ via fermionic operators $\psi_{i, L}$, $\psi_{i,R}$. Noting that in terms of the fermionic operators the vacuum averages become
	\begin{gather}
	\langle \cos \phi_i \rangle =
	\frac{1}{2} \langle \psi^\dagger_{i,R} \psi_{i,L} + h.c. \rangle , \ \langle \cos \theta_\chi \rangle = \frac{1}{2} \langle \psi^\dagger_{i,R} \psi^\dagger_{i,L} + h.c. \rangle ,
	\end{gather}
	we obtain
	\begin{gather}
	\mathcal{H}_i = 
	-iv_i \left[ \psi^\dagger_{i,R} \partial_x \psi_{i,R} - \psi^\dagger_{i,L} \partial_x \varphi_{i,L} \right] + U_i \psi^\dagger_{i,R} \psi_{i,R} \psi^\dagger_{i,L} \psi_{i,L} \nonumber \\
	+\left[  \sum_{j \neq i}{
		G_{ij} \langle \psi^\dagger_{j,R} \psi_{j,L} + h.c. \rangle } + \widetilde{G}_i \langle \psi^\dagger_{\chi,R} \psi^\dagger_{\chi,L} + h.c. \rangle \right] \times \nonumber\\  
	( \psi^\dagger_{i,R} \psi_{i,L} + h.c.)
	\end{gather}
	for $i = \sigma, l$ and
	\begin{gather}
	\mathcal{H}_\chi =\nonumber\\
	-i v_\chi \left[ \psi^\dagger_{\chi,R} \partial_x \psi_{\chi,R} - \psi^\dagger_{\chi,L} \partial_x \varphi_{\chi,L} \right] + U_\chi \psi^\dagger_{\chi,R} \psi_{\chi,R} \psi^\dagger_{\chi,L} \psi_{\chi,L} 
	\nonumber \\
	+ \sum_{j }{ \langle \psi^\dagger_{j,R} \psi_{j,L} + h.c. \rangle \left[  G_{\chi j} \psi^\dagger_{\chi,R} \psi_{\chi,L} 
		+ \widetilde{G}_j  \psi^\dagger_{\chi,R} \psi^\dagger_{\chi,L} + h.c.\right] 
	}  \ \ .
	\end{gather}
	The Hamiltonian densities contain  quadratic terms $\propto \psi^\dagger_{i,R} \psi_{i,L} + h.c.$ which describe pairing of particles and holes, and $\mathcal{H}_\chi$ contains a particle-particle pairing interaction $\propto \psi^\dagger_{\chi,R} \psi^\dagger_{\chi,L} + h.c.$. In addition, quartic interactions appear in all sectors which describe the fermionic self-interactions. For the $\sigma-$ and $l-$ fields, these self-interactions are repulsive, which corresponds to an attractive interaction between particles and holes and thus enhances the excitonic order parameter $\langle \psi^\dagger_{i,R} \psi_{i,L}\rangle$. For the $\chi$ field, the self-interaction is attractive, which enhances the superconducting order parameter $\langle \psi^\dagger_{\chi,R} \psi^\dagger_{\chi,L} \rangle$. It is therefore appropriate at the mean-field level to replace
	\begin{gather}
	\psi^\dagger_{i,R} \psi_{i,R} \psi^\dagger_{i,L} \psi_{i,L} \rightarrow - \left[ \langle \psi^\dagger_{i,R} \psi_{i,L}\rangle \psi^\dagger_{i,L} \psi_{i,R} + h.c. \right] 
	\end{gather}
	for $i = \sigma, l$ and
	\begin{gather}
	\psi^\dagger_{\chi,R} \psi_{\chi,R} \psi^\dagger_{\chi,L} \psi_{\chi,L} \rightarrow 
	\langle \psi^\dagger_{\chi,R} \psi^\dagger_{\chi,L} \rangle \psi_{\chi,L} \psi_{\chi,R} + h.c. \ \ .
	\end{gather}

	\begin{figure}
		\includegraphics[width = 0.5\textwidth]{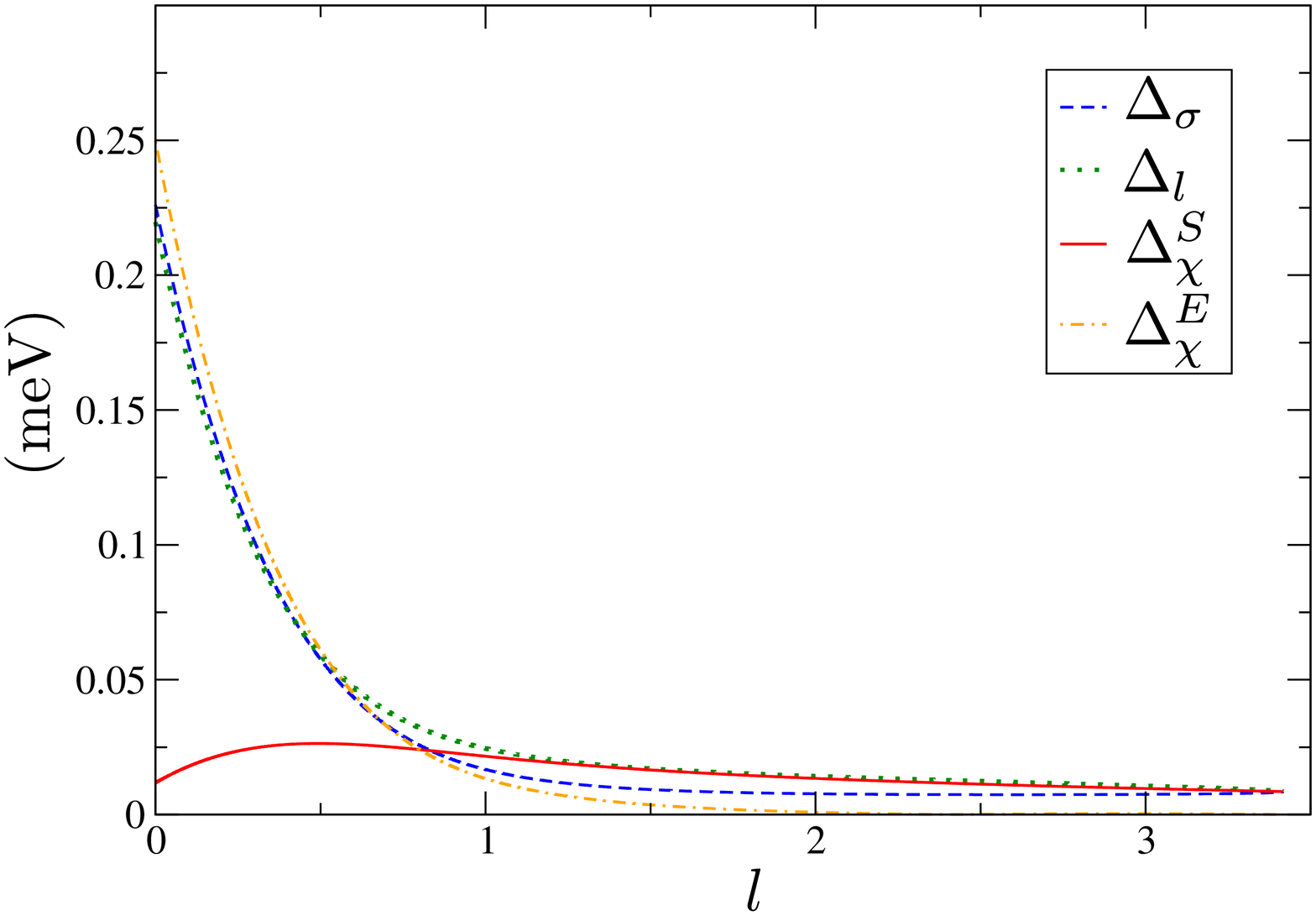}
		\caption{(\emph{Color online}) The mean-field solution for the gaps $\Delta_\sigma$ (blue, dashed), $\Delta_l$ (green, dotted), $\Delta_\chi^S$ (red, solid), and $\Delta_\chi^E$ (orange, dashed and dotted). Parameters correspond to a GaAs nanowire with $R = 25~\text{nm}$ and $\epsilon_F = 0.3E_l$.}
		\label{fig:MF}
	\end{figure}

	Thus $\mathcal{H}_i$ take the form of Bogoliubov-de Gennes Hamiltonians
	\begin{gather}
	\mathcal{H}_i = -i v_i \left[ \psi^\dagger_{i,R} \partial_x \psi_{i,R} - \psi^\dagger_{i,L} \partial_x \varphi_{i,L} \right]
	+ \Delta_i ( \psi^\dagger_{i,R} \psi_{i,L} + h.c.)
	\end{gather} 
	for $i = \sigma, l$ and
	\begin{gather}
	\mathcal{H}_\chi = -i v_\chi \left[ \psi^\dagger_{\chi,R} \partial_x \psi_{\chi,R} - \psi^\dagger_{\chi,L} \partial_x \varphi_{\chi,L} \right] \nonumber \\
	+ \Delta_\chi^E ( \psi^\dagger_{\chi,R} \psi_{\chi,L} + h.c.) + \Delta_\chi^S ( \psi^\dagger_{\chi,R} \psi^\dagger_{\chi,L} + h.c.) \ \ ,
	\end{gather}
	where
	\begin{gather}
	\Delta_i = - U_i \langle \psi^\dagger_{i,R} \psi_{i,L} \rangle + 2 \sum_{j \neq i}{
		G_{ij} \langle \psi^\dagger_{j,R} \psi_{j,L} \rangle} + 2\widetilde{G}_i \langle \psi^\dagger_{\chi,R} \psi^\dagger_{\chi,L} \rangle   
\end{gather}
for $i= \sigma, l$ and
\begin{align}
	& \Delta_\chi^E = 2 \sum_{j \neq \chi}{ G_{\chi j} \langle \psi^\dagger_{j,R} \psi_{j,L} \rangle } \ \ , \nonumber \\
	& \Delta_\chi^S = U_\chi \langle \psi^\dagger_{\chi,R} \psi^\dagger_{\chi,L}\rangle + 2\sum_j{ \widetilde{G}_j \langle \psi^\dagger_{j,R} \psi_{j,L} \rangle } \ \ .
	\end{align}
	Solving the quadratic Hamiltonians, we find that in the limit $\Delta_\sigma, \Delta_l \ll \epsilon_F$ and either $\Delta_\chi^E \ll \epsilon_F$ or $\Delta_\chi^S \ll \epsilon_F$
	\begin{align}
	&\langle \psi^\dagger_{i,R} \psi_{i,L} \rangle = - \frac{\Delta_i}{2\pi v_i} \ln \frac{ 2\Lambda}{ | \Delta_i|} \  , \ \ i = \sigma, l \nonumber \\
	&\langle \psi^\dagger_{\chi,R} \psi_{\chi,L} \rangle = - \frac{1}{4\pi v_\chi} \left[
	\Delta_+ \ln\frac{2\Lambda}{|\Delta_+|} - \Delta_- \ln \frac{ 2\Lambda}{|\Delta_-|} \right]  \ \ , \nonumber \\
	&\langle \psi^\dagger_{\chi,R} \psi^\dagger_{\chi,L} \rangle = - \frac{1}{4\pi v_\chi} \left[ \Delta_+\ln \frac{ 2\Lambda}{|\Delta_+|} + \Delta_- \ln \frac{ 2\Lambda}{|\Delta_-|}
	\right] \ \ , \nonumber \\
	& \ \ \ \ \ \ \Delta_\pm = \Delta^S_\chi \pm \Delta^E_\chi  \ \ .
	\end{align}
	Assuming $\Delta_\sigma \sim \Delta_l \sim \Delta_\pm \sim\Delta $ we may approximate the logarithmic factors by $\ln ( \Lambda/\Delta)$ and $v_l, v_\chi, v_\sigma \approx v_F$ (assuming $V^{(0)}_2 \ll 2\pi v_F$), which yields the mean-field equations
	\begin{gather}
	\left(\begin{array}{c}
	\Delta_\sigma \\
	\Delta_l \\
	\Delta_\chi^S \\
	\Delta_\chi^E 
	\end{array}\right)= \nonumber \\
	- \frac{1}{2\pi v_F} \ln \frac{ 2\Lambda}{\Delta}
	\underbrace{ \left(\begin{array}{cccc}
		- U_\sigma & 2G_{\sigma l} & 2 \widetilde{G}_\sigma & 2 G_{\sigma \chi} \\
		2 G_{\sigma l} & - U_l & 2 \widetilde{G}_l & 2 G_{l \chi} \\
		2 \widetilde{G}_\sigma & 2 \widetilde{G}_l & U_\chi & 0 \\
		2 G_{\sigma \chi} & 2 G_{ l \chi} & 0 & 0 
		\end{array}\right) }_{ \bm{G}}
	\left(\begin{array}{c}
	\Delta_\sigma \\
	\Delta_l \\
	\Delta_\chi^S \\
	\Delta_\chi^E 
	\end{array}\right) \ \ .
	\label{MF_eqs}
	\end{gather}
	Denoting the largest negative eigenvalue of $\bm{G}$ by $\lambda$, the gap is then given by
	\begin{gather}
	\Delta =2 \Lambda e^{ \frac{2\pi v_F}{\lambda}} = 2\Lambda e^{- \frac{2\pi v_F}{|\lambda|}} \ \ .
	\end{gather}
	Since the RG flows exhibit a divergence at a finite scale $l = l^*$, the gap is also equal to
	\begin{gather}
	\Delta = 2\Lambda_0 e^{- l^*} \approx 2\epsilon_F e^{- l^*} \ \ .
	\end{gather}
	The solution of the mean field equations (\ref{MF_eqs}) for a GaAs nanowire with $R = 25~\text{nm}$ and $\epsilon_F = 0.3E_l =  0.26~\text{meV}$ is shown as a function of the running scale in Fig. \ref{fig:MF}. Applying mean field theory with the bare interactions ($l = 0$), we find that gaps exist in all sectors, with $|\Delta_\sigma|, |\Delta_l|, |\Delta_\chi^E| \gg |\Delta_\chi^S| \approx 0.25~\text{meV}$. Under the RG flow, we find that $\Delta_\chi^E \rightarrow 0$ and the remaining gaps converge to a value $\approx 0.017~\text{meV}$. This shows that second-order RG is necessary to resolve the competition between the excitonic and superconducting orders in the $\chi$-sector.

\section{Local Density of States}

The local density of states (LDOS) may be expressed in terms of the Matsubara Green's function,
\begin{gather}
A_{l,s}(x,\omega) = - \frac{1}{\pi} \text{Im} G_{ls}^R(x,\omega) = - \frac{1}{\pi} \text{Im} \mathcal{G}_{ls} (x, -i \omega + 0)  \ \ , \nonumber \\
-\mathcal{G}_{ls}(x,\omega) =
\int{
	\langle\mathcal{T} \psi_{ls}(x, \tau) \psi_{ls}^\dagger(x, 0) \rangle e^{i \omega \tau} d\tau 
} \ \ ,
\end{gather}
where $G^R_{ls}$ is the retarded Green's function,  $l = 0, 1, -1$ is angular momentum and $s = \uparrow, \downarrow$ is spin. Since the $l = 0$ modes are gapless, they provide a power-law contribution to the LDOS which is well known in Luttinger liquid theory. The contribution from $l = \pm 1$ states is a product of correlation functions in the $\rho, l, \sigma, \chi$ sectors
\begin{gather}
\mathcal{G}_{l = \pm 1, s}(x, \tau) = \nonumber \\ \text{sgn}(\tau)  g_{\rho,R}(x,\tau) g_{\sigma,R}(x,\tau) g_{l,R}(x,\tau) g_{\chi,R}(x,\tau) \ \ .
\end{gather}
Due to the edge modes appearing in the $\sigma, l, \chi$ sectors, their respective correlations consist of a $\tau$-independent contribution arising from the zero-energy modes as well as a $\tau$-dependent factor due to the continuum states:
\begin{gather}
g_{\rho,R}(\tau) =g_{\rho,L}(\tau) = \frac{1}{ |\tau|^\frac{\gamma}{4}} = g_\rho(\tau) \ \ , \nonumber \\
g_{i,R}(\tau) = g_{i,L}(\tau) = \kappa_i +  \frac{ e^{- \Delta_i |\tau|}}{ |\tau|^\frac{1}{4}} = g_i(\tau) \ \ ,  \ i = \sigma, l, \chi \ \ ,
\label{gamma}
\end{gather}
with $\gamma > 1$ being an interaction-dependent exponent and $\kappa_i$ are of order unity at the edge, and decay exponentially as $x$ moves into the bulk of the system. Converting to the frequency representation,  $G^R(x,\omega) = \mathcal{G}(x,-i \omega + 0)$, we obtain the following expression for the LDOS:
\begin{align}
& -\frac{1}{\pi} \text{Im} G^R(\omega)  = 
\frac{\kappa_l \kappa_\sigma \kappa_\chi}{\Gamma(\frac{\gamma}{4}) \Lambda^{\frac{ \gamma}{4} - 2}} |\omega|^{\frac{\gamma}{4} - 1}
\nonumber \\
& \ \ \  +
\kappa_l \kappa_\sigma \kappa_\chi \sum_{i}{ \frac{\Theta(|\omega |-\Delta_i)}{ \kappa_i  \Gamma( \frac{\gamma + 1}{4})\Lambda^{ \frac{ \gamma -7}{4}}}(|\omega| - \Delta_i)^{\frac{ \gamma -3}{4}}} \nonumber \\
+
 & \ \ \ \kappa_l \kappa_\sigma \kappa_\chi\sum_{i < j}{ \frac{\Theta( | \omega| - \Delta_i - \Delta_j)}{ \kappa_i \kappa_j\Gamma( \frac{\gamma+ 2}{4}) \Lambda^\frac{\gamma -6}{4}}
	( |\omega| - \Delta_i - \Delta_j)^\frac{\gamma - 2}{4}
}  \nonumber \\
+& \ \ \ \frac{\Theta( | \omega| - \Delta_\sigma - \Delta_l - \Delta_\chi)}{\Gamma( \frac{ \gamma + 3}{4})\Lambda^{\frac{\gamma -5}{4}}}( |\omega| - \Delta_\sigma - \Delta_l - \Delta_\chi)^{\frac{\gamma - 1}{4}}   \ \ ,
\label{LDOS}
\end{align}
where $\Lambda = \Lambda_0 \approx \epsilon_F$ is the high-energy cutoff. The first term in the Green's function $\propto | \omega|^{\frac{\gamma}{4} - 1}$ is singular for values of interactions satisfying $1 < \gamma  < 4.$  The remaining terms have edges at energies $\Delta_i$, $\Delta_i + \Delta_j$ and $\Delta_\sigma + \Delta_l + \Delta_\chi$, corresponding to the gaps to various incoherent multi-Luther-Emery-fermion excitations.

The calculation of the exponent $\gamma$ is significantly complicated by the presence of interactions between the $l = 0$ and $l = 1$ states, which are marginal under the RG flow but nevertheless must be included because the charge sector is gapless. We slightly change our notation and represent the field $\varphi^R_{\rho} \rightarrow \varphi^R_{\rho^+}$ to highlight that they only include densities in the upper bands. The Hamiltonian density in the charge sector is
\begin{widetext}
\begin{gather}
\mathcal{H} = \nonumber \\
\frac{1}{4\pi} \left[
(v_0 + \frac{ U_{00}}{\pi}) ( \partial_x \varphi^R_{\rho^0})^2 + ( v_1 + \frac{2 U_{11} + 2U_{1,-1} + V_2}{2\pi})( \partial_x \varphi^R_{\rho^+})^2
+ \frac{ 2 \sqrt{2} U_{01}}{\pi} \partial_x \varphi^R_{\rho^0} \partial_x \varphi^R_{\rho^+} + R \rightarrow L
\right] \nonumber \\
- \frac{1}{4\pi^2}(
(2 U_{00} - \widetilde{U}_{00})\partial_x \varphi^R_{\rho^0} \partial_x \varphi^L_{\rho^0}
+ (2U_{11} + 2U_{1,-1} - \widetilde{U}_{11} - \widetilde{U}_{1,-1}) \partial_x \varphi^R_{\rho^+} \partial_x \varphi^L_{\rho^+} \nonumber \\
+ 2\sqrt{2} U_{01} \left[ \partial_x \varphi^R_{\rho^0} \partial_x \varphi^L_{\rho^+} + L \leftrightarrow R \right] ) \ \ ,
\label{hamilcharge}
\end{gather}
\end{widetext}
where the fields 
\begin{align}
&\varphi^R_{\rho^+} = \frac{ \varphi^R_{ l = 1, \uparrow} + \varphi^R_{l = 1, \downarrow} + \varphi^R_{l = -1, \uparrow} + \varphi^R_{l = -1, \downarrow}}{2} \ \ , \nonumber \\
&\varphi^R_{\rho^0} = \frac{\varphi^R_{l = 0, \uparrow} + \varphi^R_{l = 0, \downarrow}}{\sqrt{2}}
\end{align}
and $\uparrow, \downarrow$ are the spin indices. The interactions are given by
\begin{gather}
U_{l_1 l_2} = \frac{ 2e^2}{\epsilon_r} \ln \frac{2L}{R} = U_0 \ \ , \ \ 
\widetilde{U}_{ll} = \frac{2 e^2}{\epsilon_r} I_0(2k_l R) K_0 (2k_l R)  \ \ , \nonumber \\
\widetilde{U}_{1,-1} = \frac{2 e^2}{\epsilon_r} I_2(2k_1R) K_2(2k_1 R)
\end{gather}
where $R$ is the radius of the wire, $k_l$ are the Fermi momenta in bands $l = 0, \pm 1$ and $L \gg R$ is the distance to the nearest screening plane.

\begin{figure}
	\includegraphics[width = 0.5\textwidth]{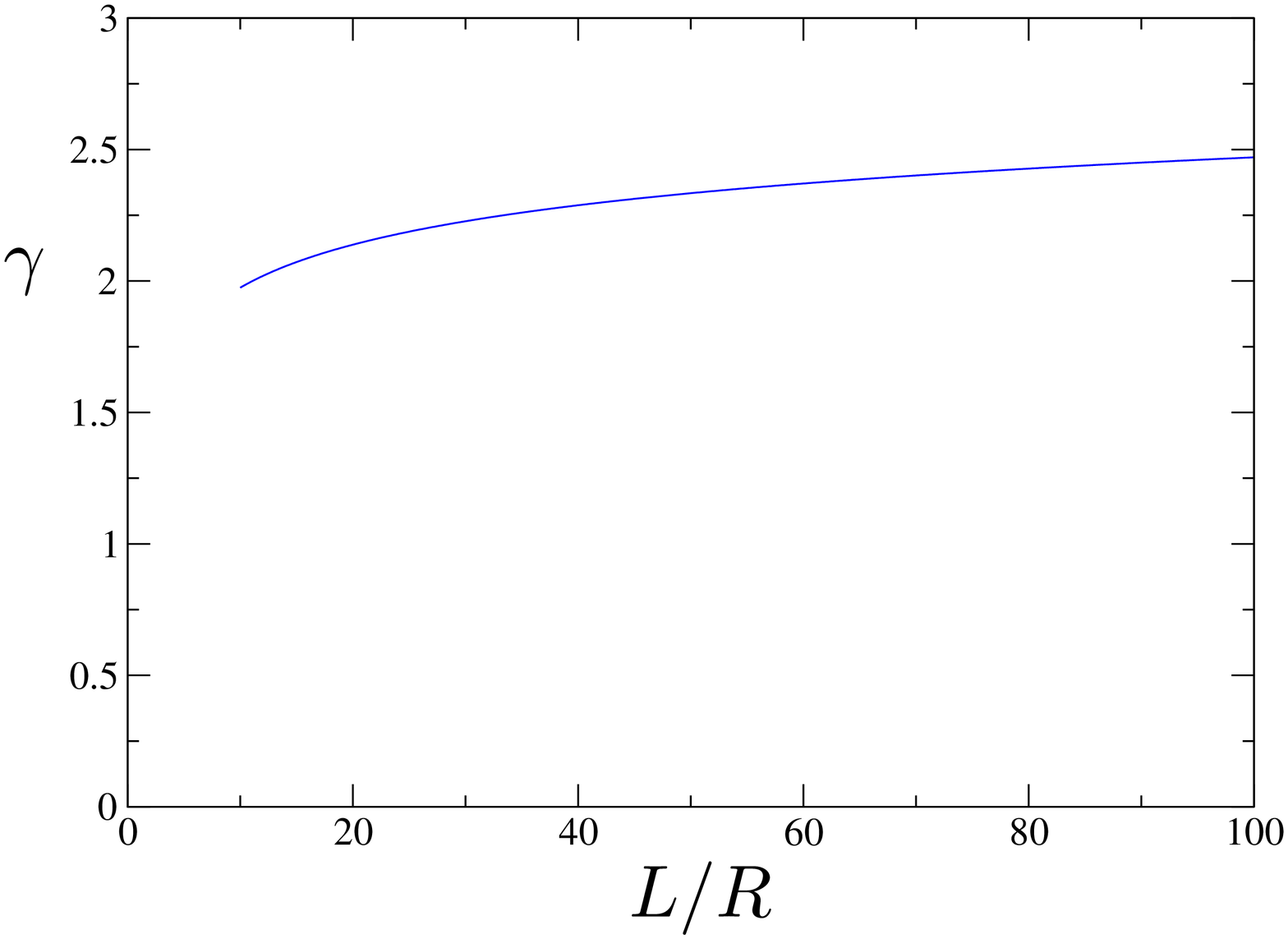}
	\caption{Calculated values of the exponent $\gamma$ appearing in Eq. (\ref{gamma}) as a function of the ratio $L/R$ where $L$ is the distance to the screening plane and $R$ is the radius of the wire.}
	\label{fig:gamma}
\end{figure}

The Hamiltonian density must be diagonalized via canonical transformation, which gives
\begin{gather}
\mathcal{H} = \sum_{\gamma = 1, 2}{ \frac{v_\gamma}{4\pi} \left[ ( \partial_x \widetilde{\varphi}^R_\gamma)^2 + ( \partial_x \widetilde{\varphi}^L_\gamma)^2 \right] } 
\end{gather}
where 
\begin{align}
& \partial_x \widetilde{\varphi}^R_\gamma = \sum_\alpha{ U_\gamma^{\alpha} \partial_x \varphi^R_\alpha - V_\gamma^\alpha \partial_x \varphi^L_\alpha} \ \ , \nonumber \\
& \partial_x \widetilde{\varphi}^L_\gamma =  \sum_\alpha{- V_\gamma^\alpha \partial_X \varphi^R_\alpha + U_\gamma^\alpha \partial_x \varphi^L_\alpha } \ \ , \nonumber \\
\alpha = ( \rho^0, \rho^+)
\end{align}
and the coefficients satisfy
\begin{widetext}
\begin{gather}
\left(\begin{array}{cccc}
2\pi v_0 + 2 U_0 & 2\sqrt{2} U_0 & - 2U_0 + \widetilde{U}_{00} &- 2 \sqrt{2} U_0 \\
2\sqrt{2} U_0 & 2\pi v_1 + 4U_0 + V_2 & - 2 \sqrt{2} U_0 & -4U_0 + \widetilde{U}_{11} + \widetilde{U}_{1,-1}\\
2 U_0 - \widetilde{U}_{00} & 2 \sqrt{2} U_0 & - 2\pi v_0 - 2U_0 & -2\sqrt{2} U_0 \\
2\sqrt{2} U_0 & 4 U_0 - \widetilde{U}_{11} - \widetilde{U}_{1,-1} & - 2 \sqrt{2} U_0 & - 2\pi v_1 - 4 U_0 - V_2
\end{array}\right) \left(\begin{array}{c}
U_{\gamma}^{ \rho^0} \\ U_{\gamma}^{\rho^+} \\ V_{\gamma, \rho^0} \\ V_{\gamma}^{ \rho^+}
\end{array}\right) 
= 
2\pi v_\gamma  \left(\begin{array}{c}
U_{\gamma}^{ \rho^0} \\ U_{\gamma}^{\rho^+} \\ V_{\gamma}^{ \rho^0} \\ V_{\gamma}^{\rho^+}
\end{array}\right)
\end{gather}
with $v_\gamma > 0$.
\end{widetext}
Since $\widetilde{\varphi}_\gamma^R, \widetilde{\varphi}_\gamma^L$ are free fields, their correlation functions are given by
\begin{gather}
\langle e^{i\kappa (\varphi^R_\gamma(x, \tau) - \widetilde{\varphi}^R_\gamma(x,0))} \rangle = \langle e^{i \kappa ( \varphi^L_\gamma(x, \tau) - \widetilde{\varphi}^L_\gamma(x,0))} \rangle = \frac{1}{|\tau|^{\kappa^2}}.
\end{gather}
The correlation function
\begin{gather}
\langle e^{\frac{ i }{2} ( \varphi^R_{\rho^+}(x, \tau) - \varphi^R_{\rho^+}(x, 0))} \rangle  = \frac{1}{|\tau|^\frac{ \gamma}{4}}
\end{gather}
may then be calculated straightforwardly by transformation from the $\varphi^R_{\rho^+}, \varphi^R_{\rho^0}$ fields to the free bosonic fields. $\gamma$ is plotted as a function of $L/R$ in Fig. \ref{fig:gamma}. We obtain $\gamma \approx 2-3$ in the range $10 <L/R < 100$. From Eq. (\ref{LDOS}) we therefore see that the LDOS exhibits a peak at zero frequency.

\section{Diagrammatic representation of the RG equations}

\begin{figure}[t]
	\begin{tabular}{cc}
		\fbox{\includegraphics[width = 0.3\textwidth]{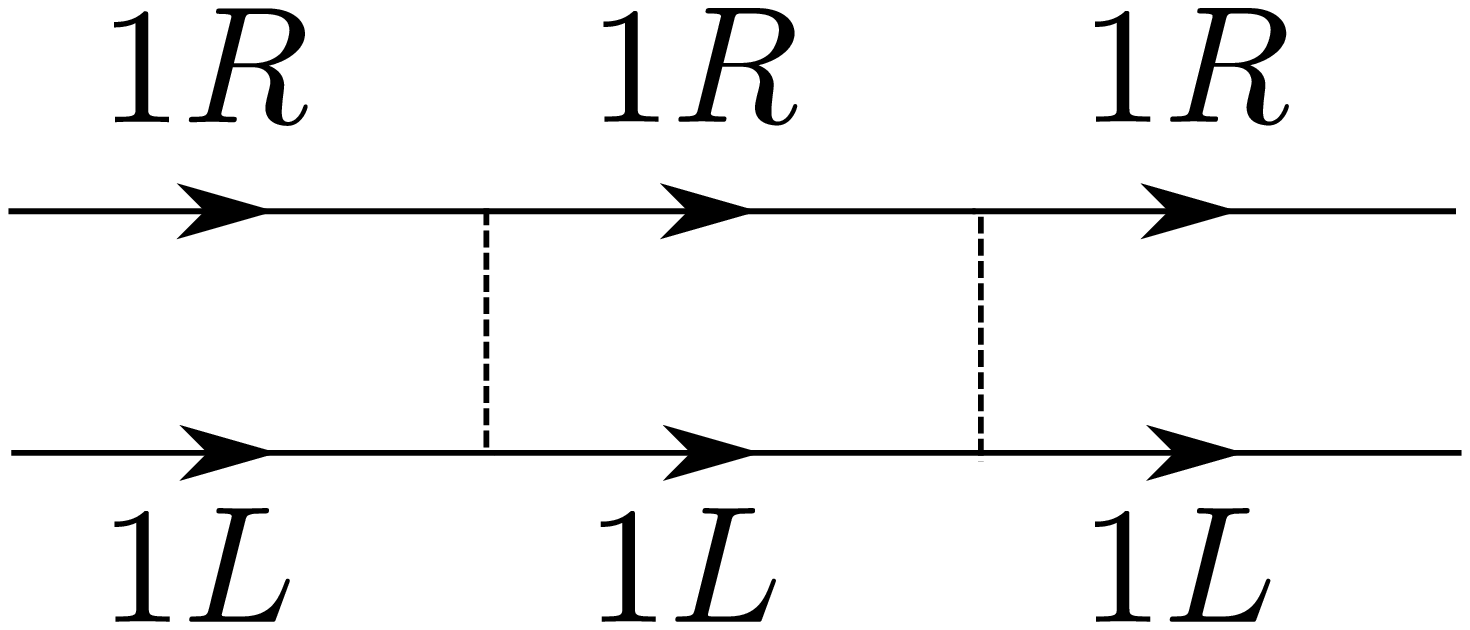}} &
		\fbox{\includegraphics[width = 0.3\textwidth]{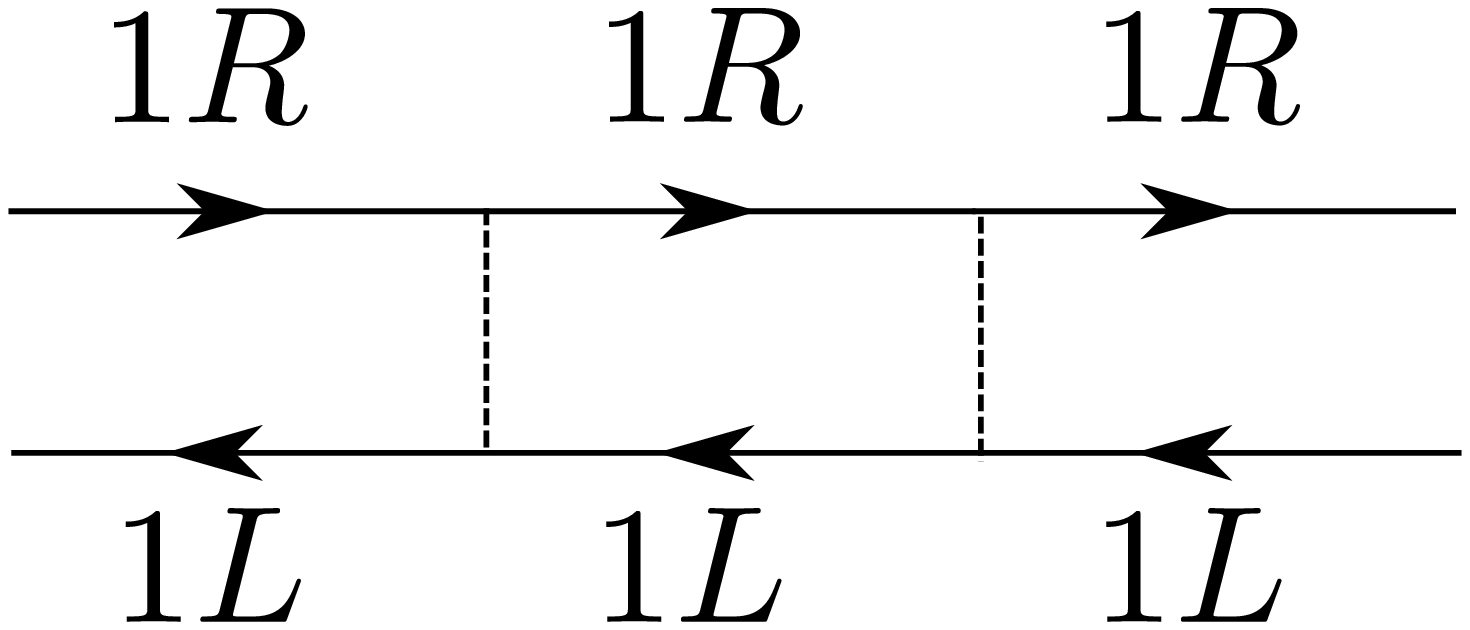}} \\ ~ \\
		\fbox{\includegraphics[width = 0.3\textwidth]{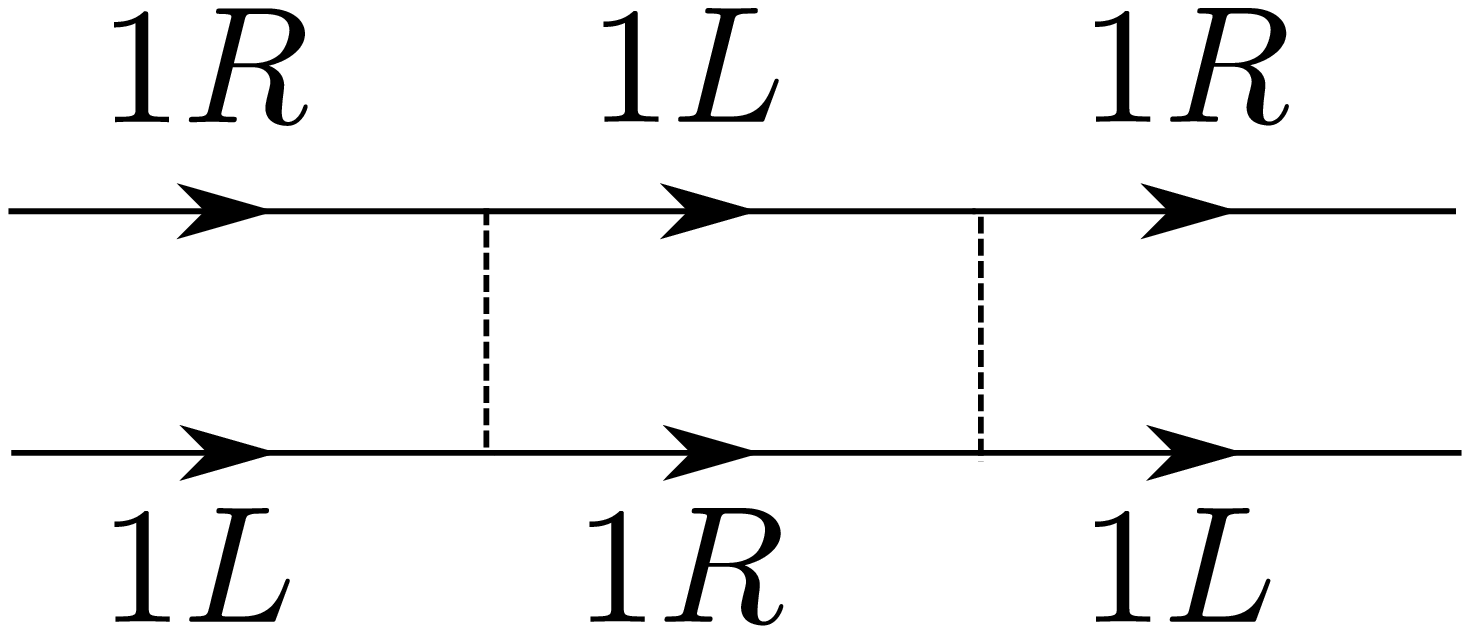}} &
		\fbox{\includegraphics[width = 0.3\textwidth]{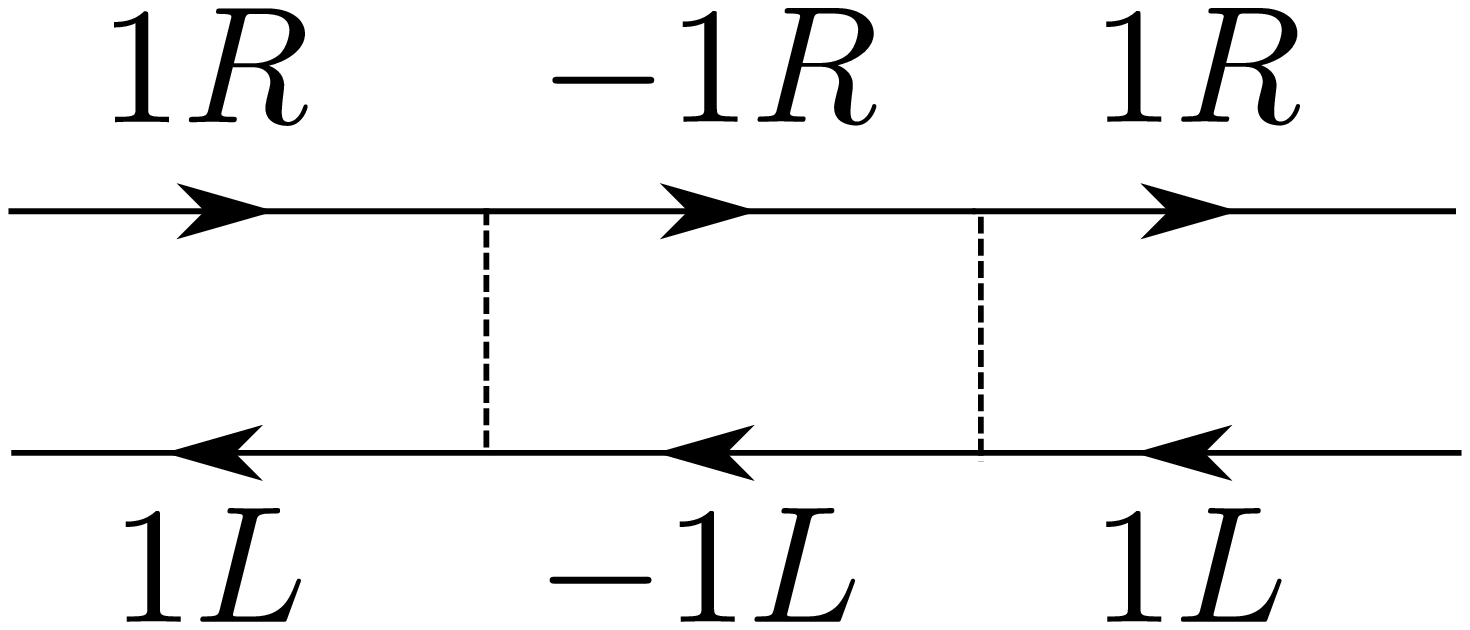}} 
	\end{tabular}
	\caption{Diagrams contributing to the flow of the forward-scattering interaction $U_{11}$. The forward (backward) scattering interactions contribute in the Cooper (electron-hole) channels. In the Cooper channel the total angular momentum is $L = 2$ in the loop, while $L = 0$ in the particle-hole channel.}
	\label{U1}
\end{figure}

\begin{figure}
	\begin{tabular}{ccc}
		\fbox{\includegraphics[width = 0.3\textwidth]{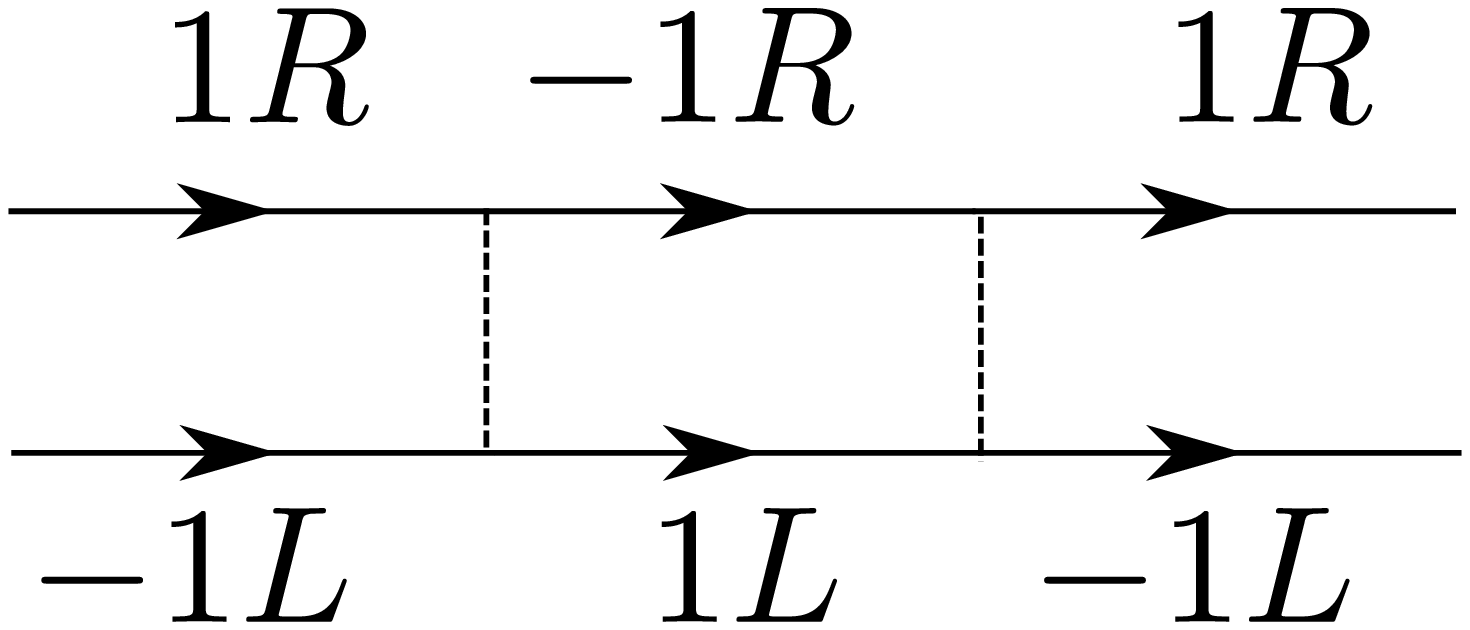}} &
		\fbox{\includegraphics[width = 0.3\textwidth]{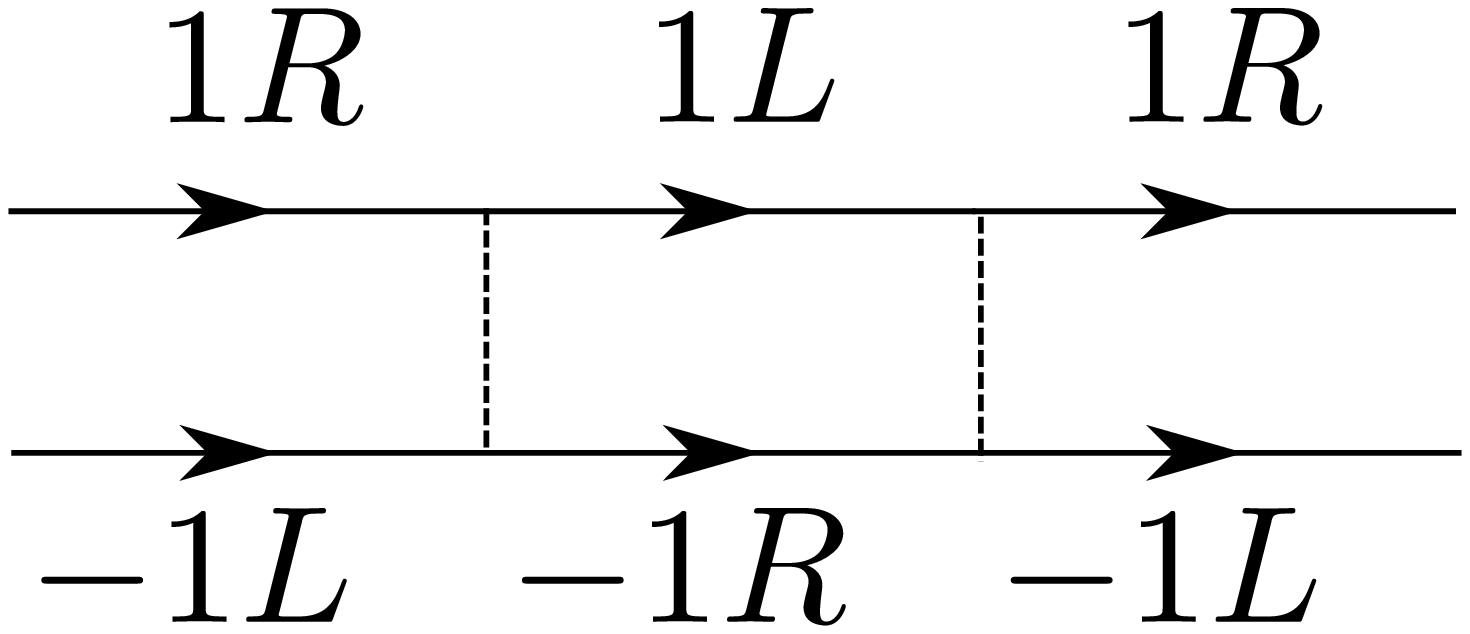}} &
		\fbox{\includegraphics[width = 0.3\textwidth]{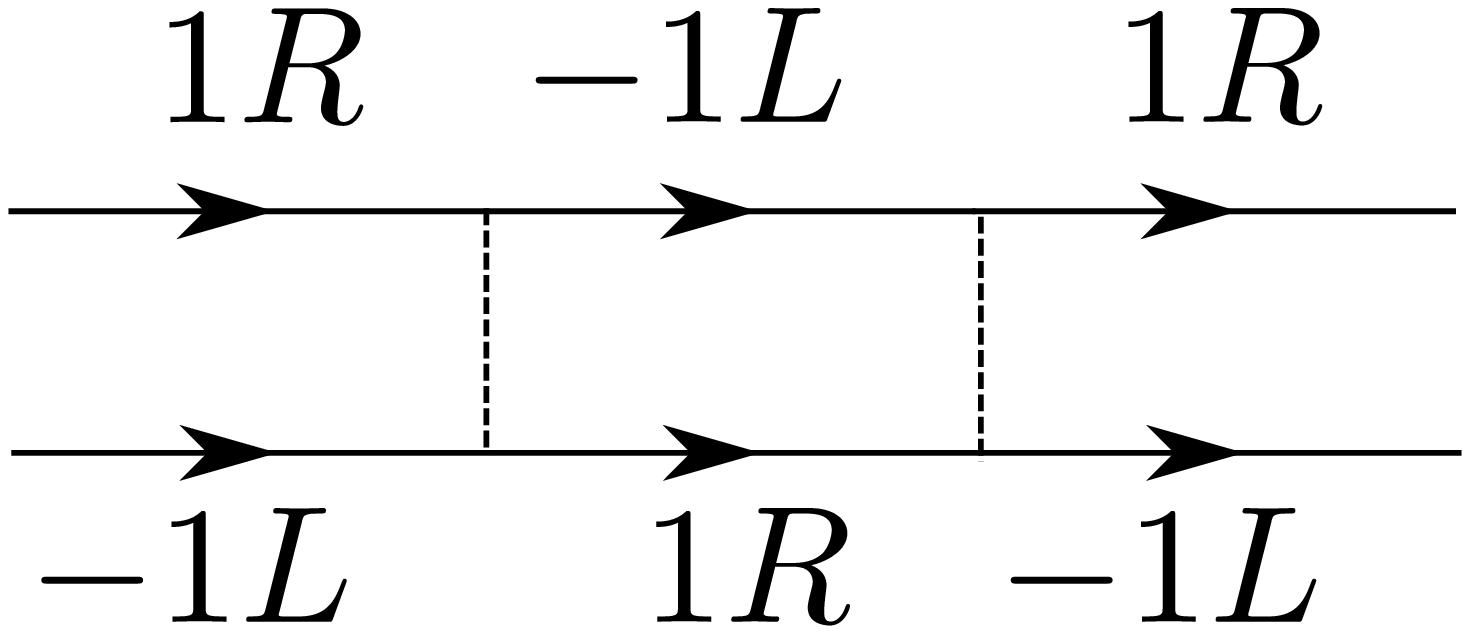}}
	\end{tabular}
	\caption{Diagrams contributing to the flow of the forward-scattering interaction $U_{1,-1}$. In the intermediate states $L = 0$.}
	\label{U2}
\end{figure}

\begin{figure}[b]
	\begin{tabular}{ccc}
		\fbox{\includegraphics[width = 0.3\textwidth]{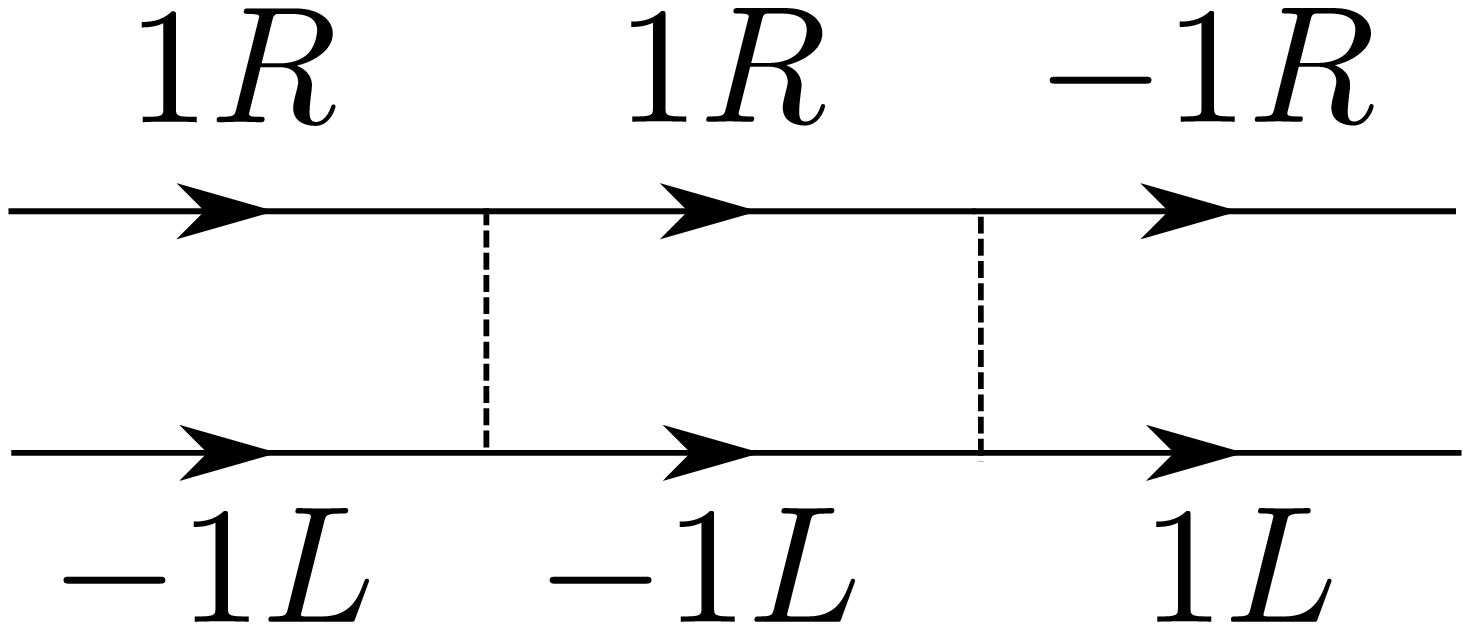}} &
		\fbox{\includegraphics[width = 0.3\textwidth]{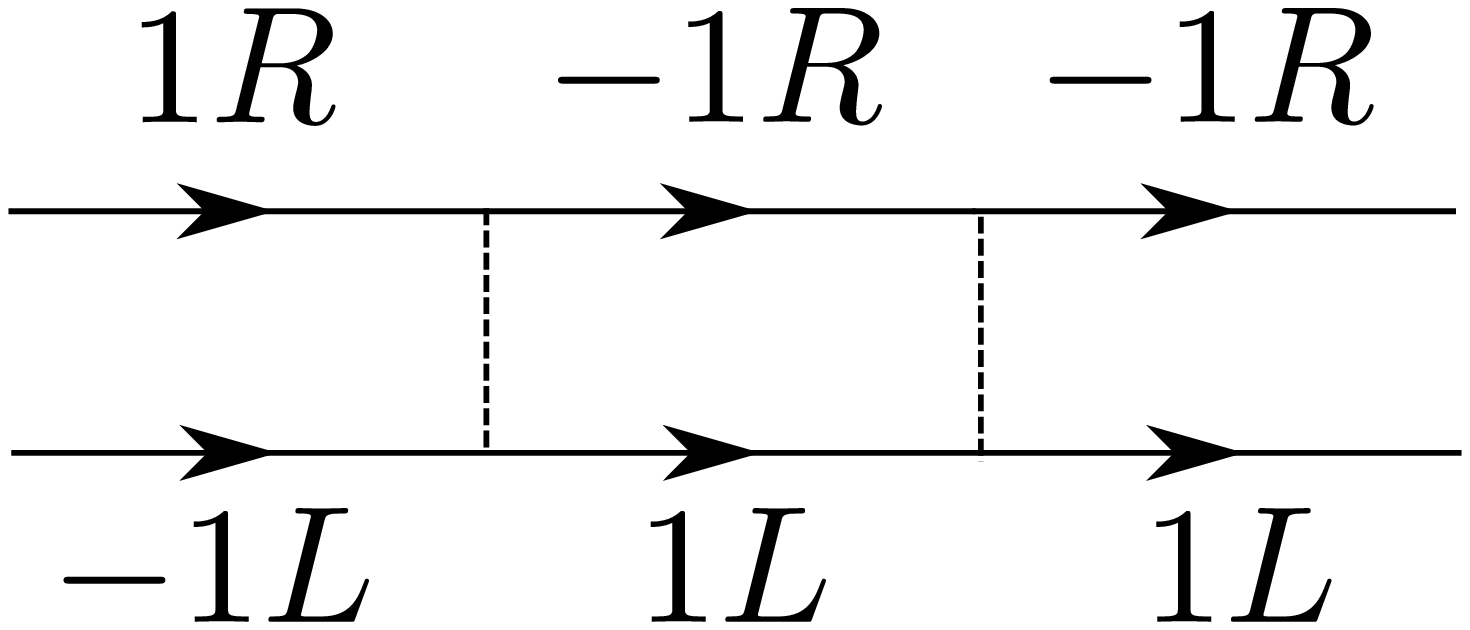}}  &
		\fbox{\includegraphics[width = 0.3\textwidth]{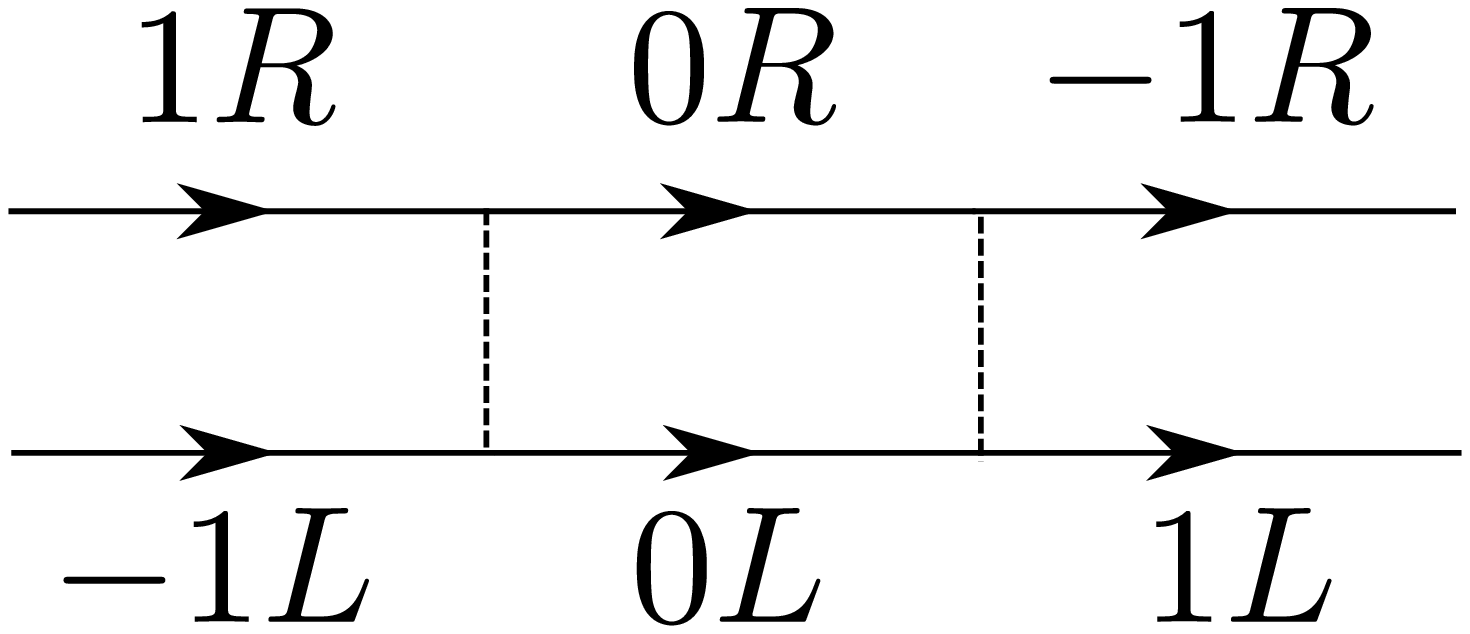}} \\ ~ \\
		\fbox{\includegraphics[width = 0.3\textwidth]{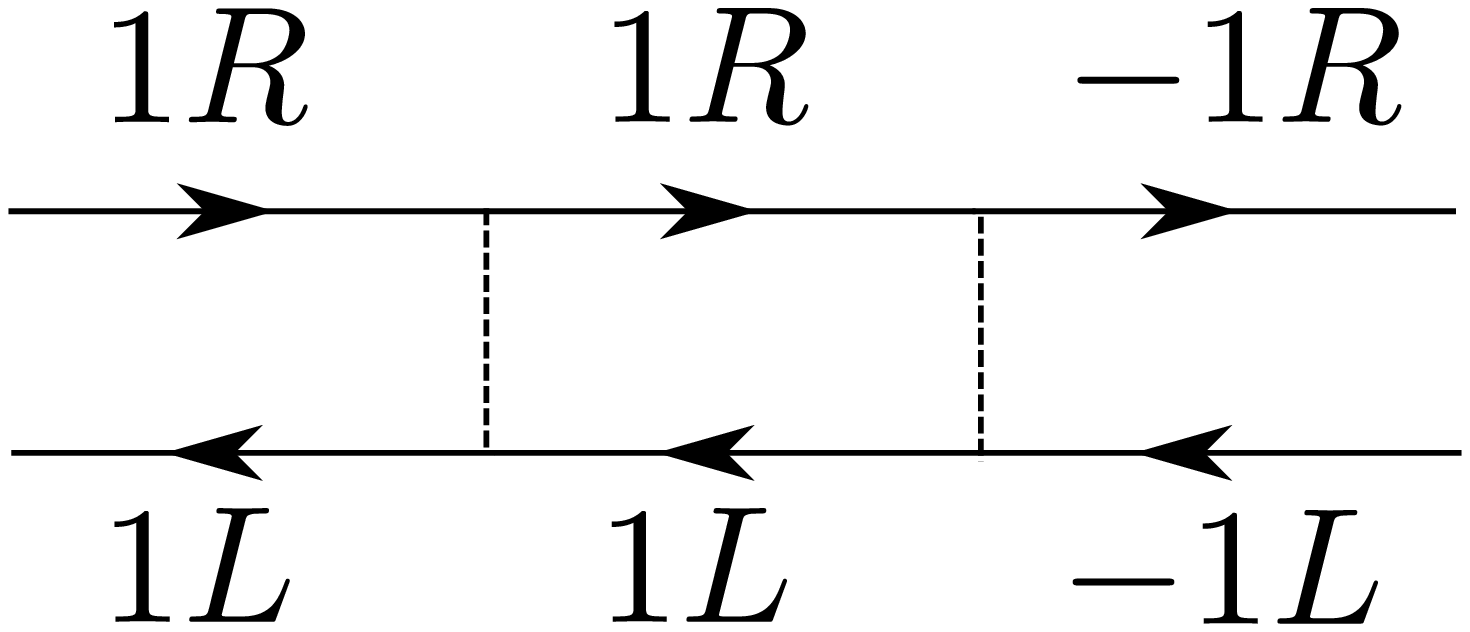}} & 
		\fbox{\includegraphics[width = 0.3\textwidth]{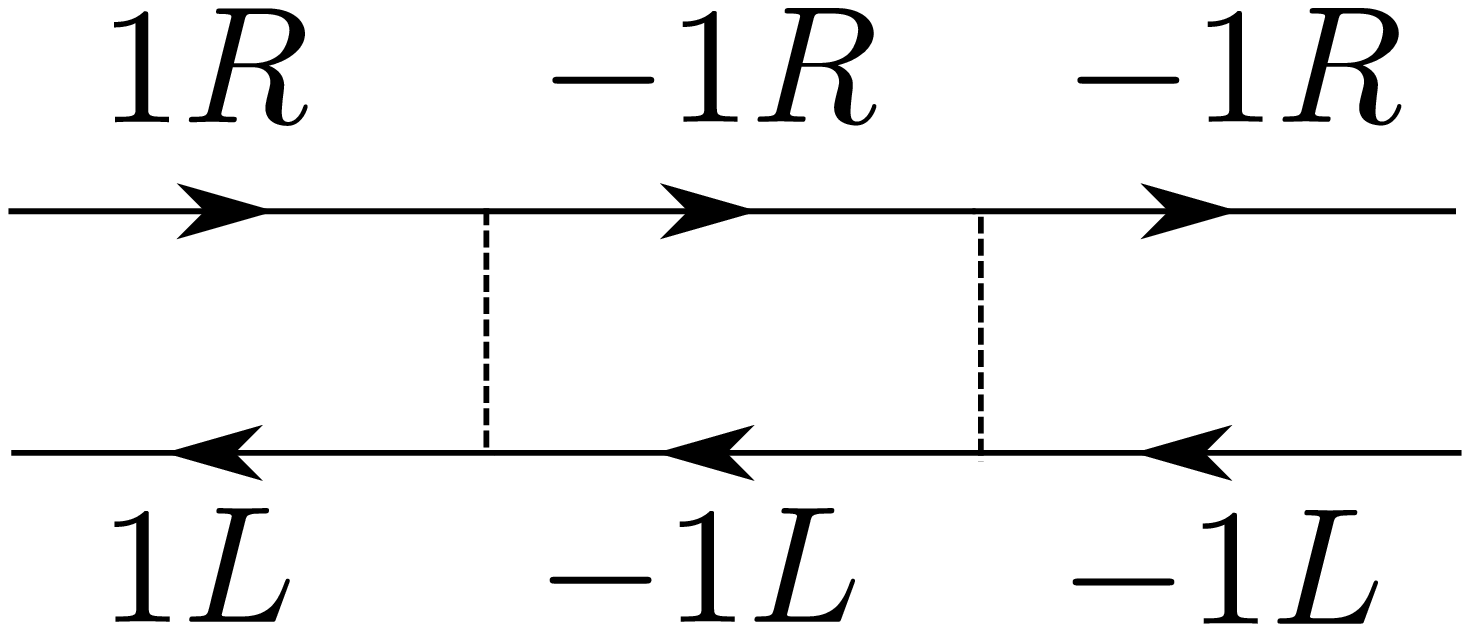}}  & \\ ~ \\
		\fbox{\includegraphics[width = 0.3\textwidth]{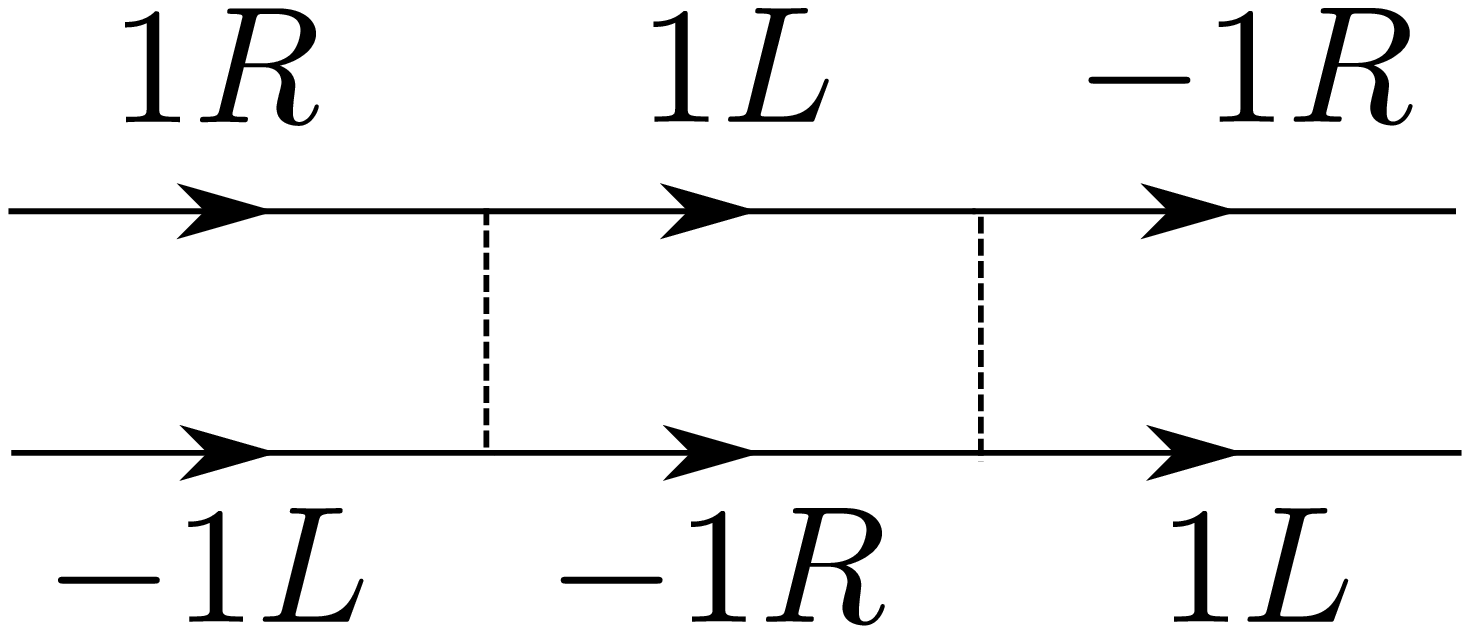}} &
		\fbox{\includegraphics[width = 0.3\textwidth]{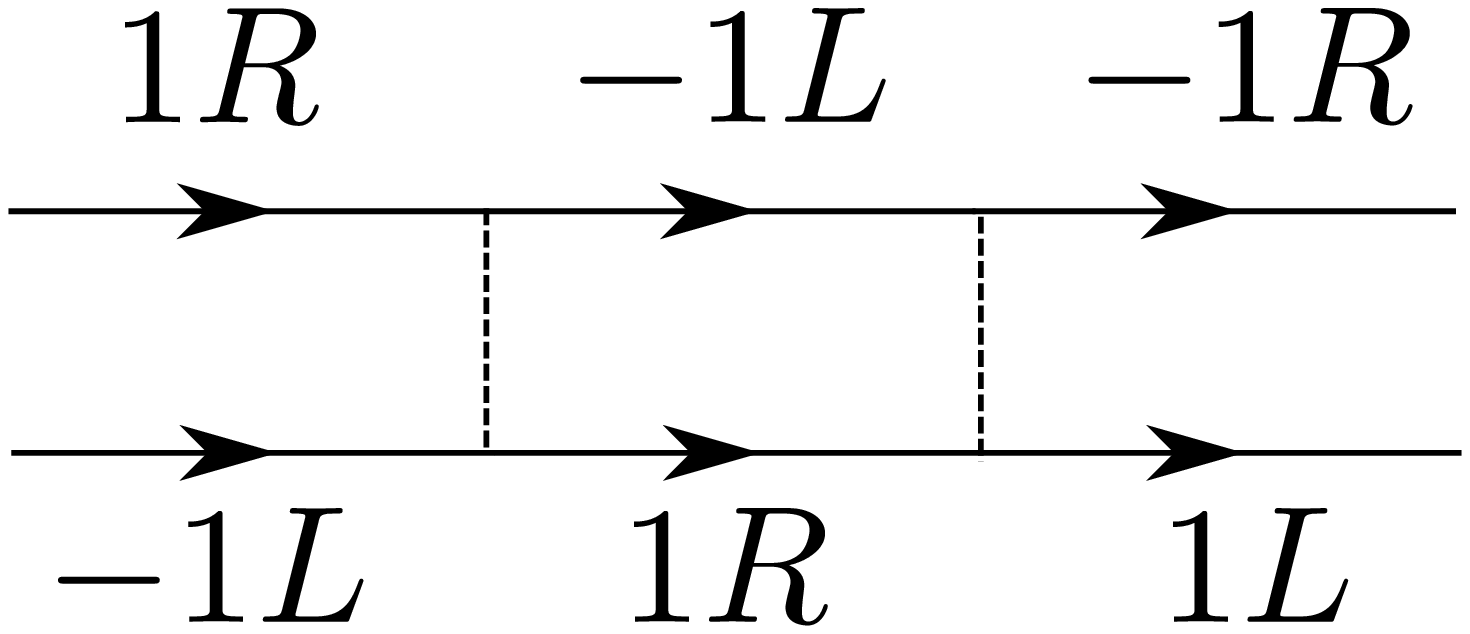}}  &
		\fbox{\includegraphics[width = 0.3\textwidth]{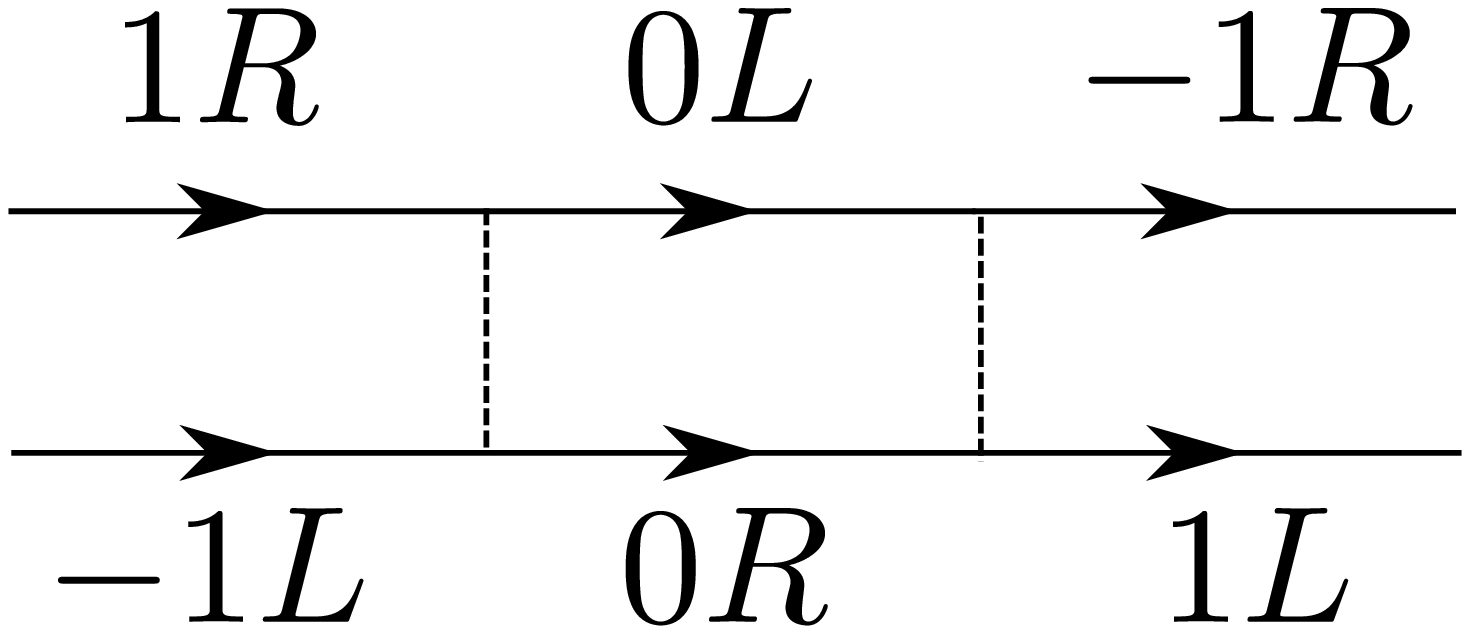}}  
	\end{tabular}
	\caption{Diagrams contributing to the flow of the forward-scattering interaction $V_2$. The first two lines show contributions from virtual forward-scattering, while the last line shows contributions from virtual backscattering. $L = 0$ in all intermediate states. }
	\label{U3}
\end{figure}

\begin{figure}
	\begin{tabular}{ccc}
		\fbox{\includegraphics[width = 0.3\textwidth]{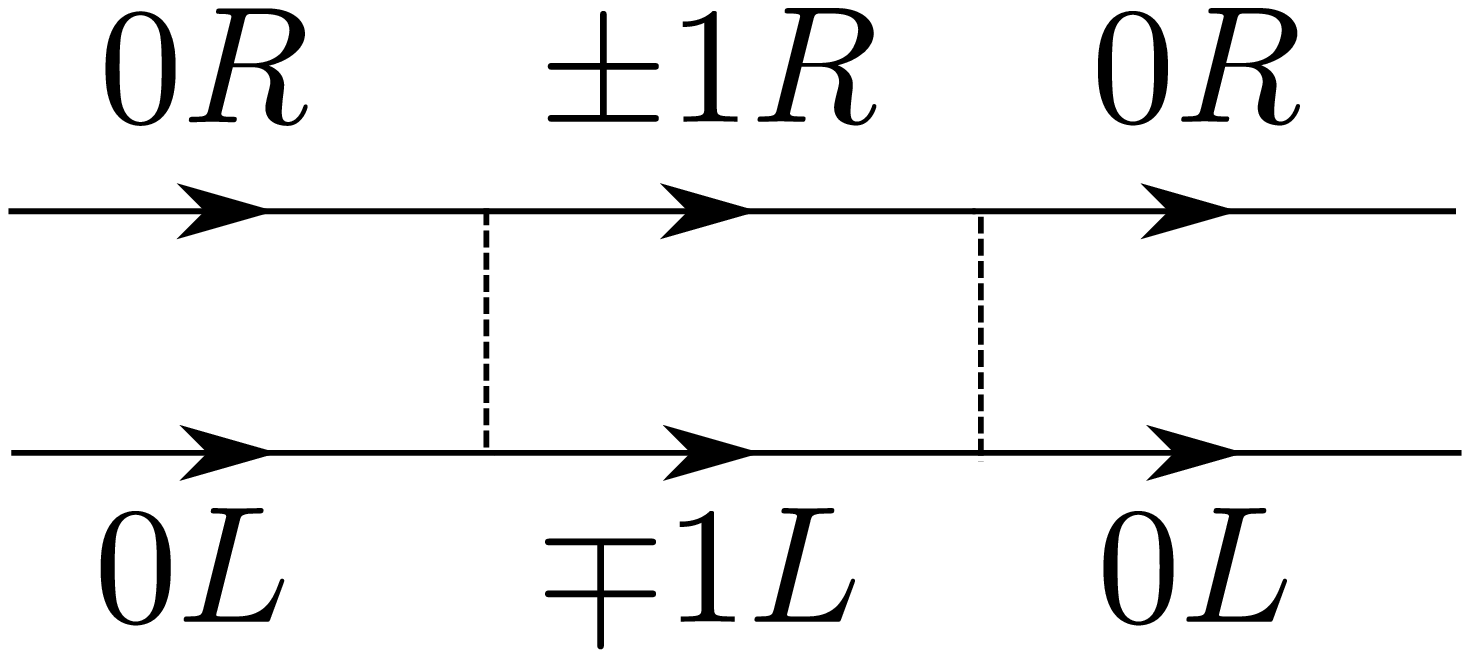}} &
		\fbox{\includegraphics[width = 0.3\textwidth]{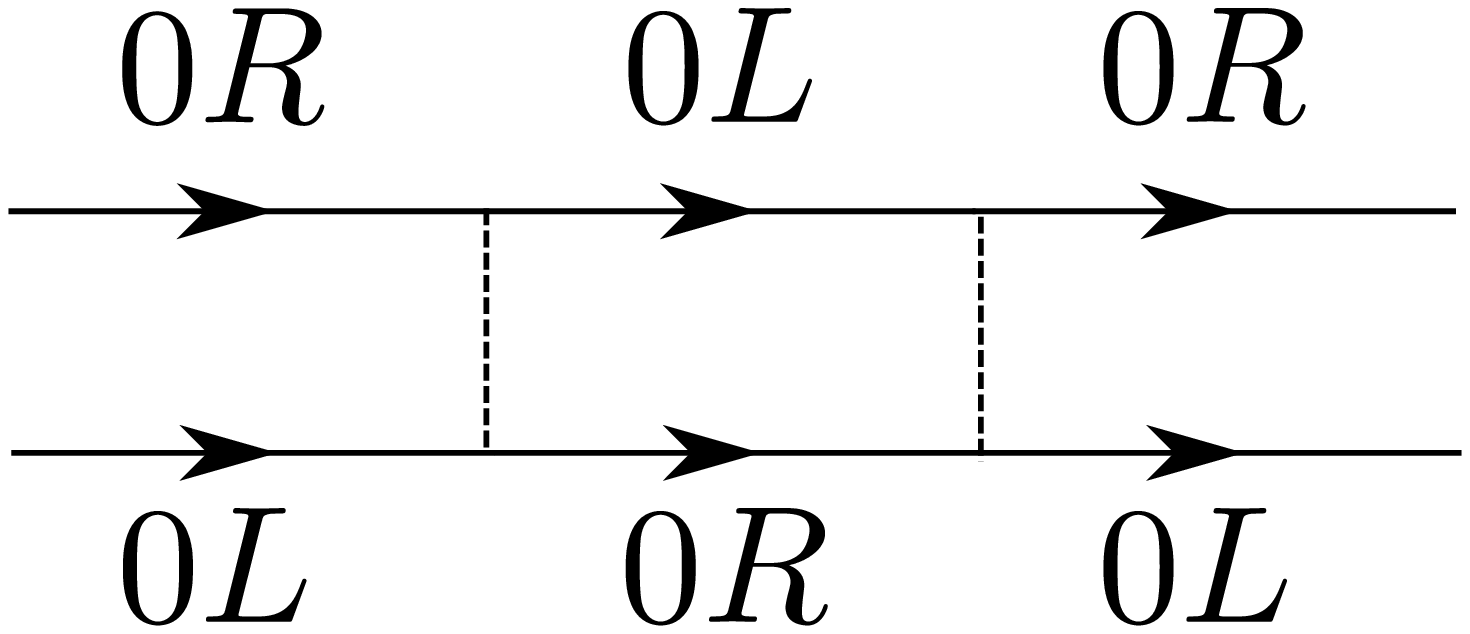}}
		\fbox{\includegraphics[width = 0.3\textwidth]{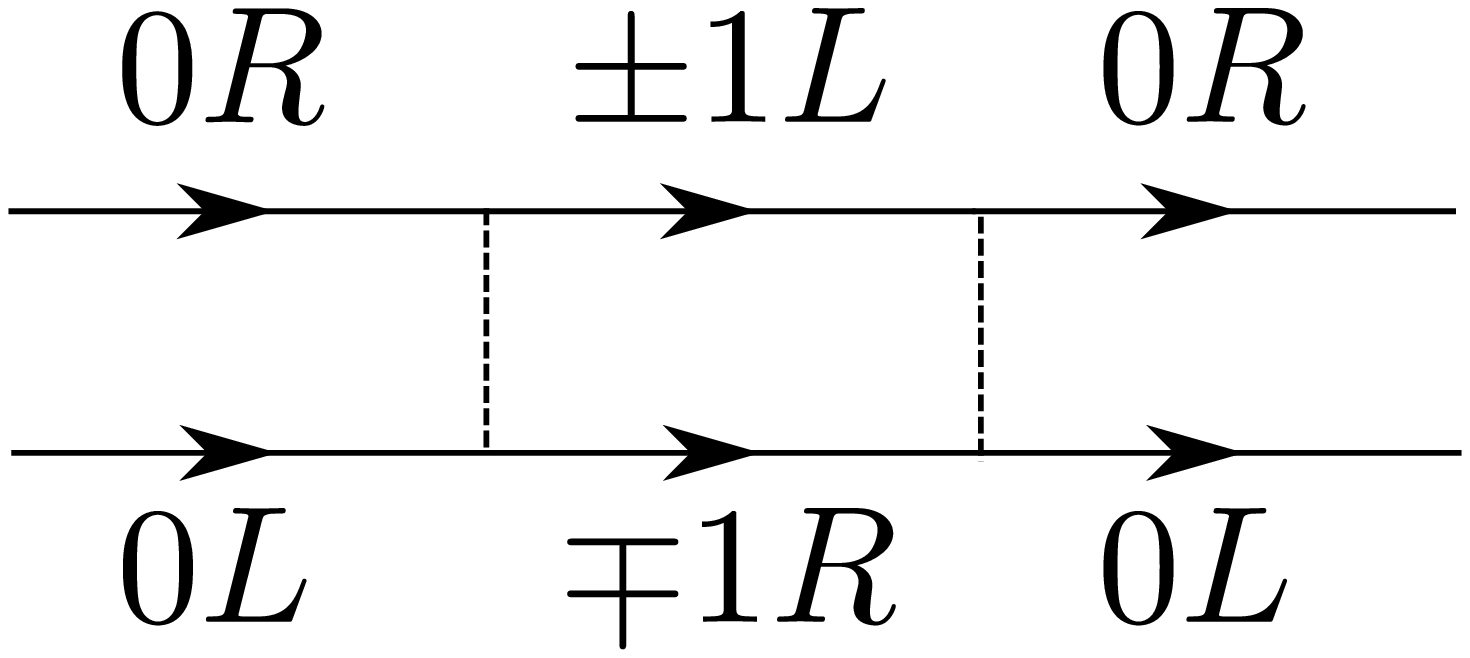}} 
	\end{tabular}
	\caption{Diagrams contributing to the flow of the interaction $U_{00}$. In the intermediate states $L = 0$.}
	\label{U4}
\end{figure}

\begin{figure}
	\begin{tabular}{cc}
		\fbox{\includegraphics[width = 0.3\textwidth]{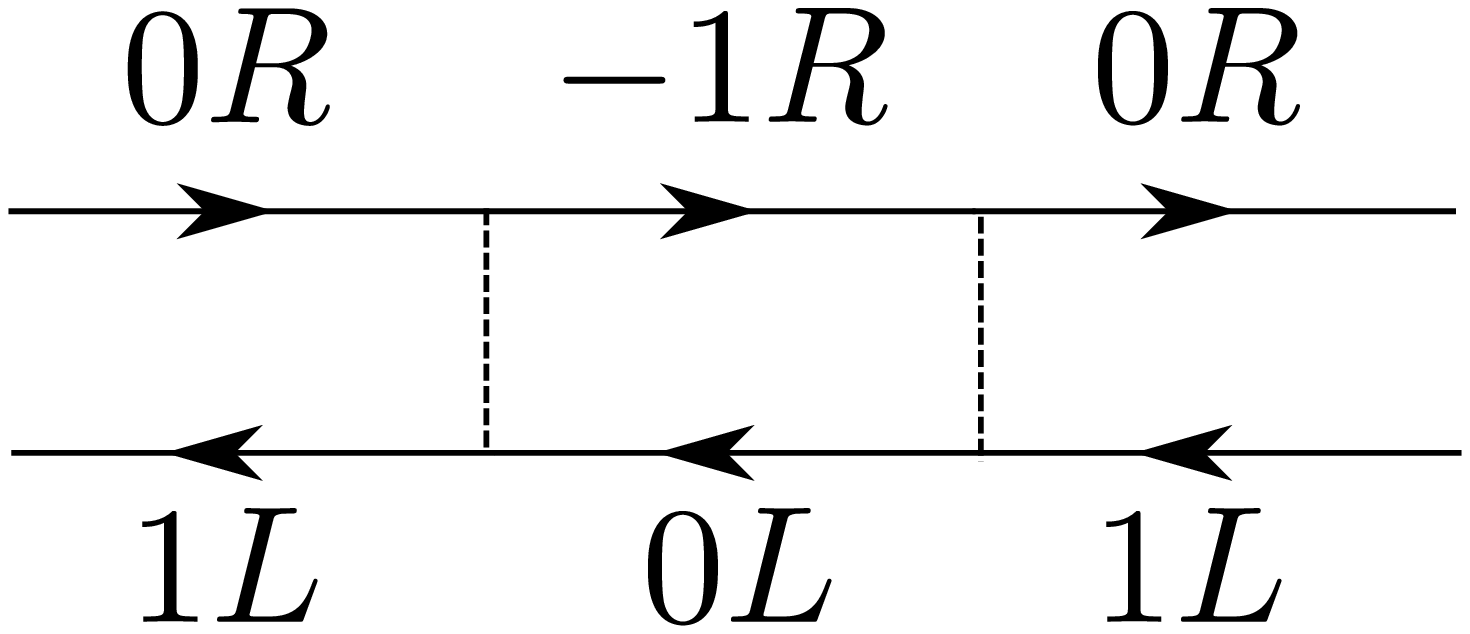}} &
		\fbox{\includegraphics[width = 0.3\textwidth]{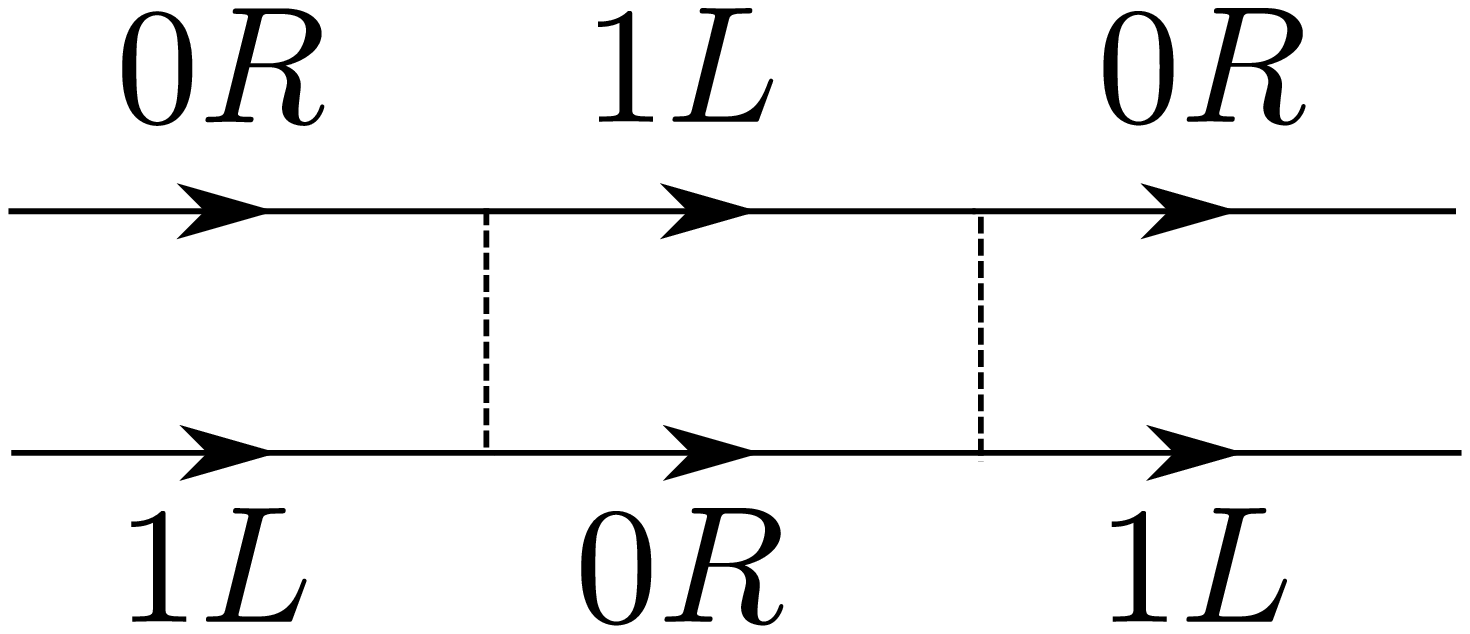}}
	\end{tabular}
	\caption{Diagrams contributing to the flow of the interaction $U_{01}$. }
	\label{U5}
\end{figure}

\begin{figure}
	\begin{tabular}{ccc}
		\fbox{\includegraphics[width = 0.3\textwidth]{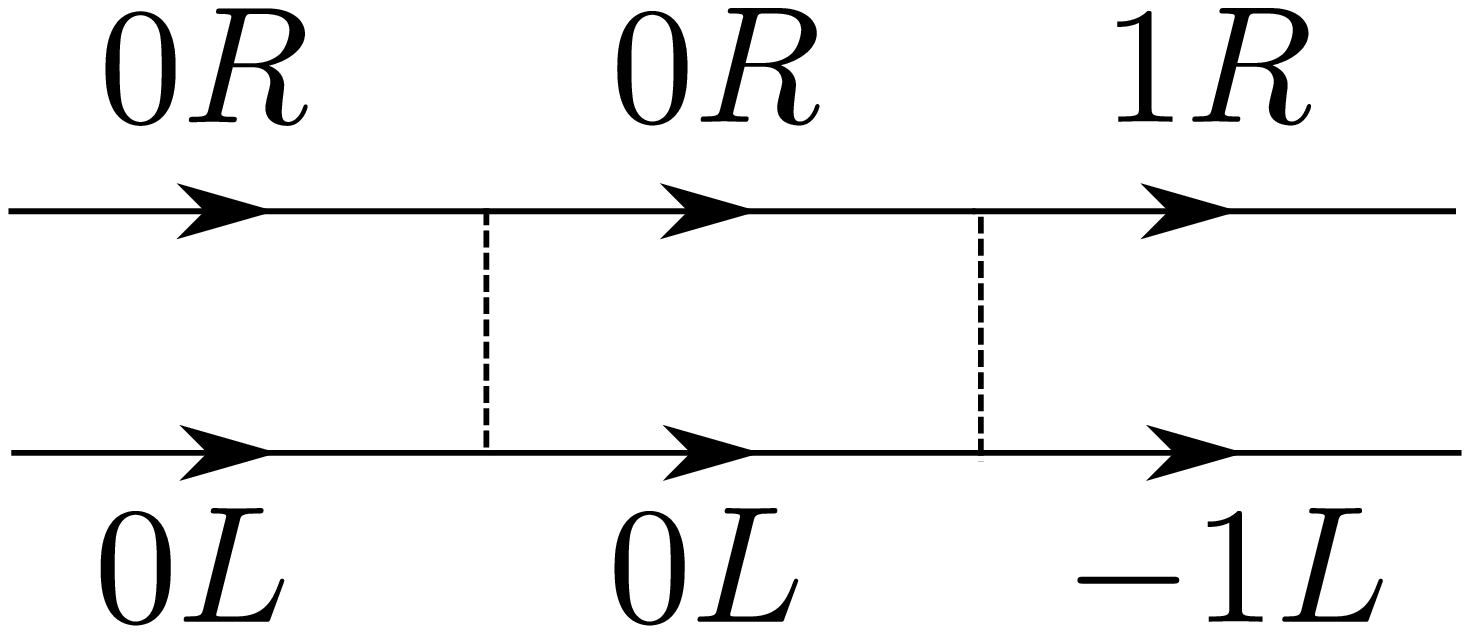}} &
		\fbox{\includegraphics[width = 0.3\textwidth]{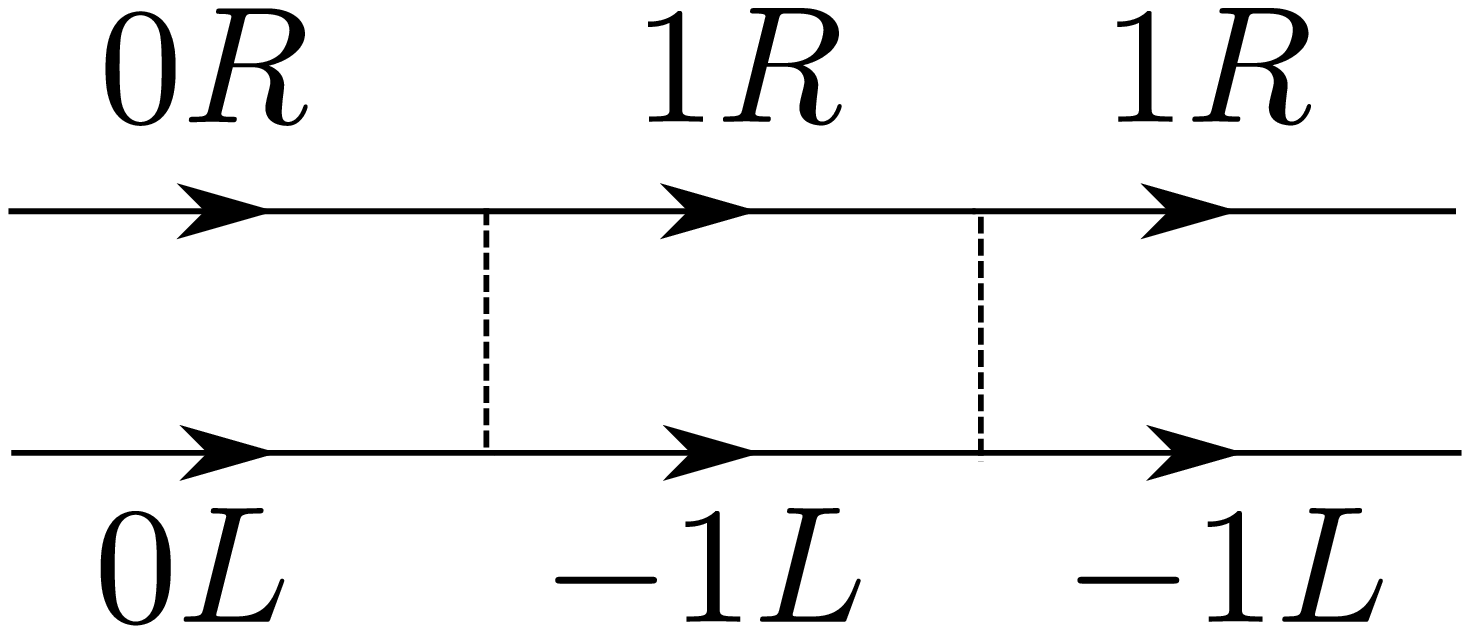}}  &
		\fbox{\includegraphics[width = 0.3\textwidth]{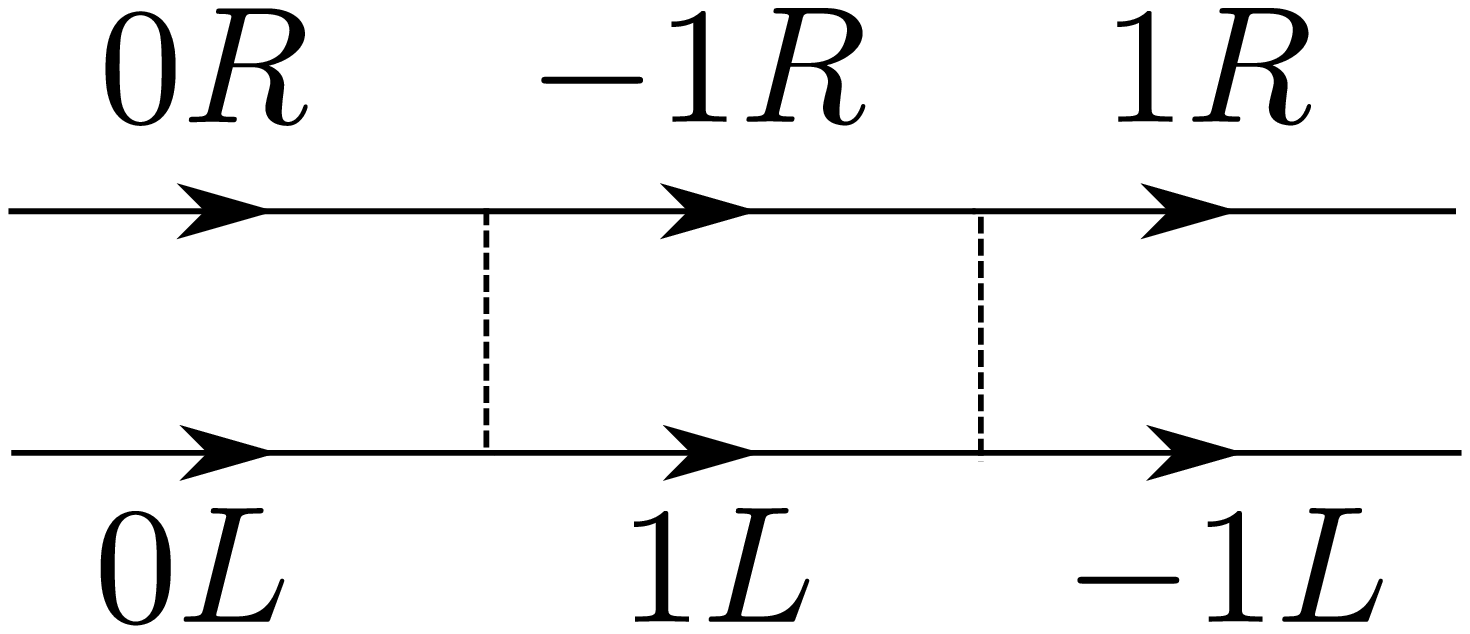}} \\ ~ \\
		\fbox{\includegraphics[width = 0.3\textwidth]{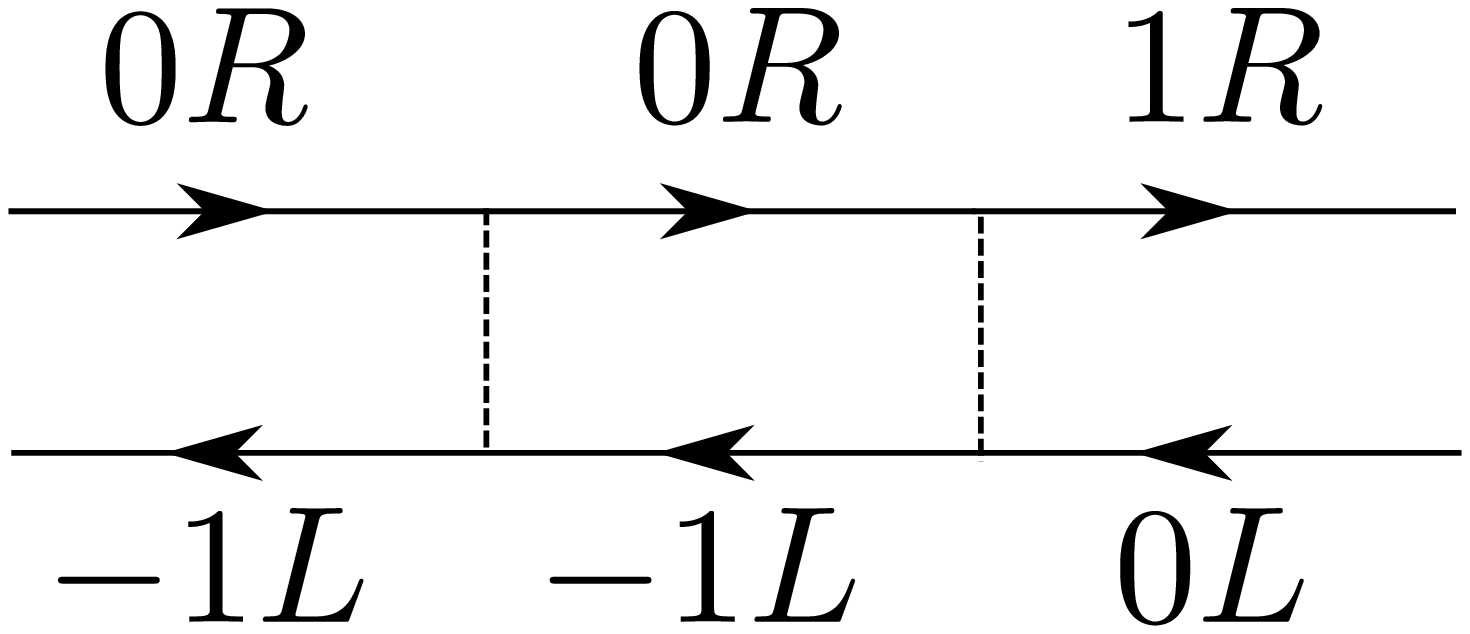}} &
		\fbox{\includegraphics[width = 0.3\textwidth]{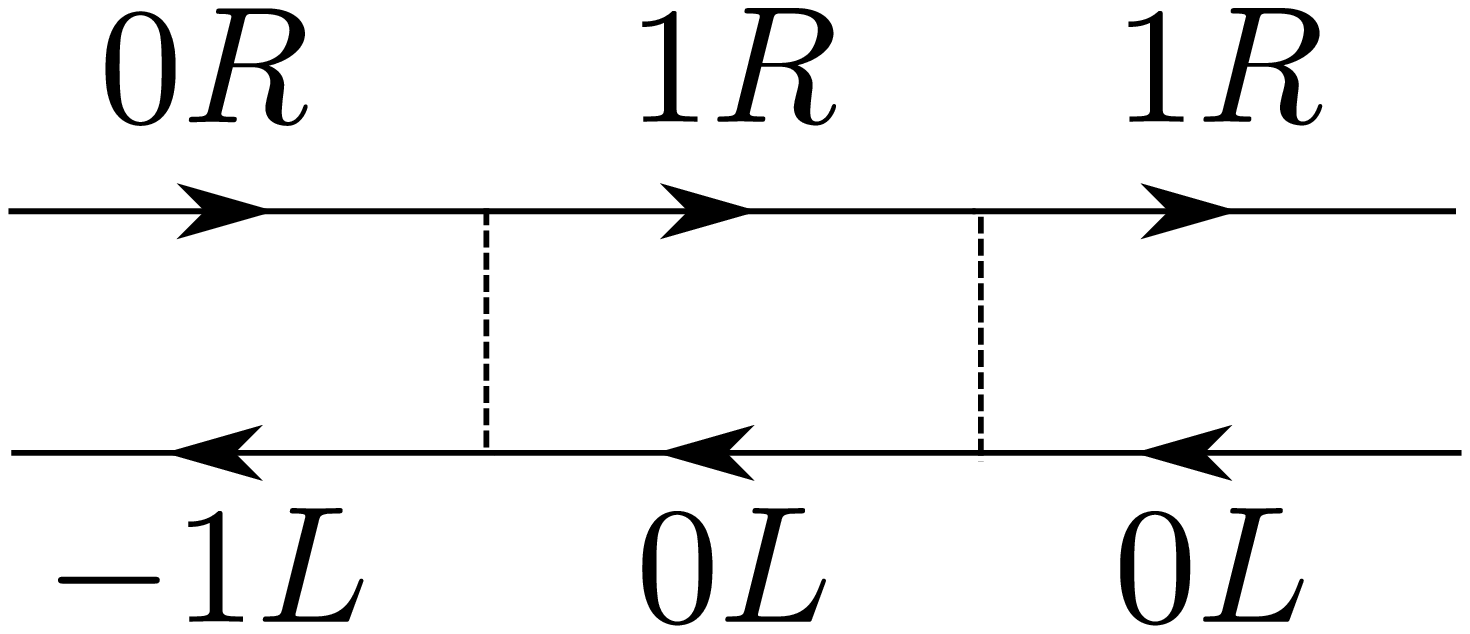}} & \\ ~ \\
		\fbox{\includegraphics[width = 0.3\textwidth]{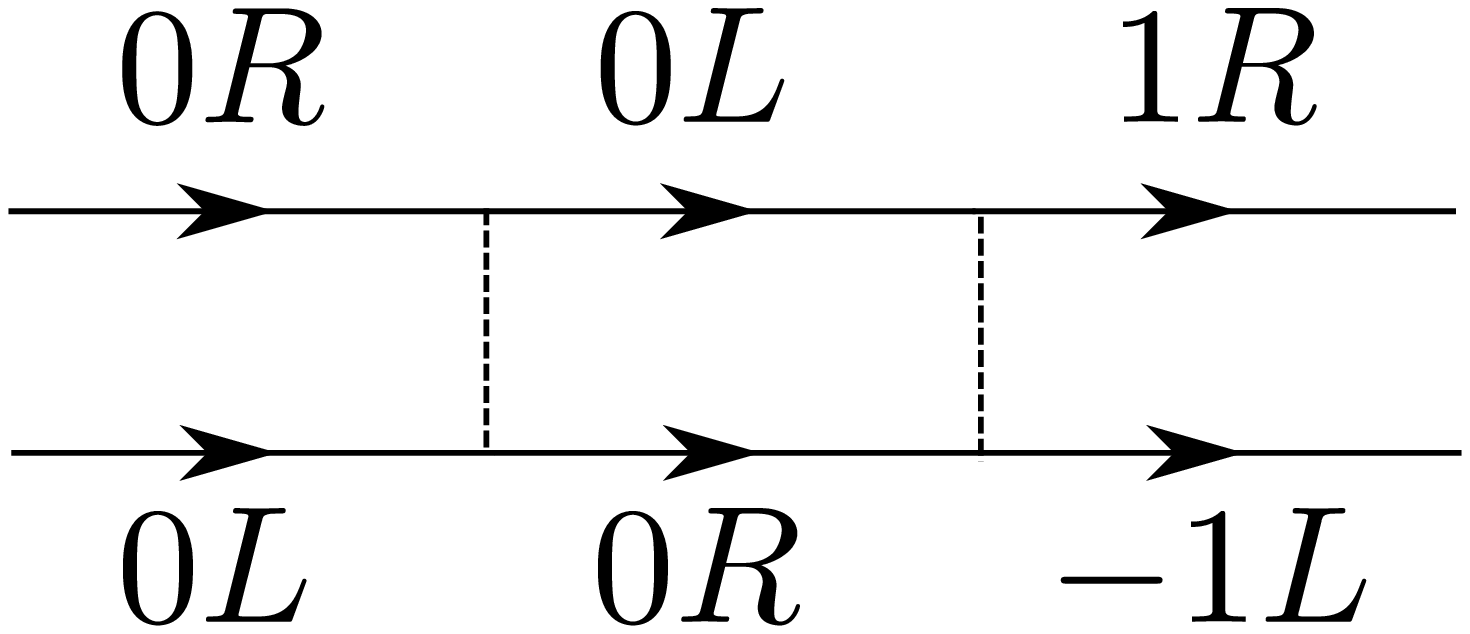}} &
		\fbox{\includegraphics[width = 0.3\textwidth]{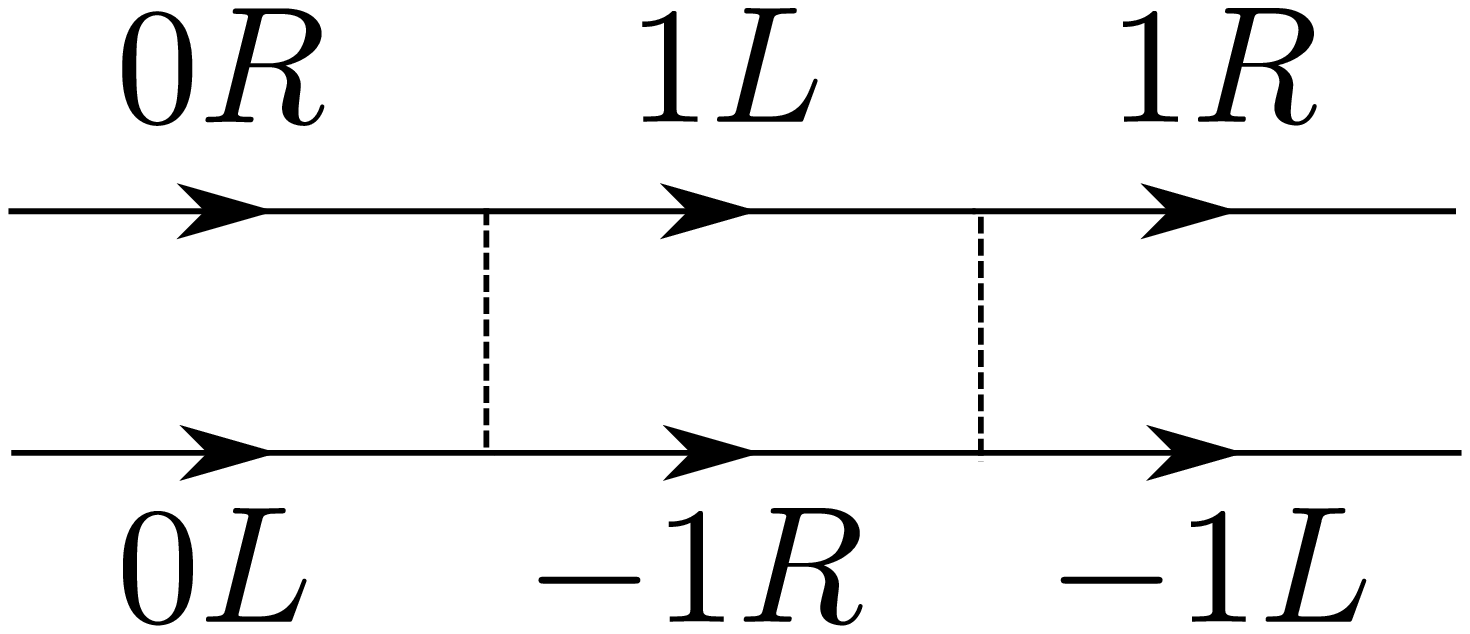}}  &
		\fbox{\includegraphics[width = 0.3\textwidth]{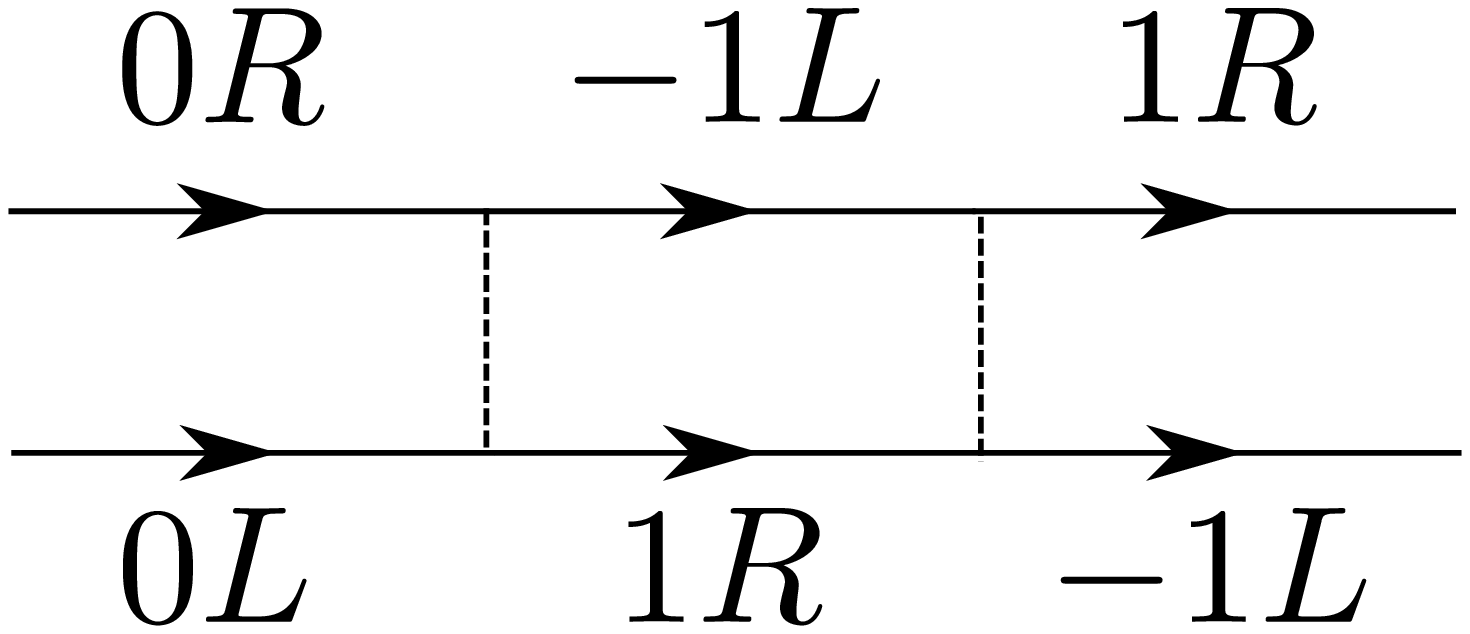}}
	\end{tabular}
	\caption{Diagrams contributing to the flow of the interaction $V_1$. The upper two lines show the contributions from virtual forward scattering, while the lowest line shows the contributions from virtual backscattering. In the Cooper channels, $L = 0$ in the intermediate states, while $L = 1$ in the intermediate states for the electron-hole channels.}
	\label{U6}
\end{figure}

\begin{figure}
	\begin{tabular}{cc}
		\fbox{\includegraphics[width = 0.3\textwidth]{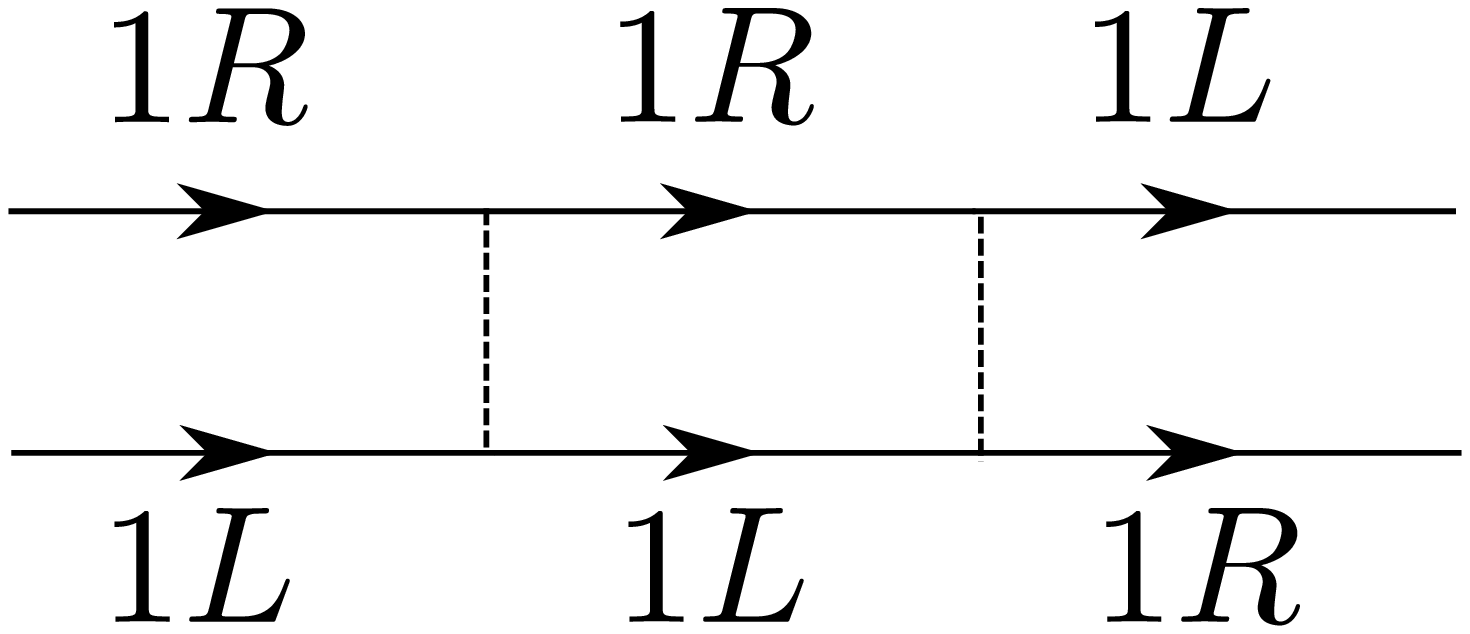}} &
		\fbox{\includegraphics[width = 0.3\textwidth]{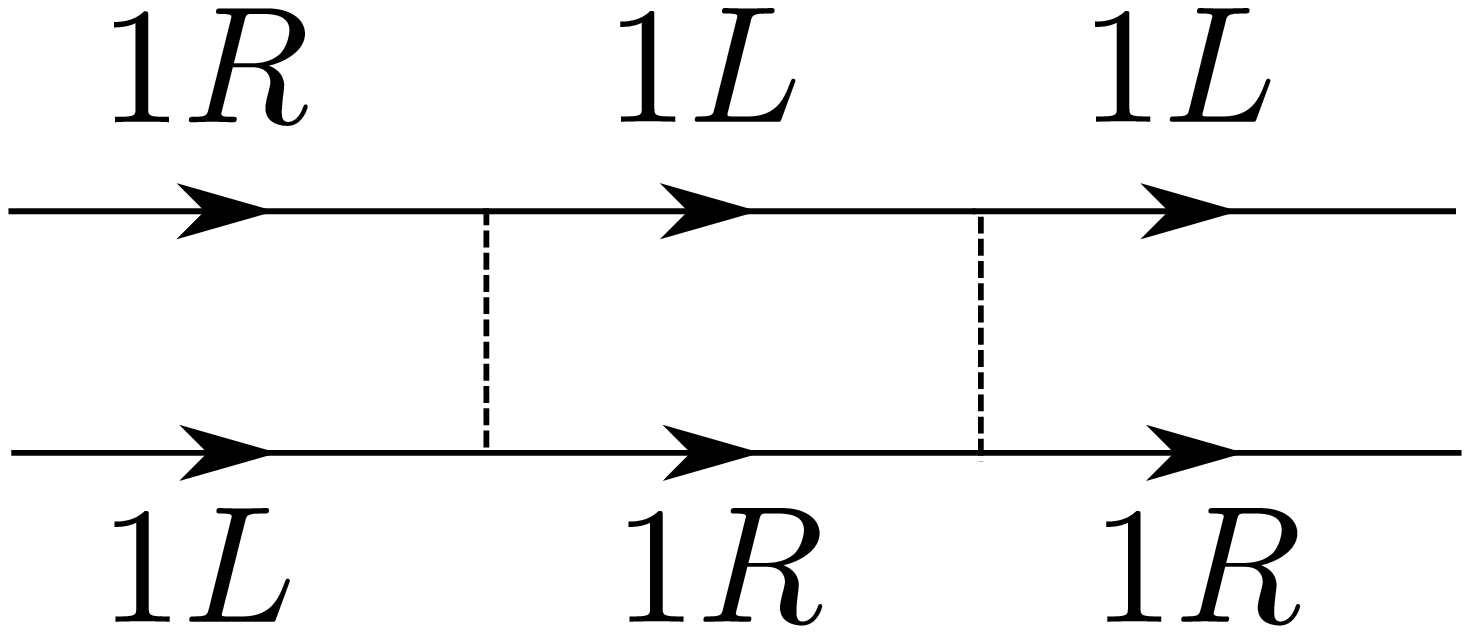}}  
		\\ ~ \\
		\fbox{\includegraphics[width = 0.3\textwidth]{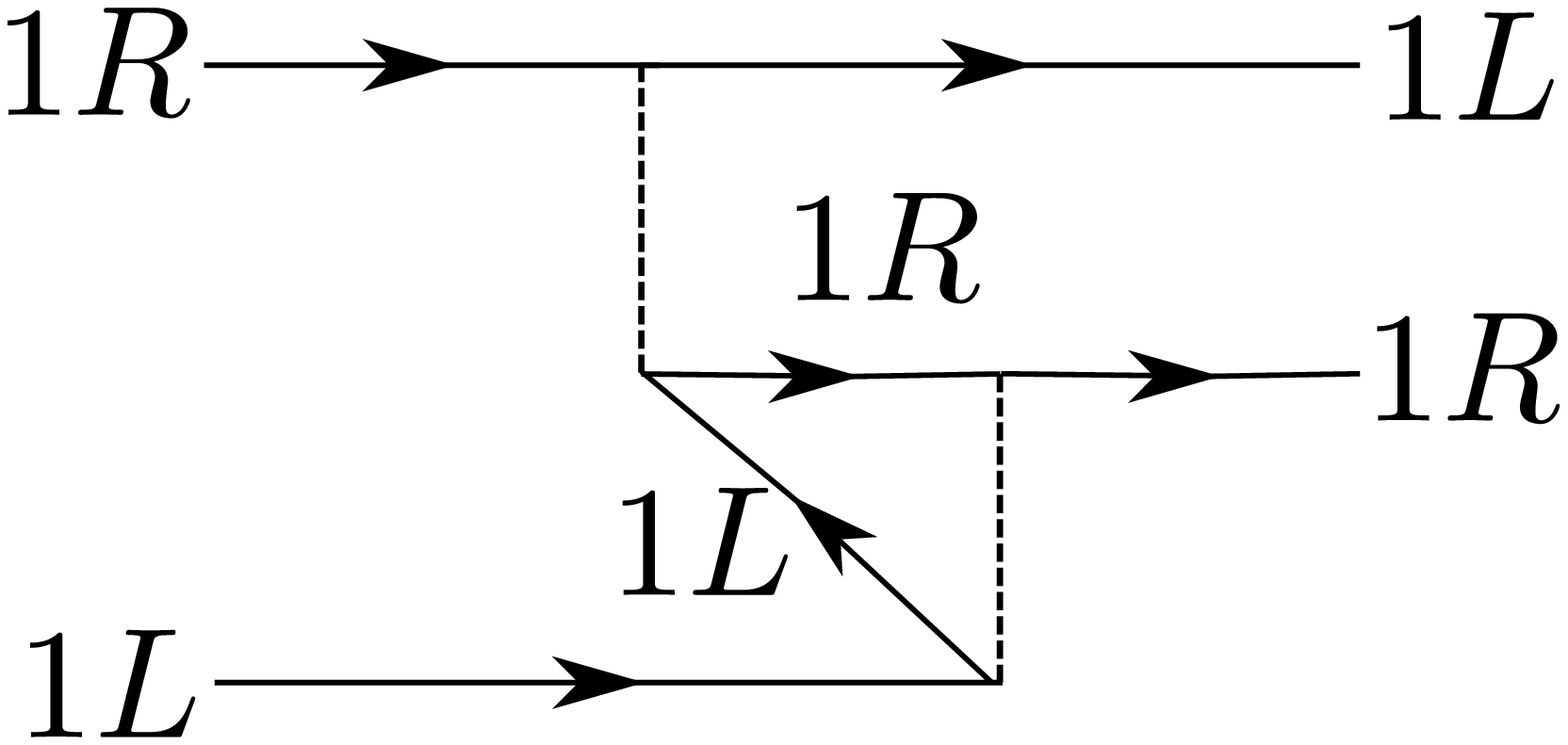}} &
		\fbox{\includegraphics[width = 0.3\textwidth]{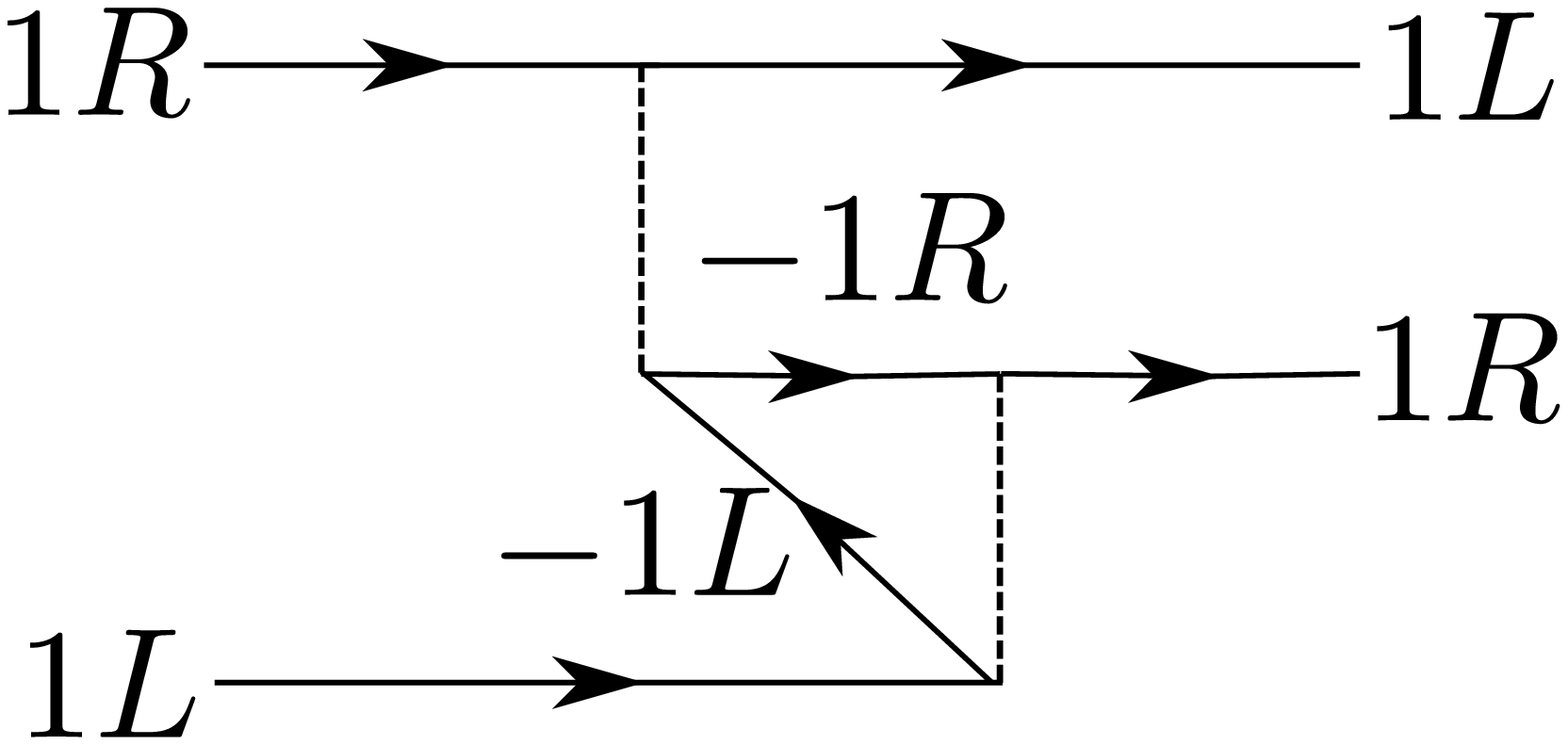}} \\ ~ \\
		\fbox{\includegraphics[width = 0.3\textwidth]{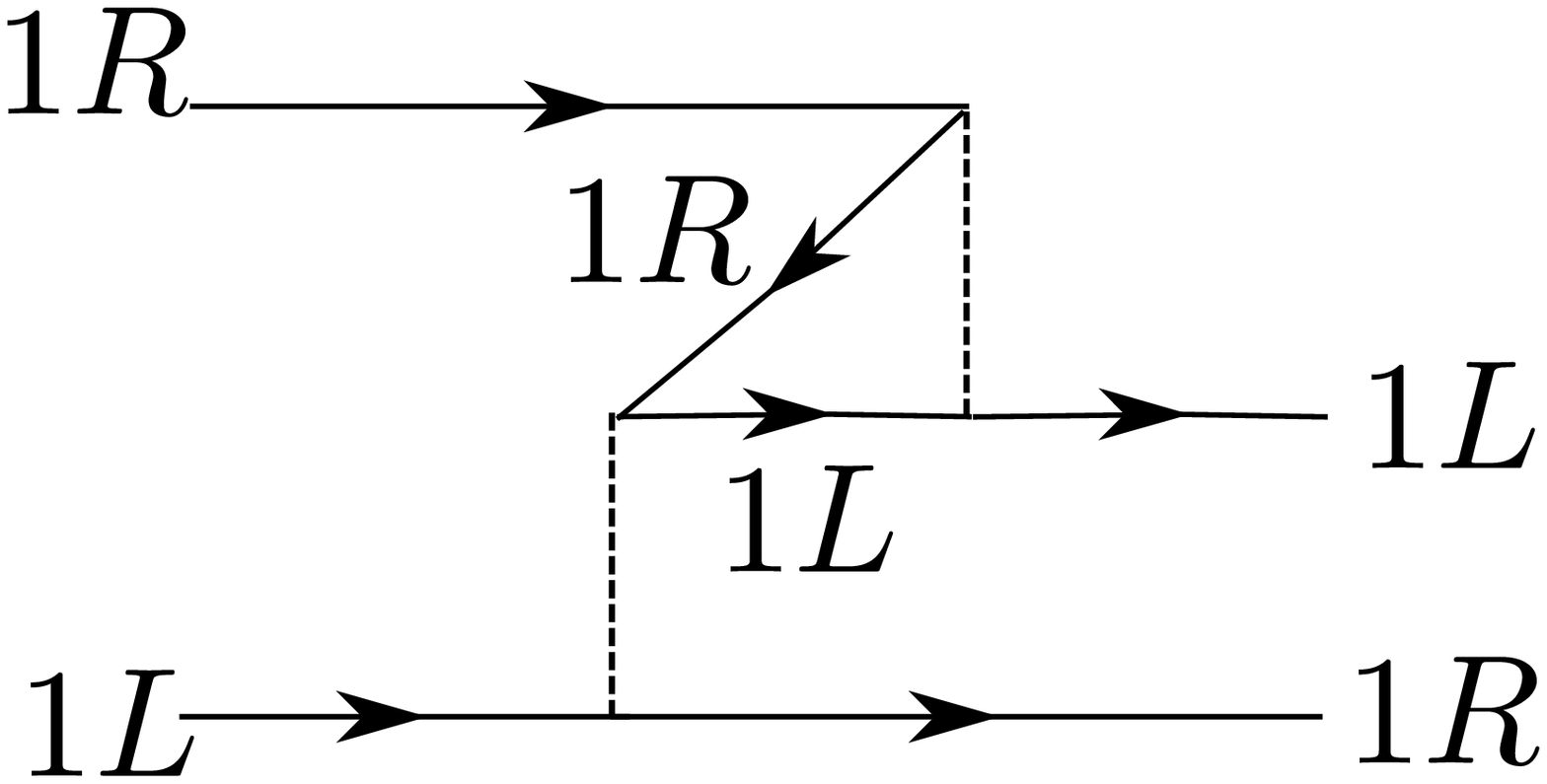}} &
		\fbox{\includegraphics[width = 0.3\textwidth]{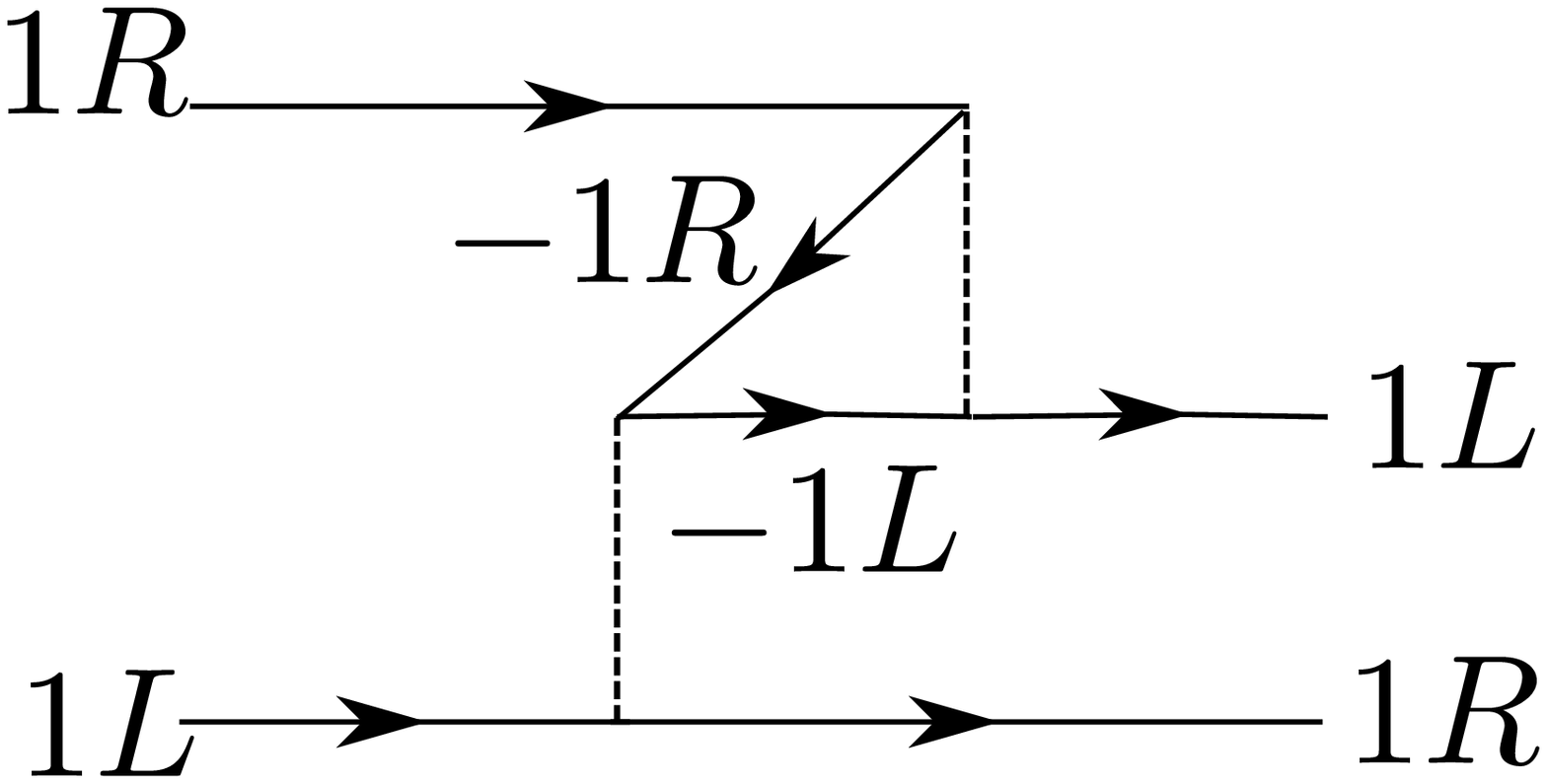}} \\ ~ \\
		\fbox{\includegraphics[width = 0.3\textwidth]{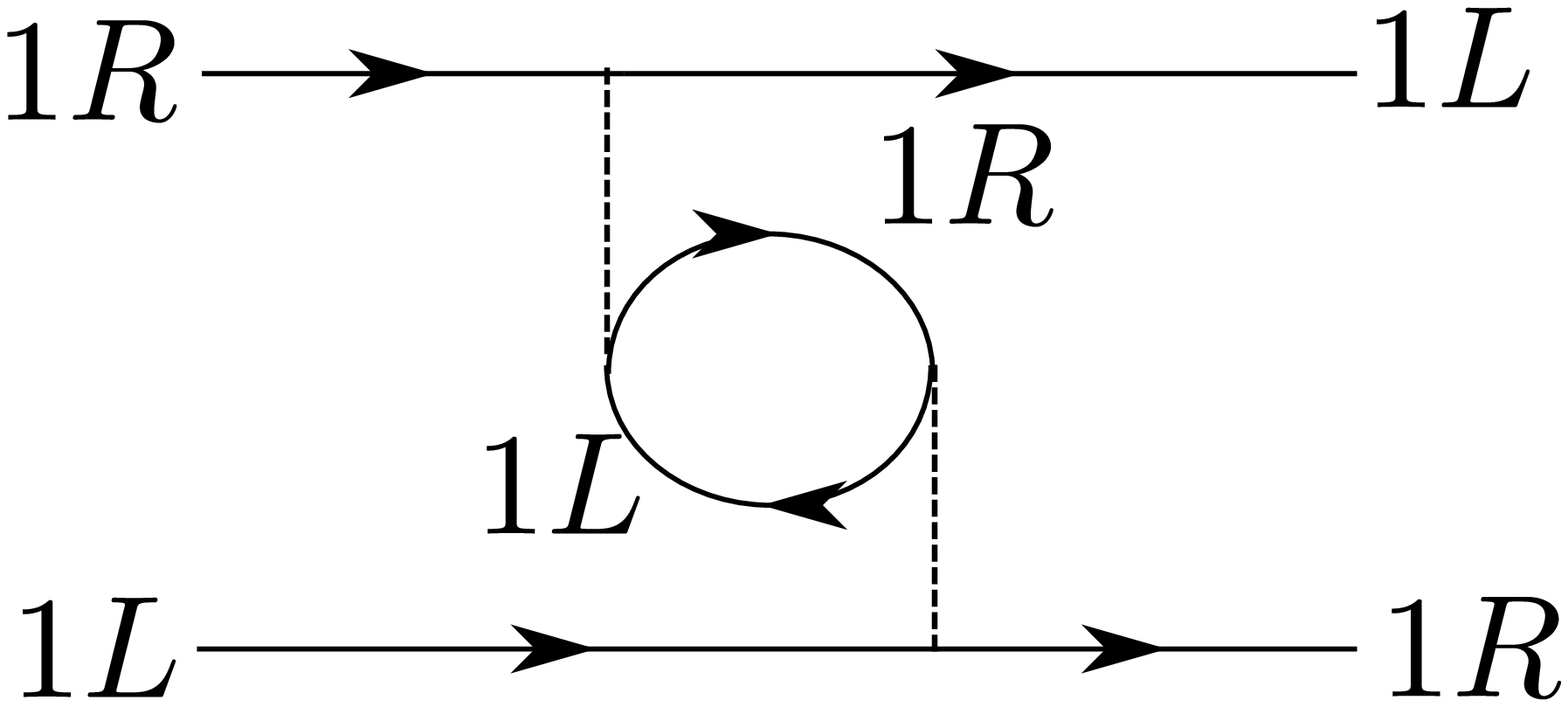}} &
		\fbox{\includegraphics[width = 0.3\textwidth]{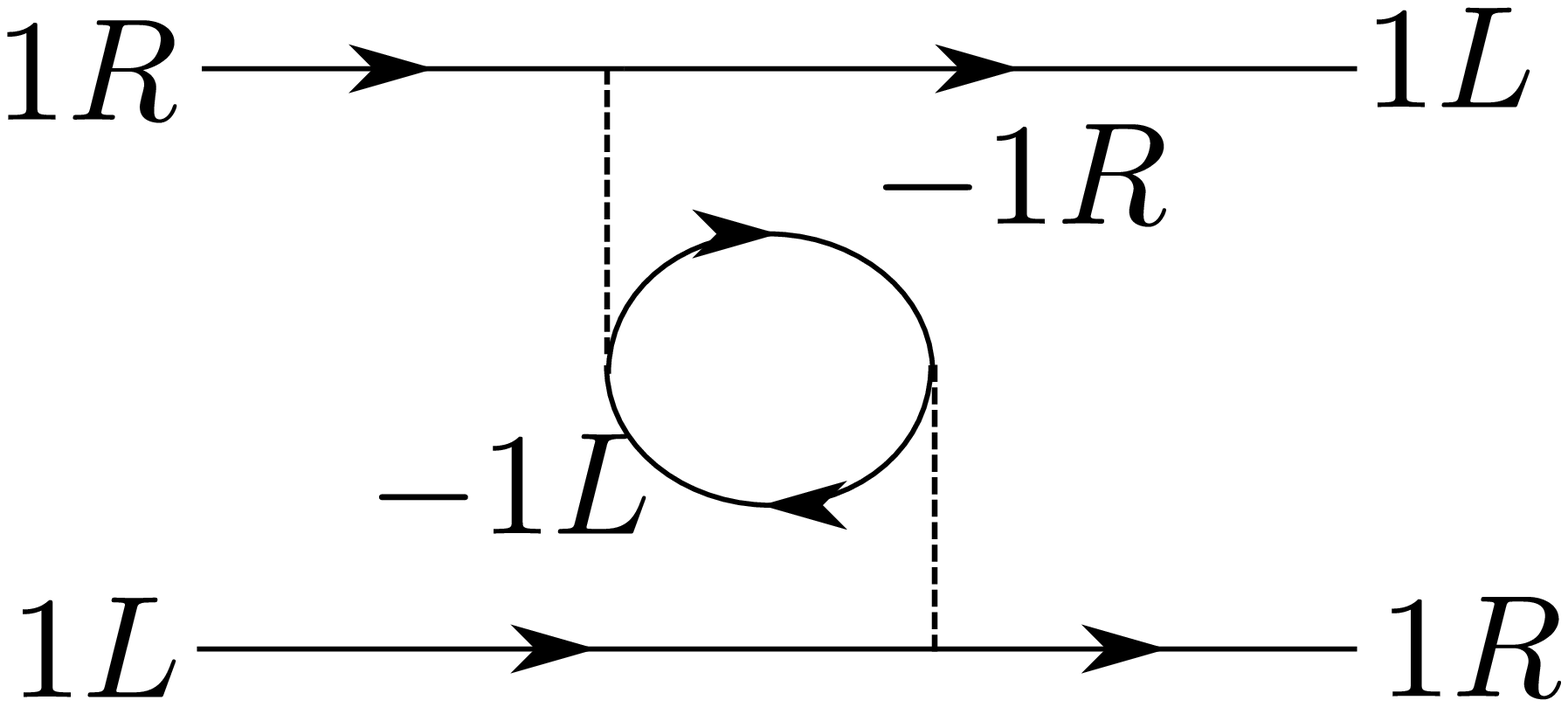}}  
	\end{tabular}
	\caption{Diagrams contributing to the flow of the backscattering interaction $\widetilde{U}_{11}$.In the Cooper channels, $L = 2$, while in the electron-hole channels $L = 0$.}
	\label{U7}
\end{figure}

\begin{figure}
	\begin{tabular}{ccc}
		\fbox{\includegraphics[width = 0.3\textwidth]{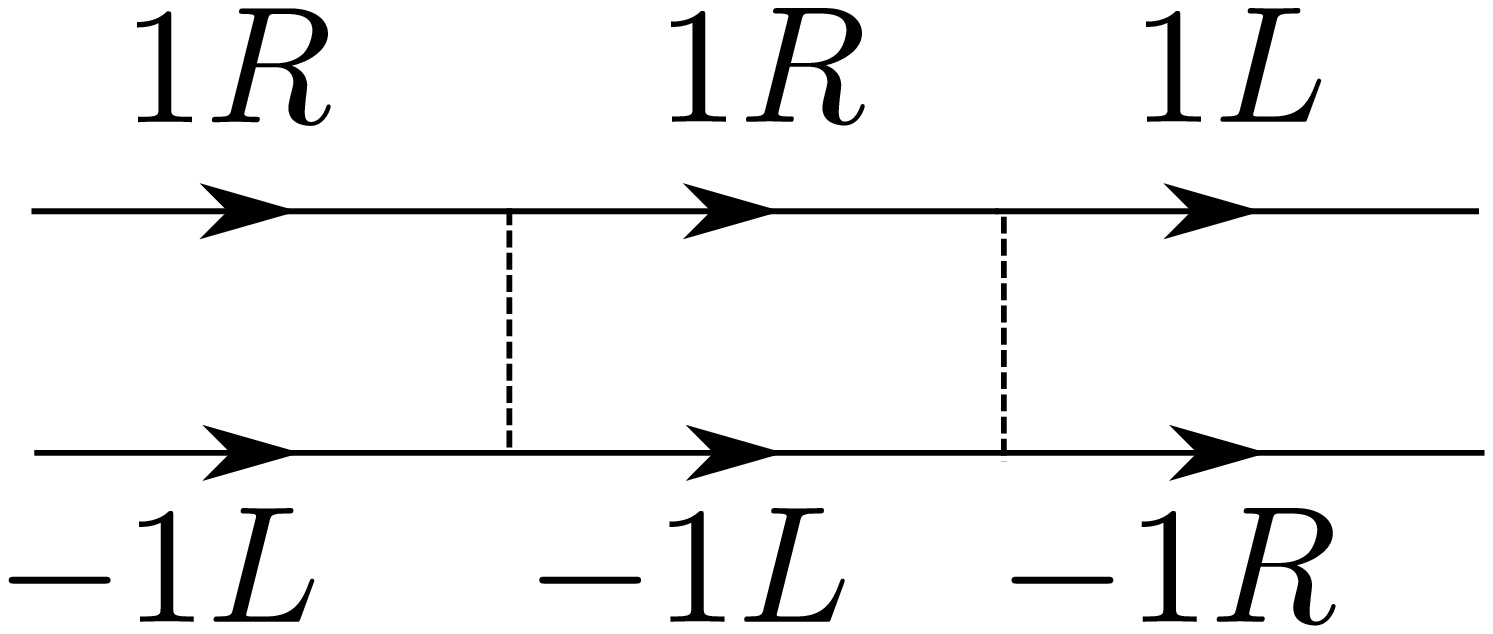}} &
		\fbox{\includegraphics[width = 0.3\textwidth]{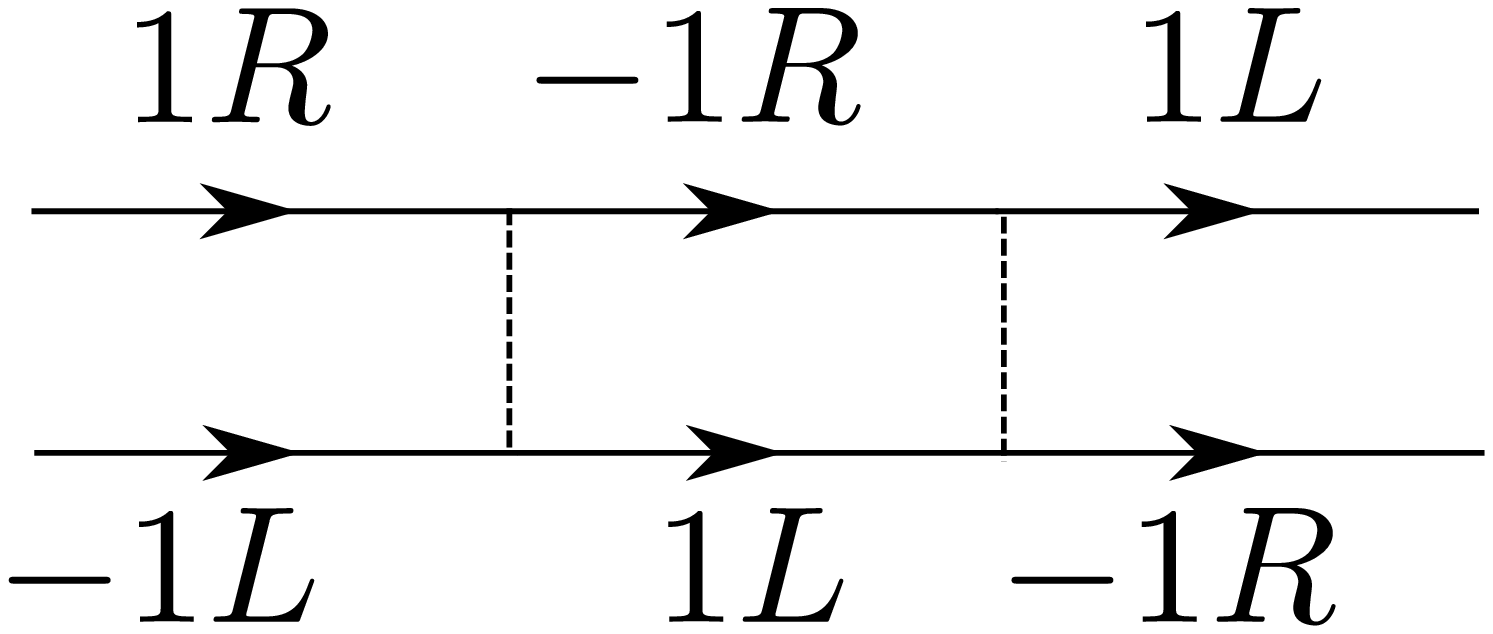}}  &
		\fbox{\includegraphics[width = 0.3\textwidth]{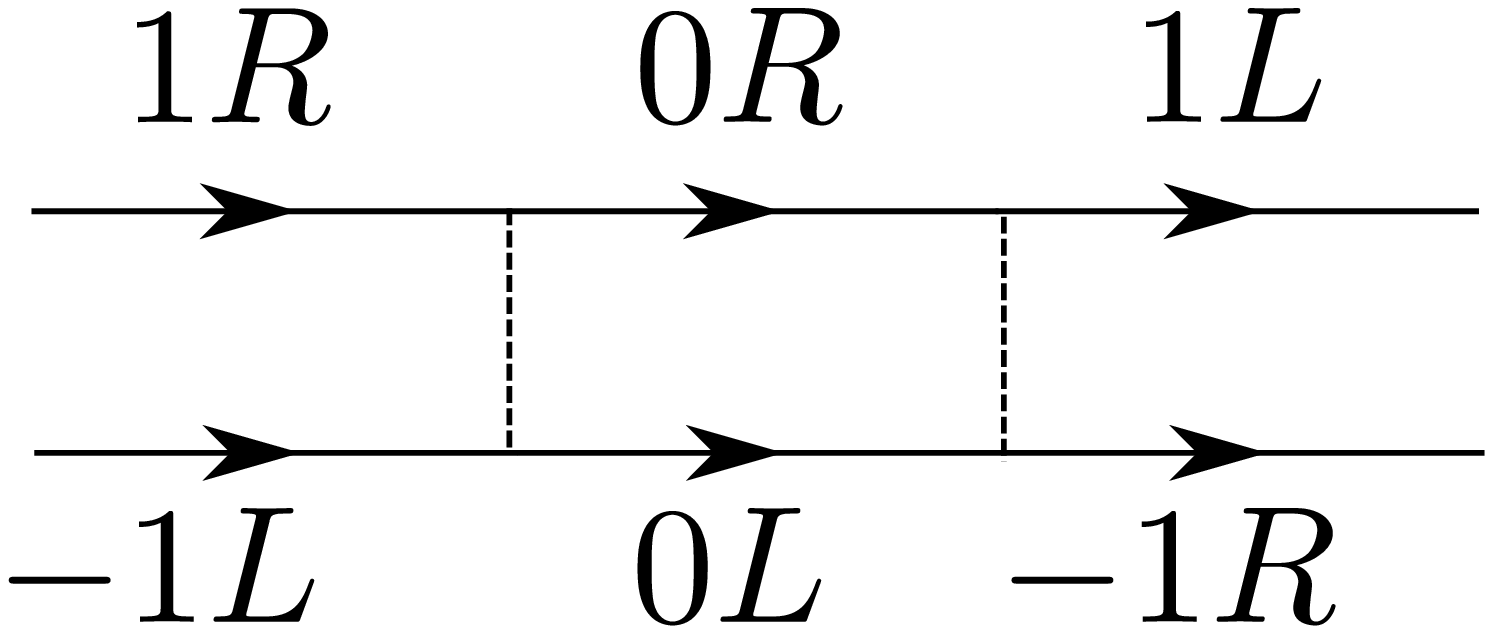}} \\ ~ \\
		\fbox{\includegraphics[width = 0.3\textwidth]{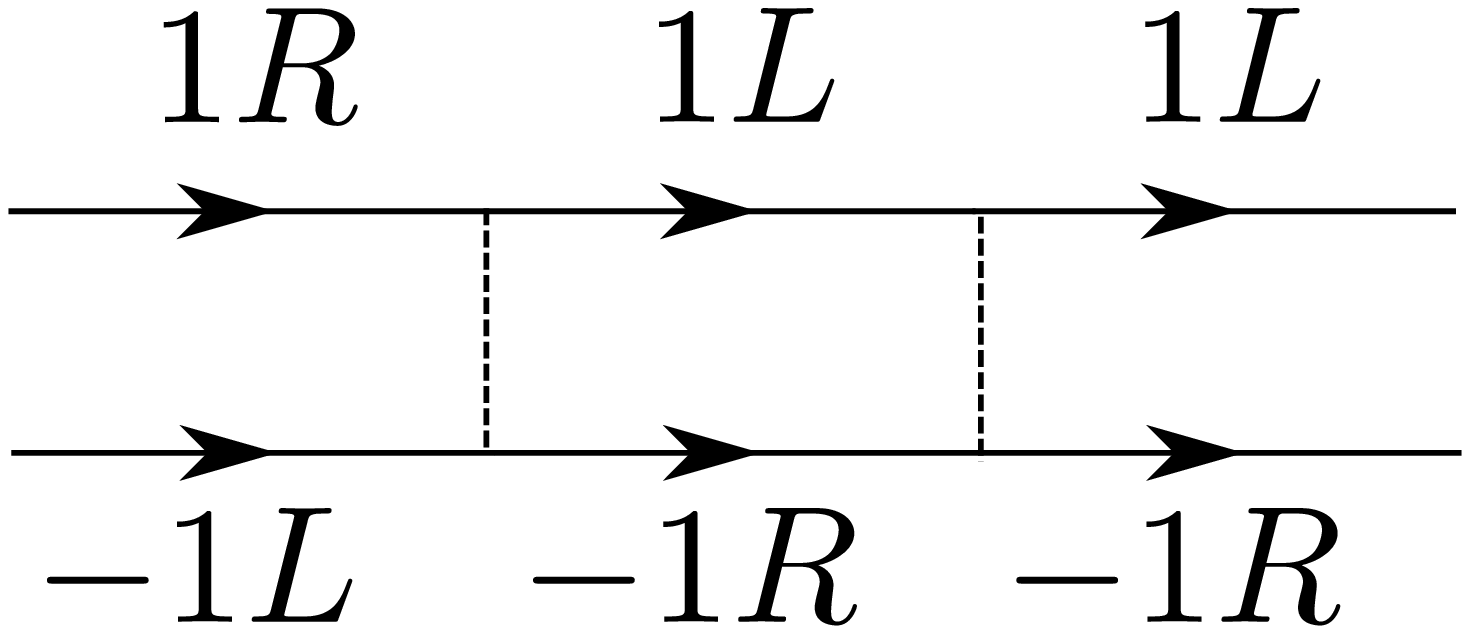}} &
		\fbox{\includegraphics[width = 0.3\textwidth]{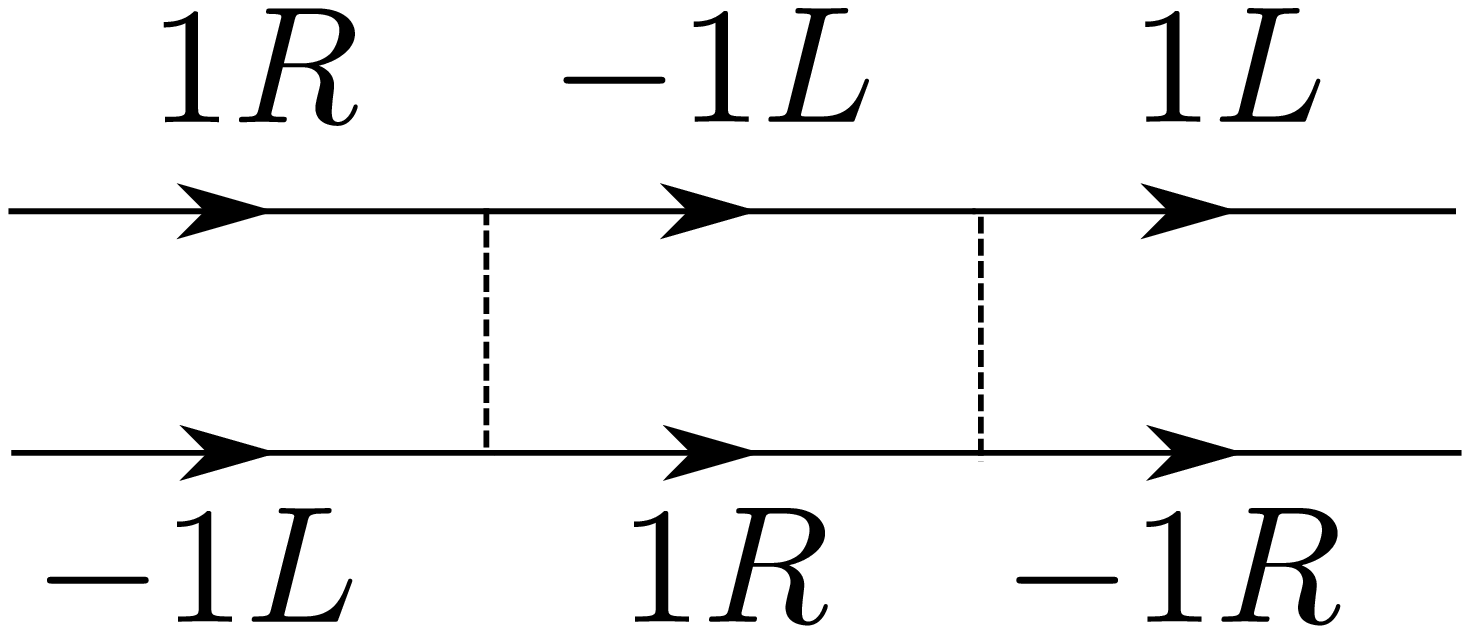}}  &
		\fbox{\includegraphics[width = 0.3\textwidth]{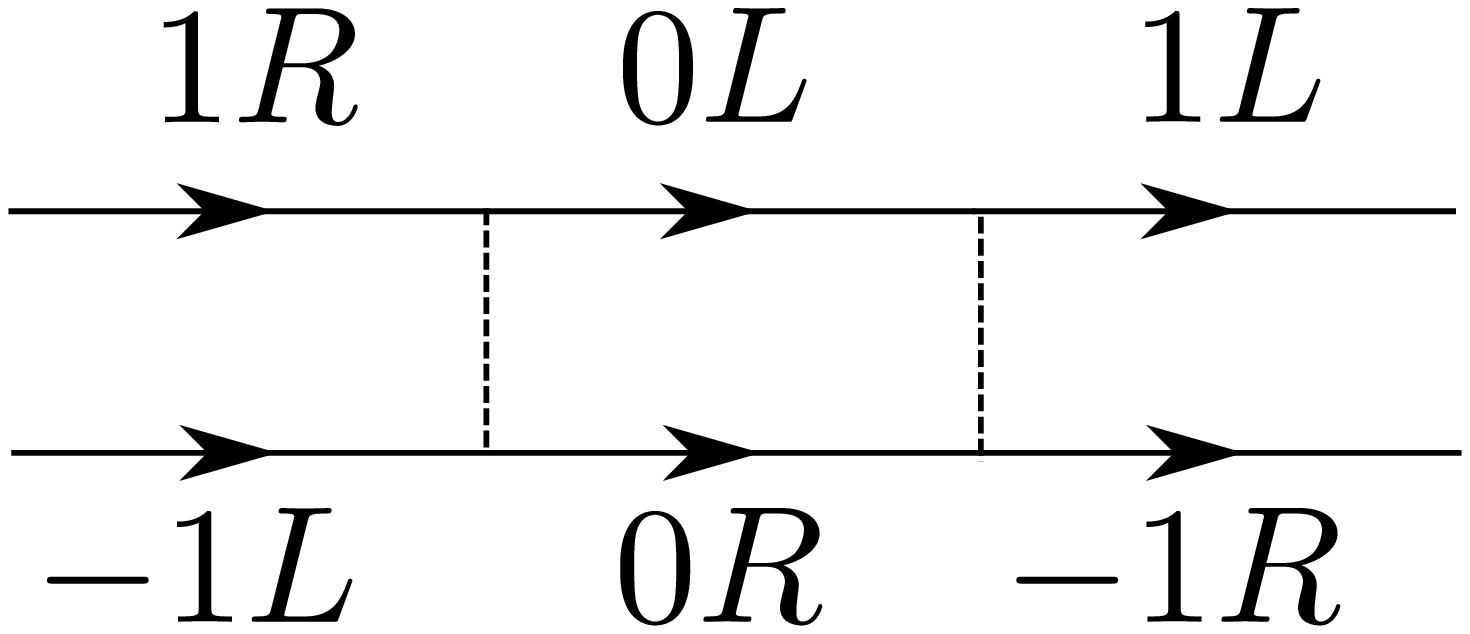}}
		\\ ~ \\
		\fbox{\includegraphics[width = 0.3\textwidth]{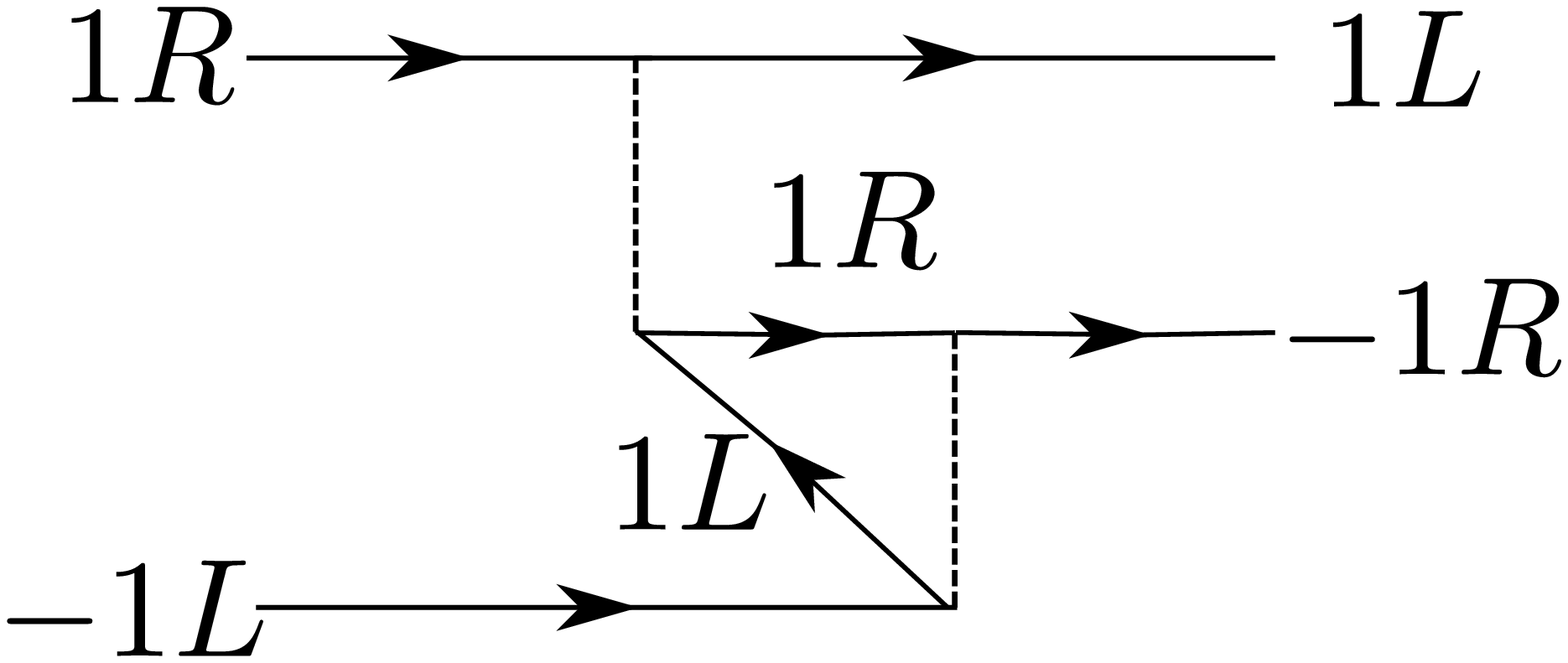}} &
		\fbox{\includegraphics[width = 0.3\textwidth]{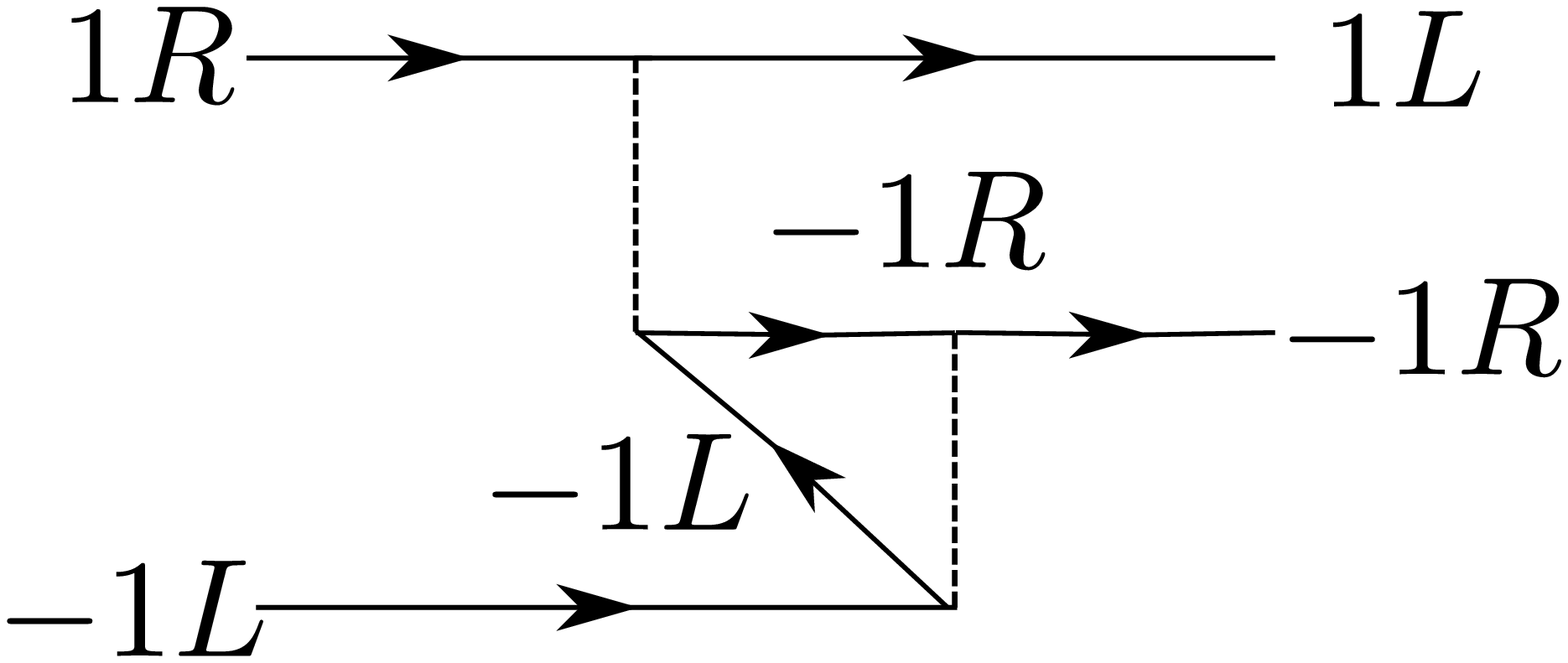}}  &
		\\ ~ \\
		\fbox{\includegraphics[width = 0.3\textwidth]{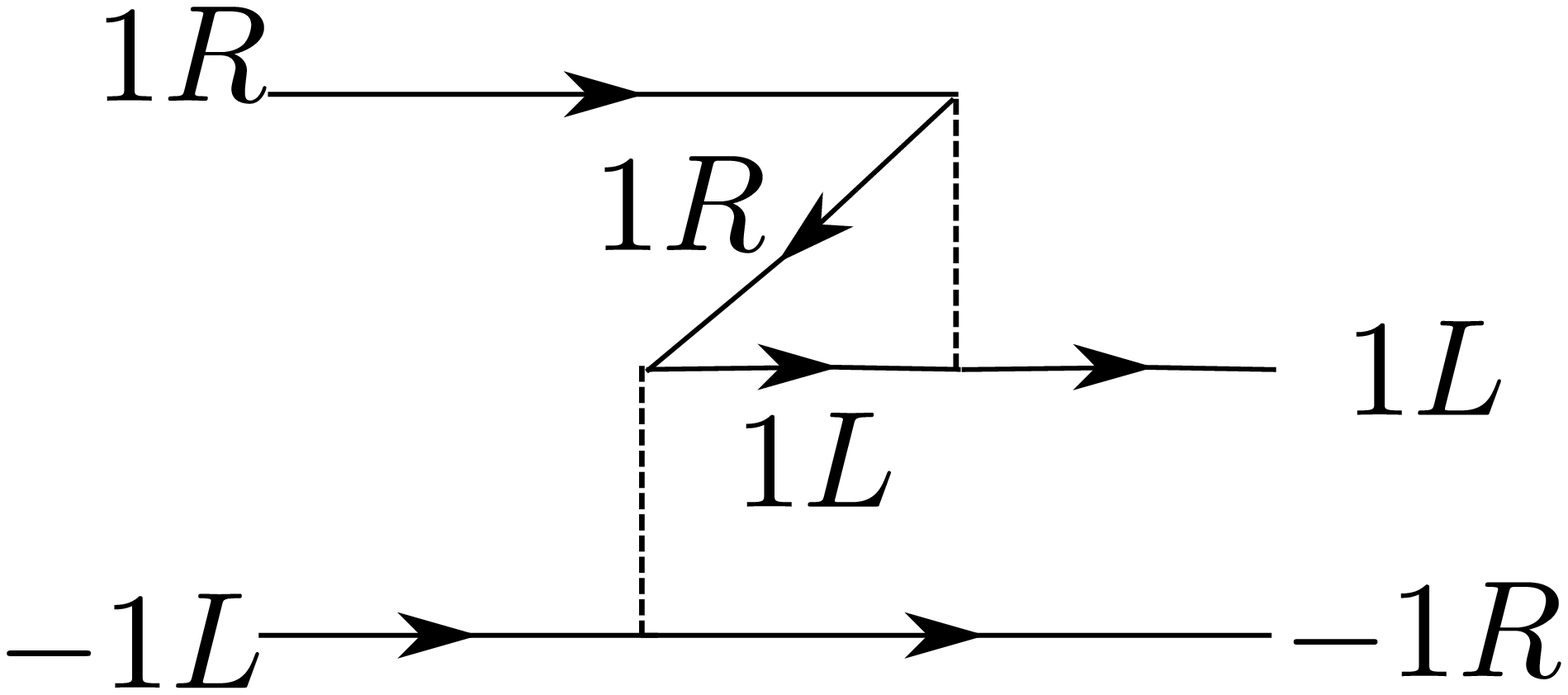}} &
		\fbox{\includegraphics[width = 0.3\textwidth]{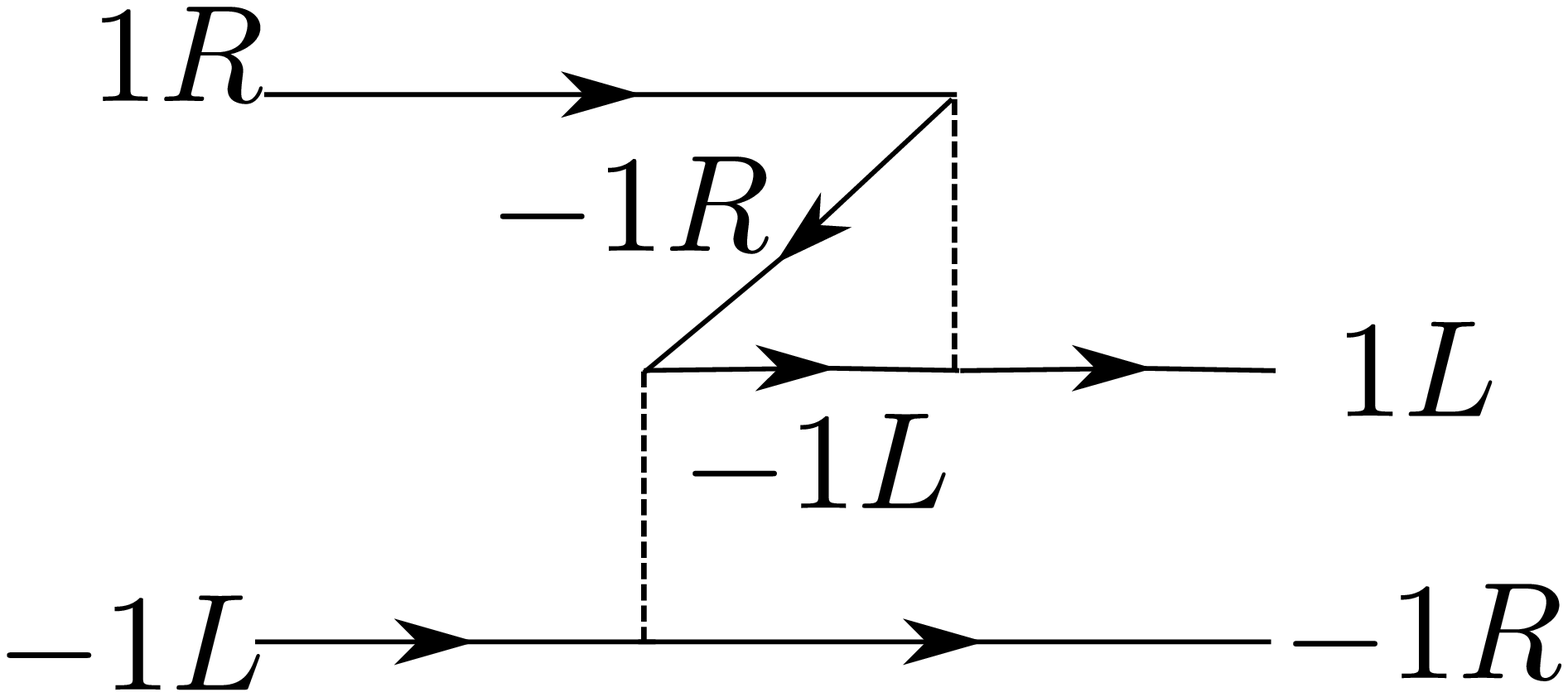}}  &
		\\ ~ \\
		\fbox{\includegraphics[width = 0.3\textwidth]{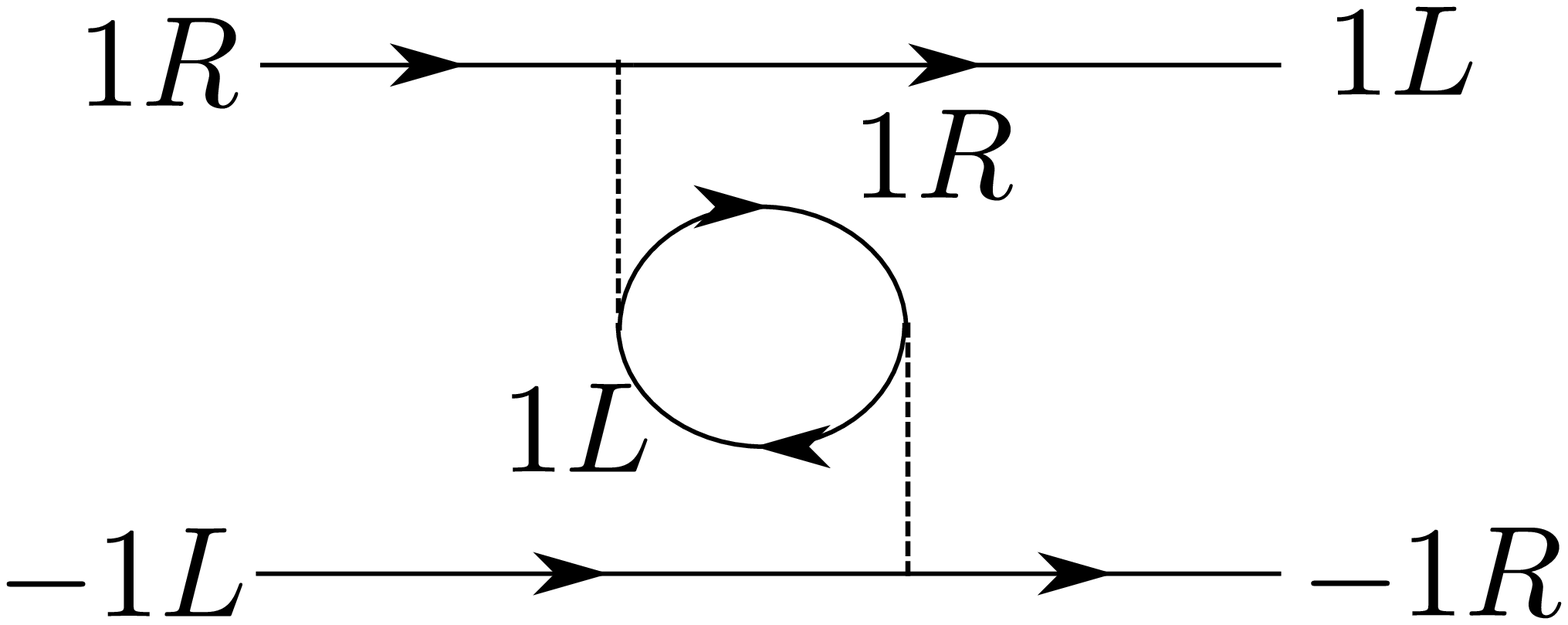}} &
		\fbox{\includegraphics[width = 0.3\textwidth]{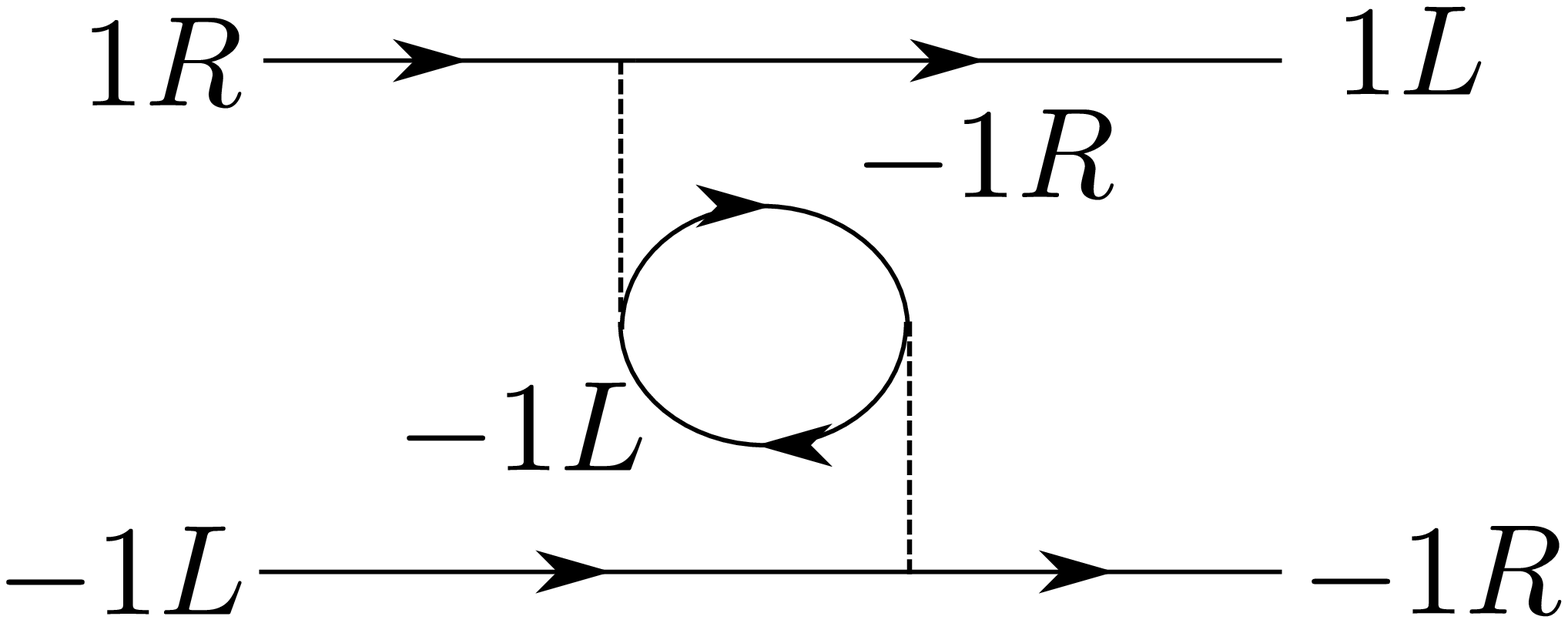}}  &
	\end{tabular}
	\caption{Diagrams contributing to the flow of the interaction $\widetilde{V}_2$. In all channels $L = 0$.
		\label{U8}
	}
\end{figure}

\begin{figure}
	\begin{tabular}{ccc}
		\fbox{\includegraphics[width = 0.3\textwidth]{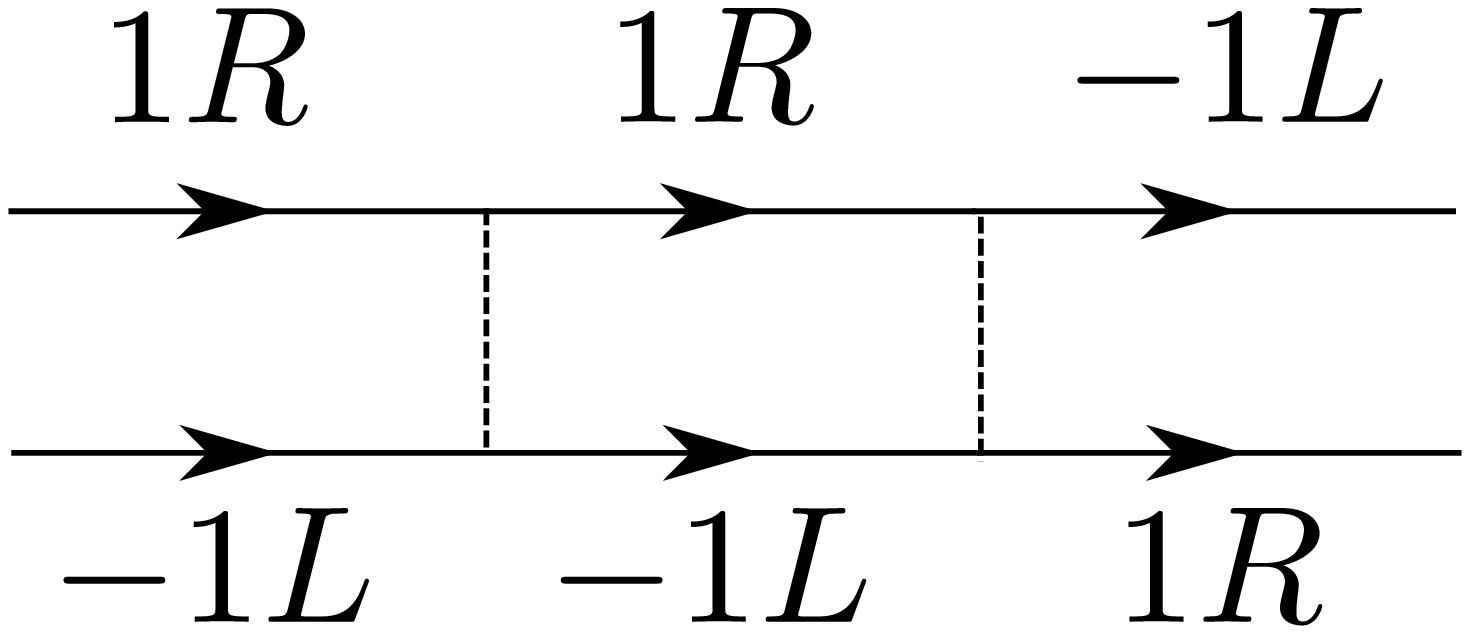}} &
		\fbox{\includegraphics[width = 0.3\textwidth]{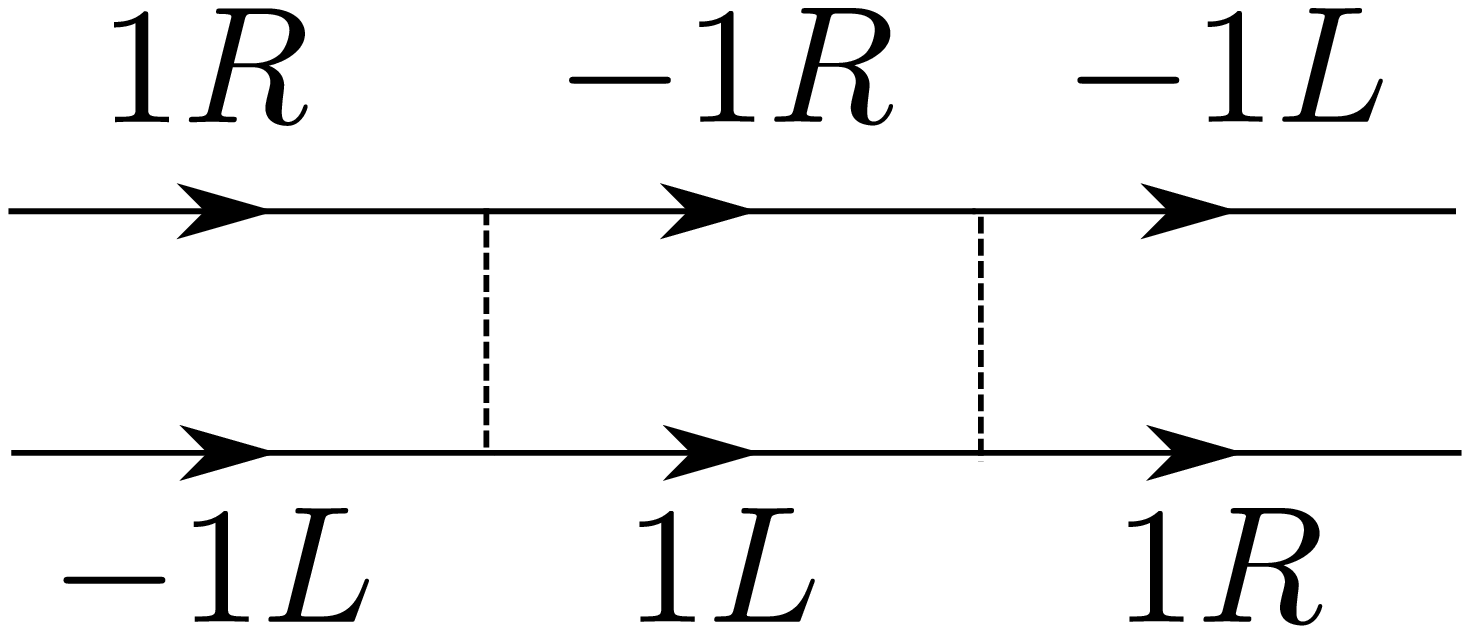}}  &
		\fbox{\includegraphics[width = 0.3\textwidth]{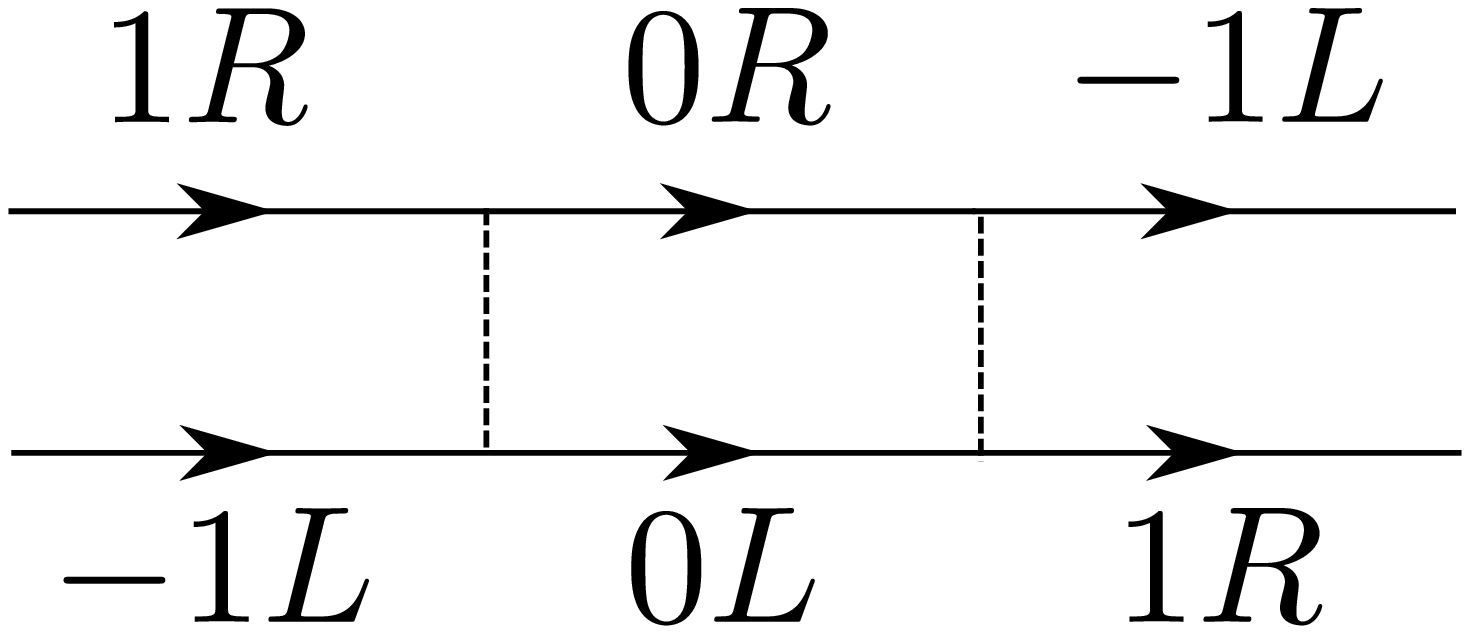}} \\ ~ \\
		\fbox{\includegraphics[width = 0.3\textwidth]{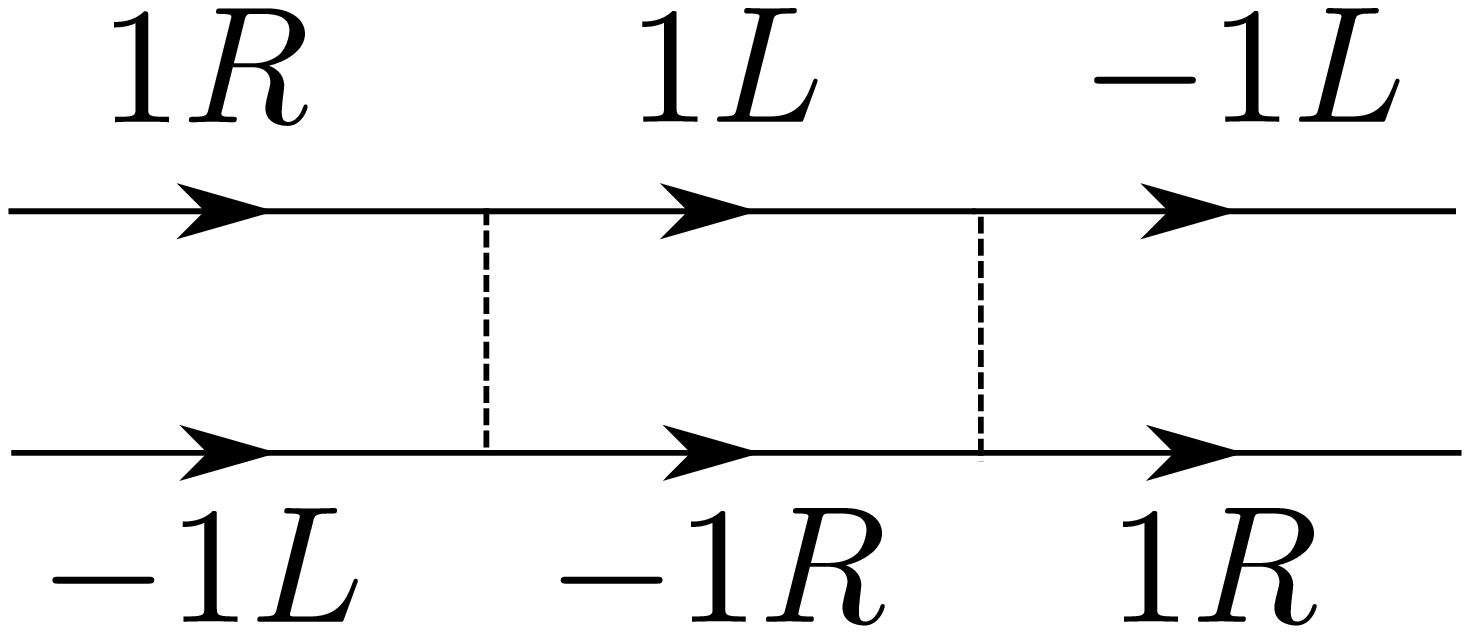}} &
		\fbox{\includegraphics[width = 0.3\textwidth]{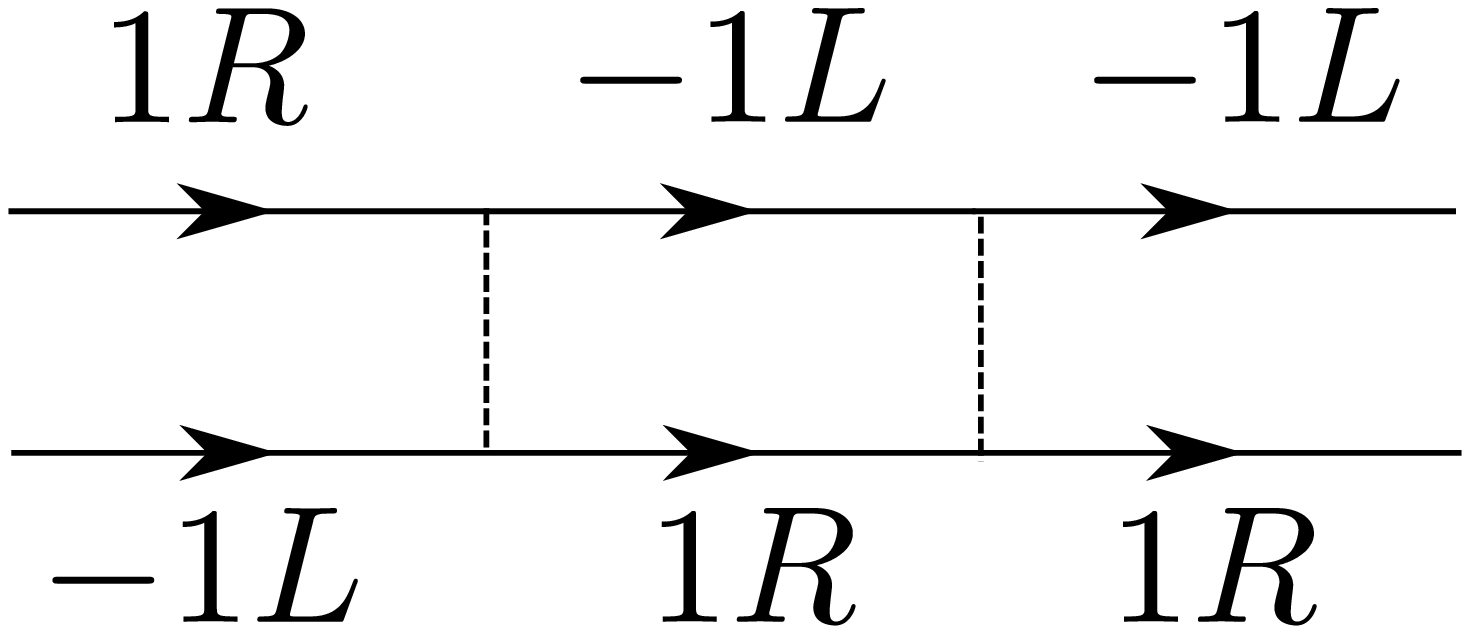}}  &
		\fbox{\includegraphics[width = 0.3\textwidth]{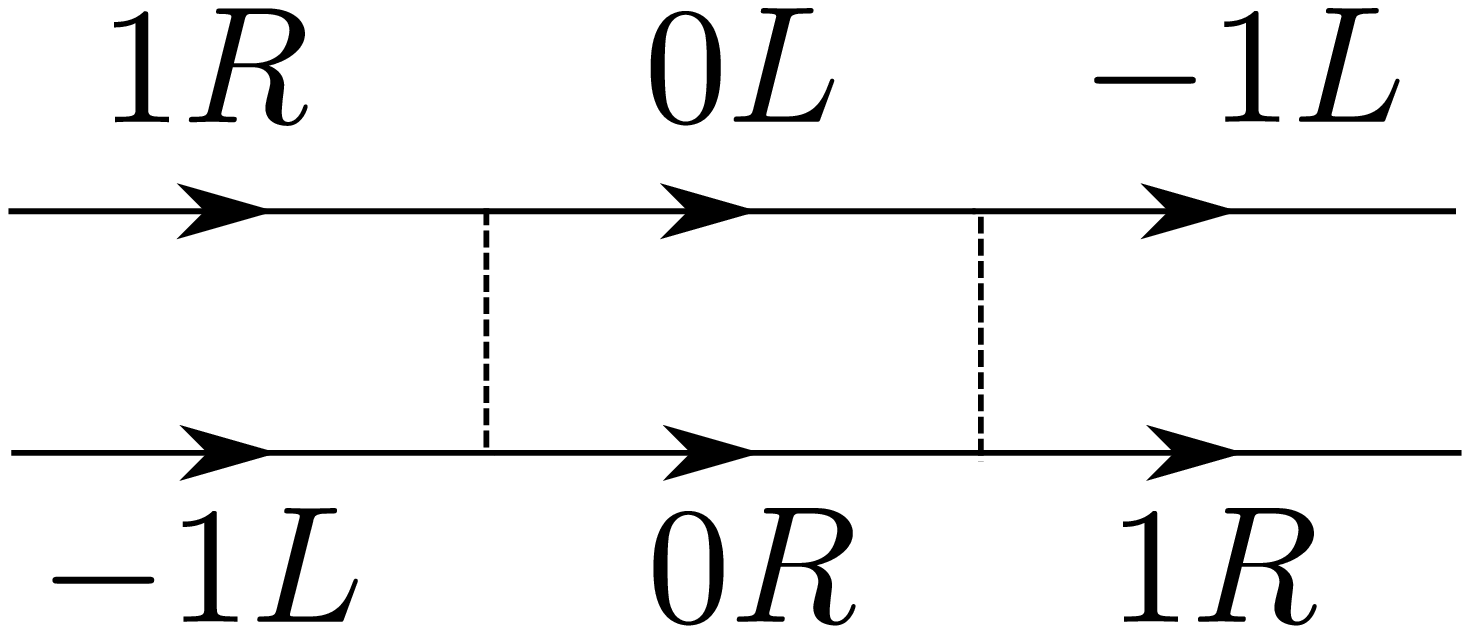}}
		\\ ~ \\
		\fbox{\includegraphics[width = 0.3\textwidth]{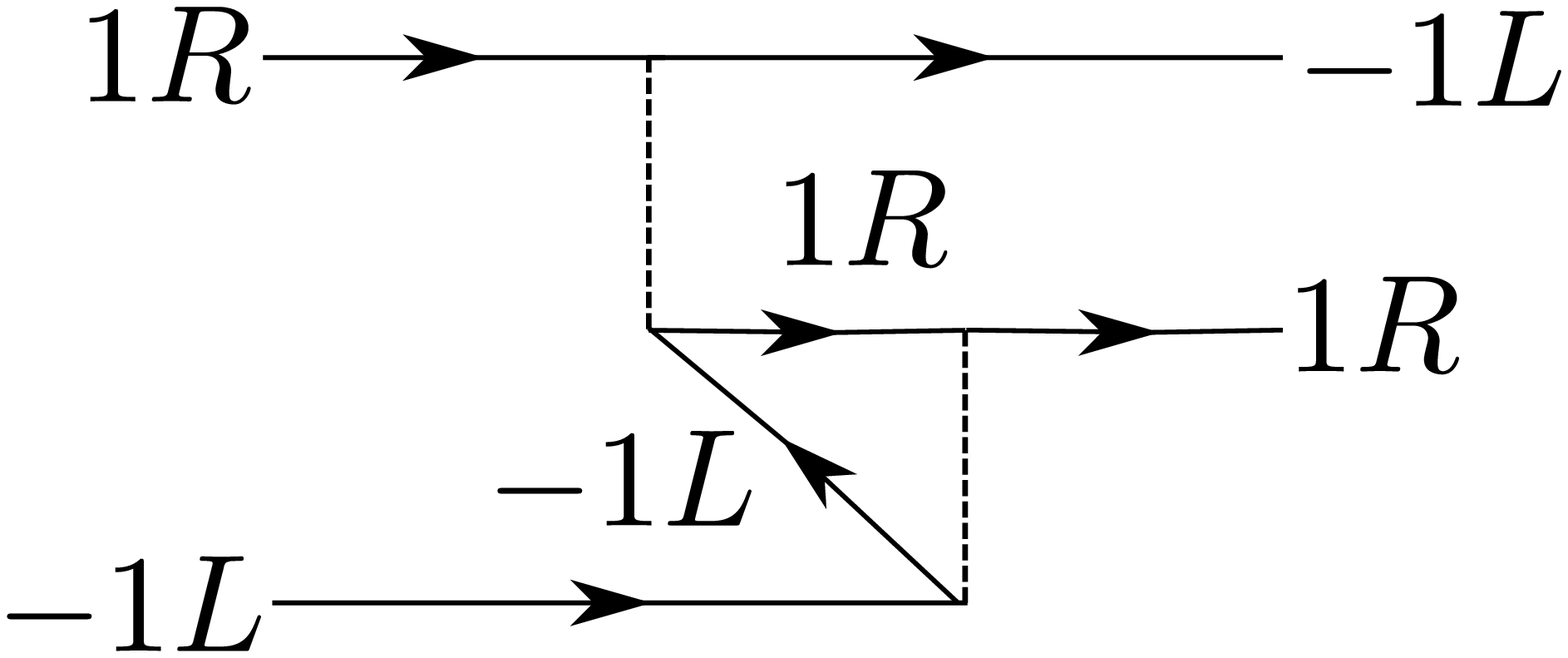}} &
		\fbox{\includegraphics[width = 0.3\textwidth]{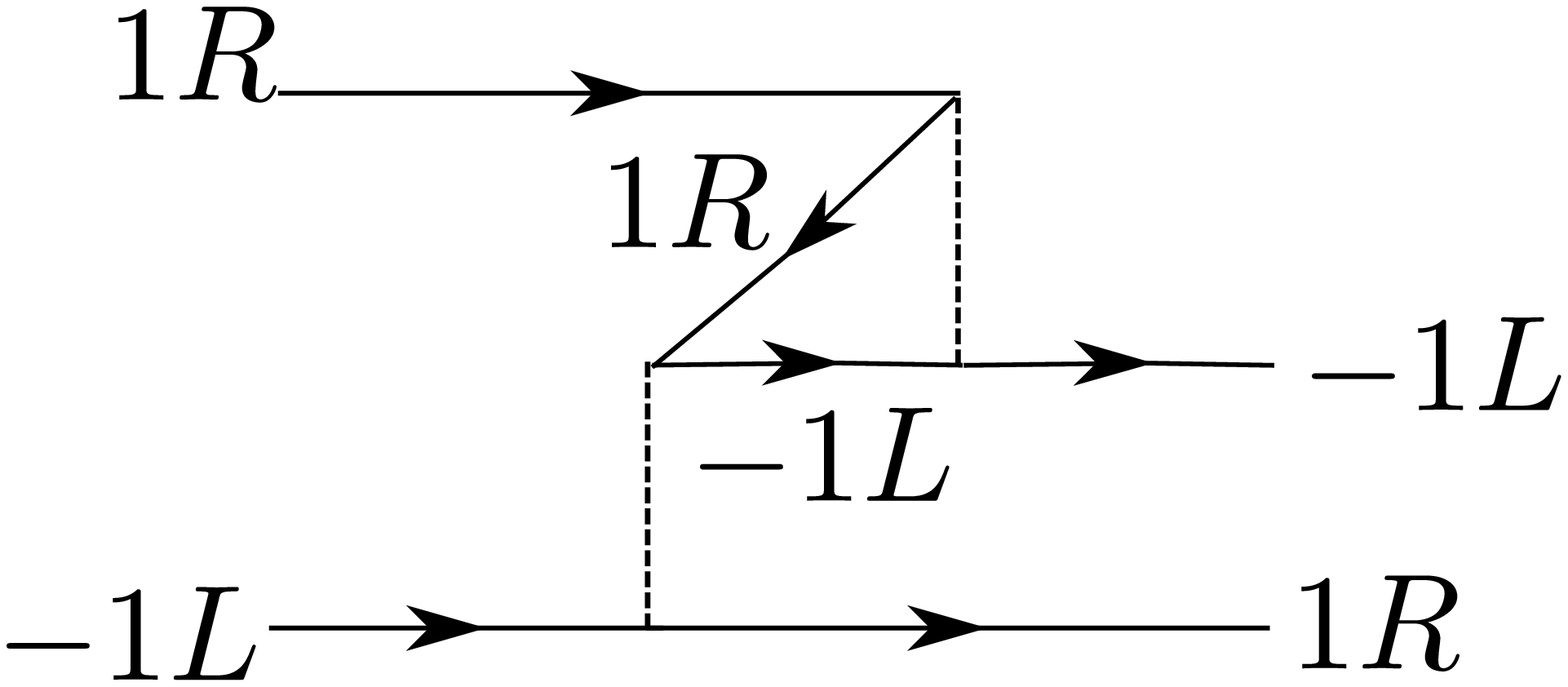}}  &
		\fbox{\includegraphics[width = 0.3\textwidth]{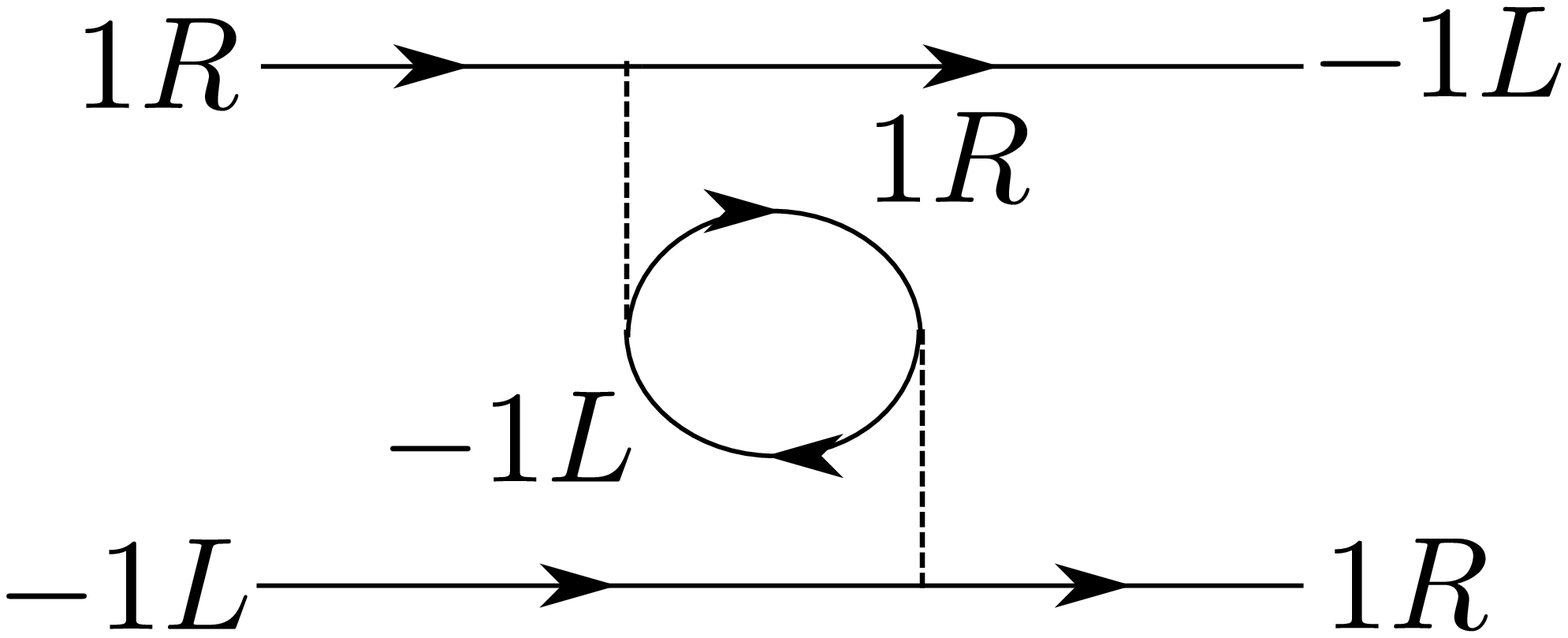}}
	\end{tabular}
	\caption{Diagrams contributing to the flow of the interaction $\widetilde{U}_{1,-1}$. In the Cooper channels $L = 0$, but in the electron-hole channels $L = \pm 2$.
		\label{U9}
	}
\end{figure}

\begin{figure}
	\begin{tabular}{cc}
		\fbox{\includegraphics[width = 0.3\textwidth]{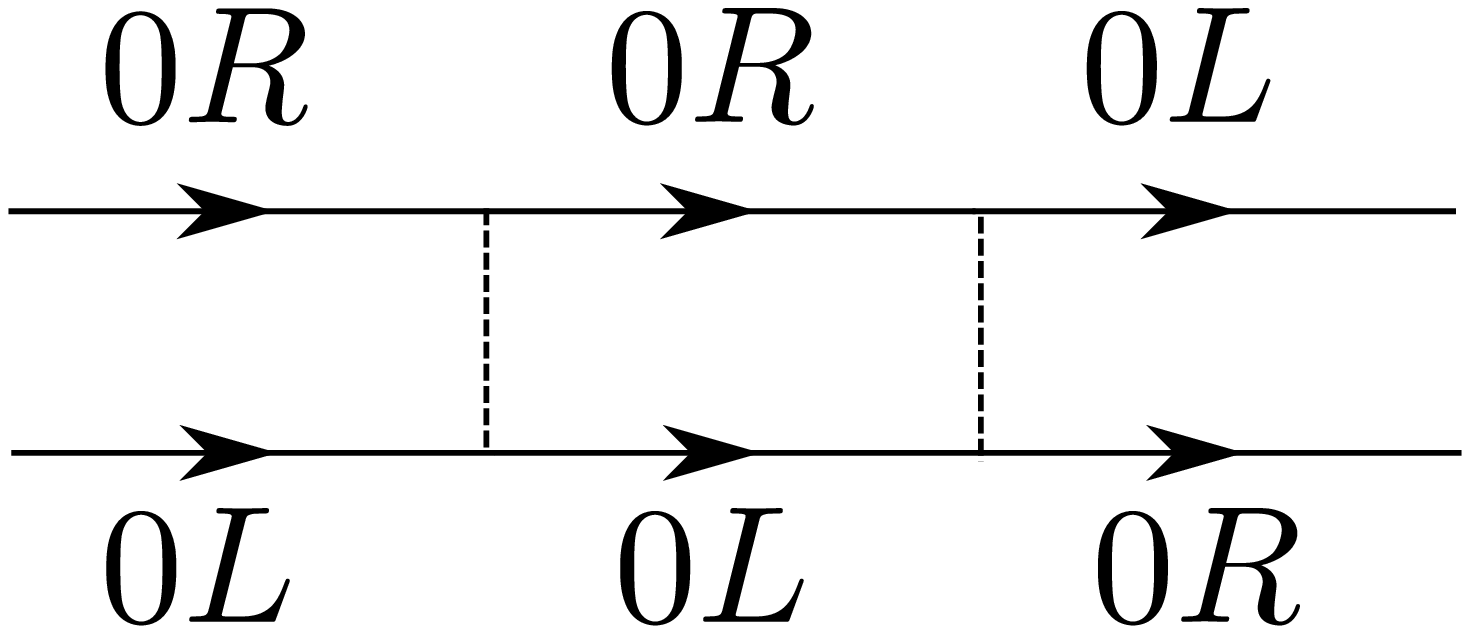}} &
		\fbox{\includegraphics[width = 0.3\textwidth]{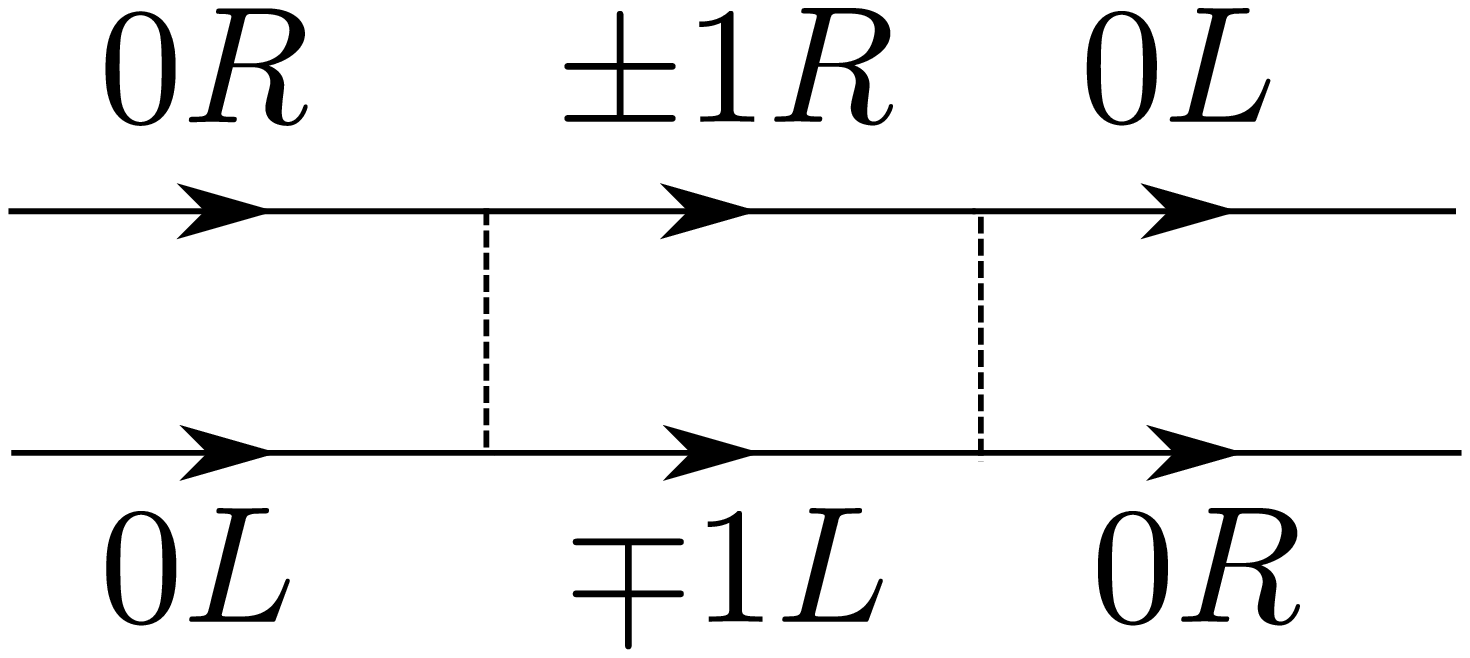}} \\ ~ \\
		\fbox{\includegraphics[width = 0.3\textwidth]{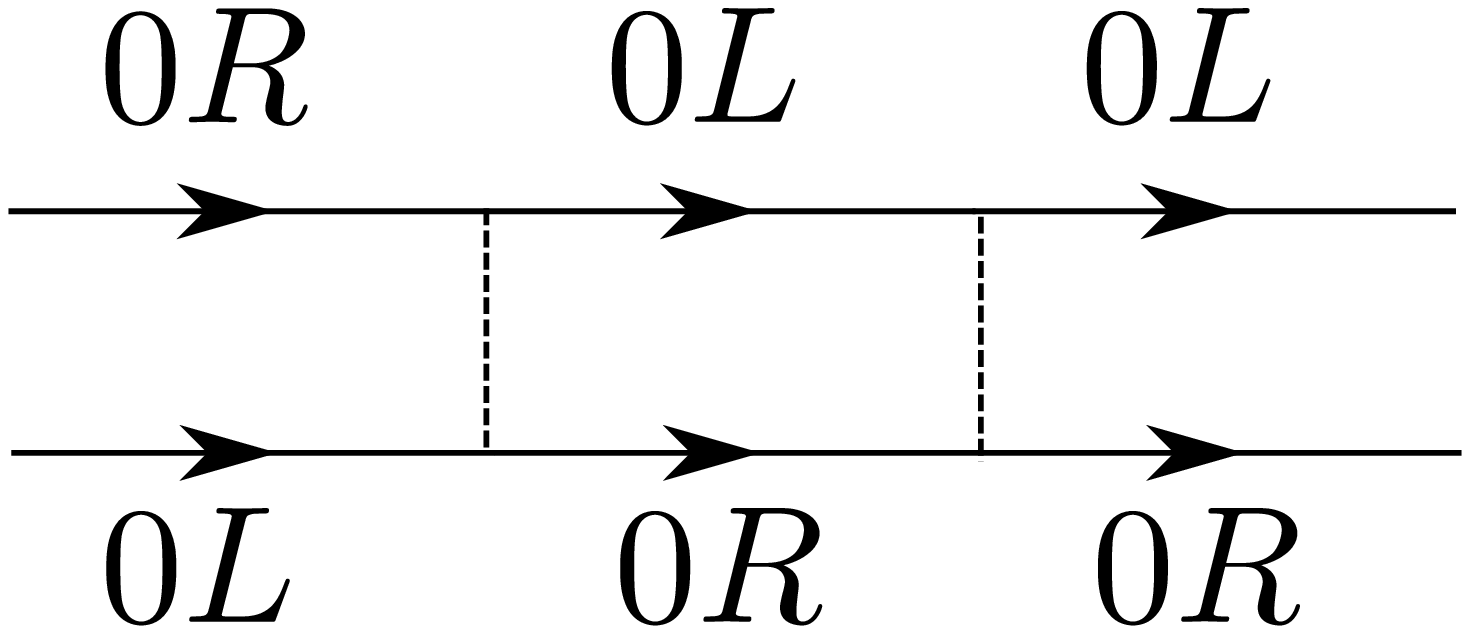}} &
		\fbox{\includegraphics[width = 0.3\textwidth]{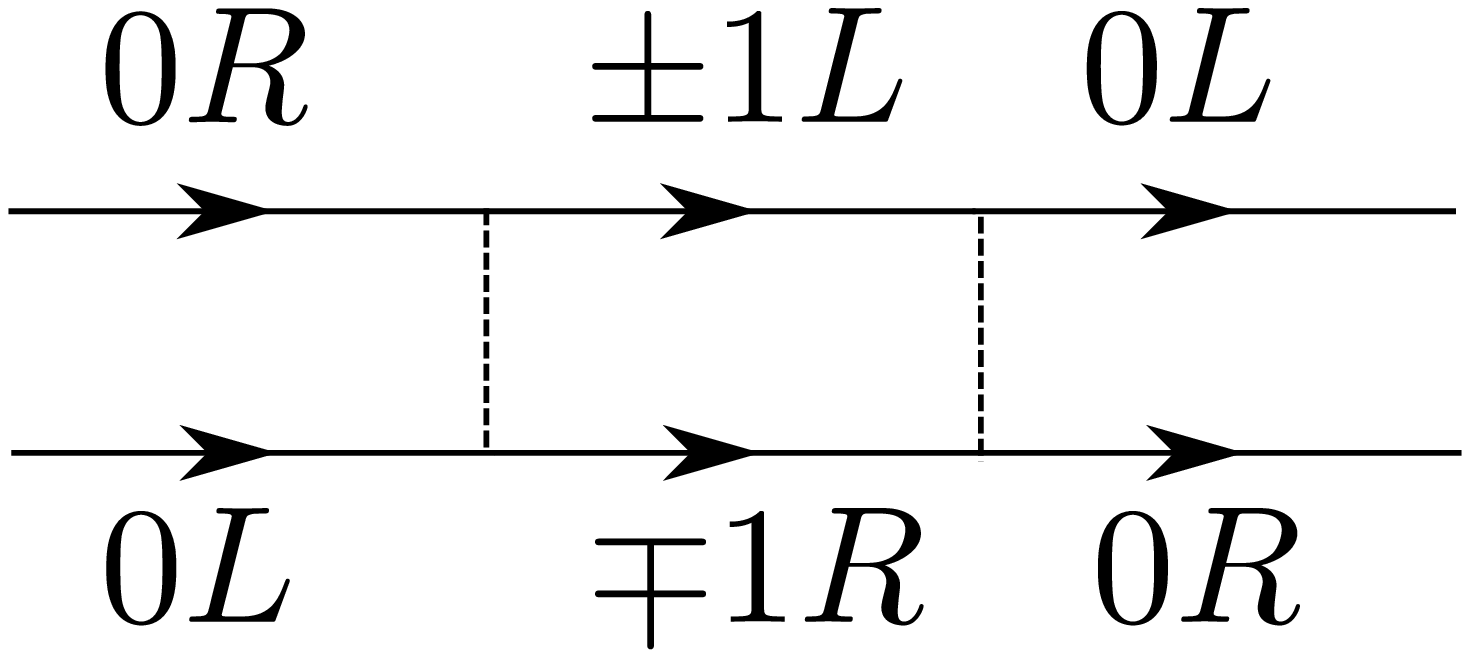}} \\ ~ \\
		\fbox{\includegraphics[width = 0.3\textwidth]{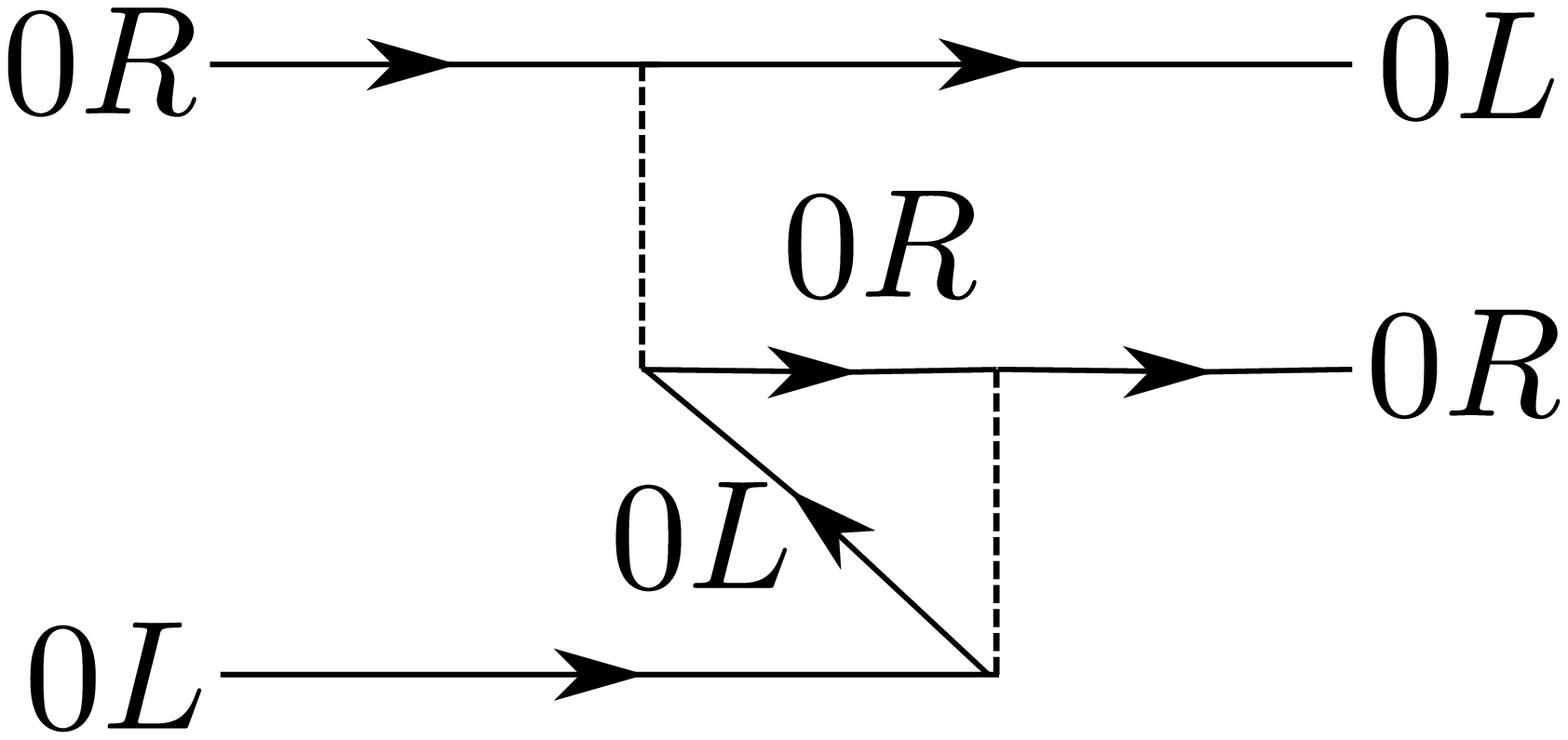}} &
		\fbox{\includegraphics[width = 0.3\textwidth]{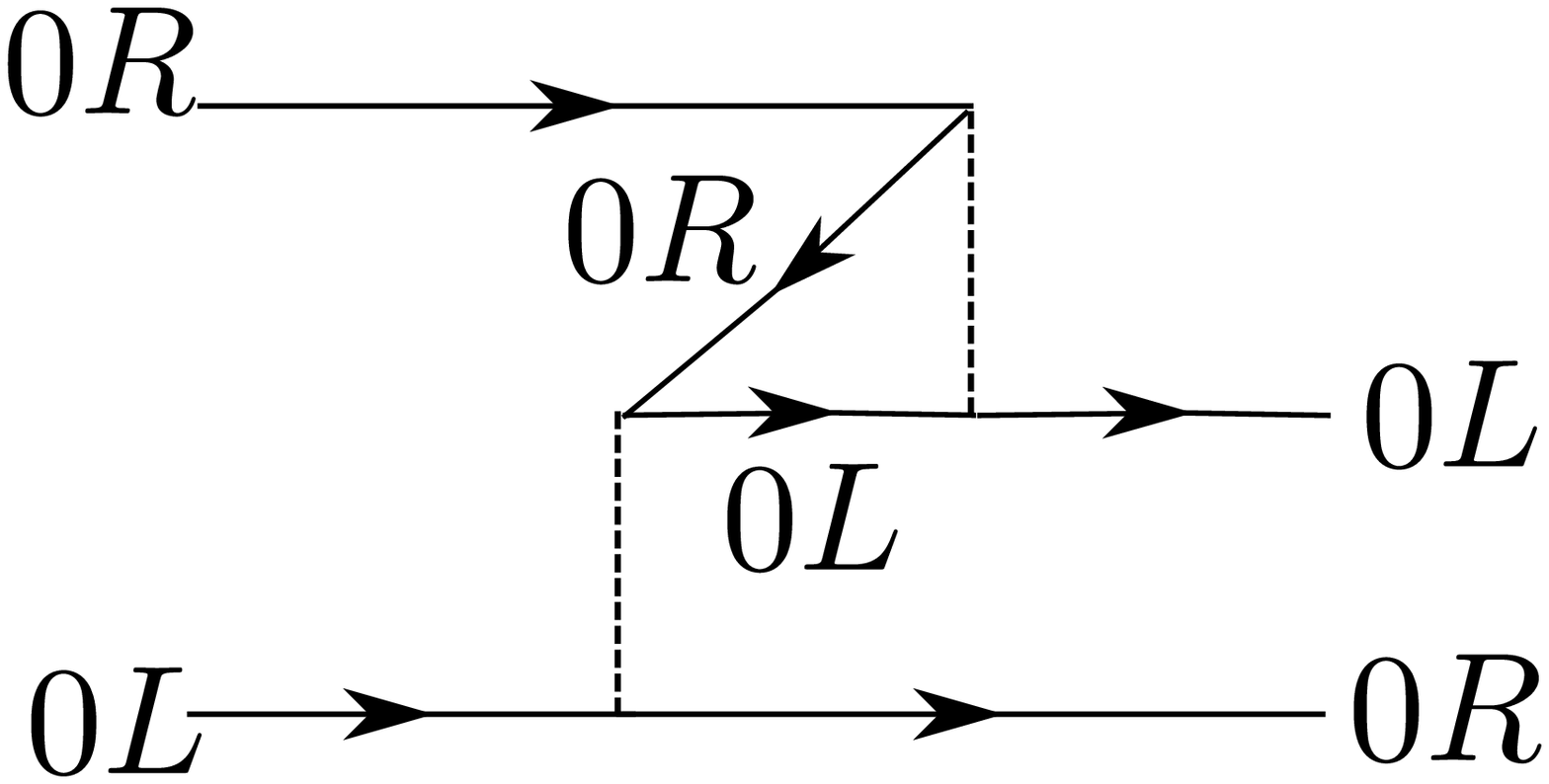}} \\ ~ \\
		\fbox{\includegraphics[width = 0.3\textwidth]{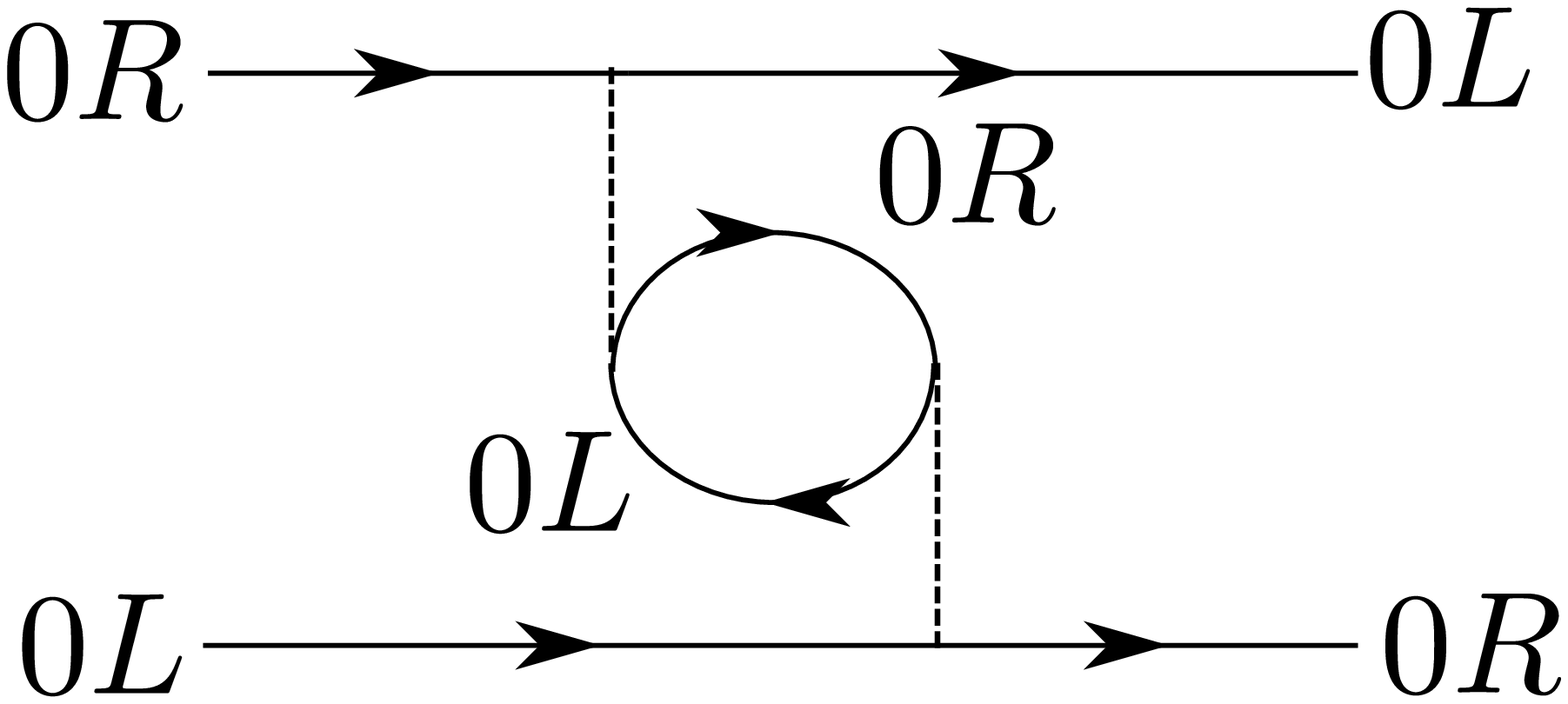}} &
	\end{tabular}
	\caption{Diagrams contributing to the flow of the interaction $\widetilde{U}_{00}$. In the intermediate states $L = 0$.}
	\label{U10}
\end{figure}

\begin{figure}
	\begin{tabular}{cc}
		\fbox{\includegraphics[width = 0.3\textwidth]{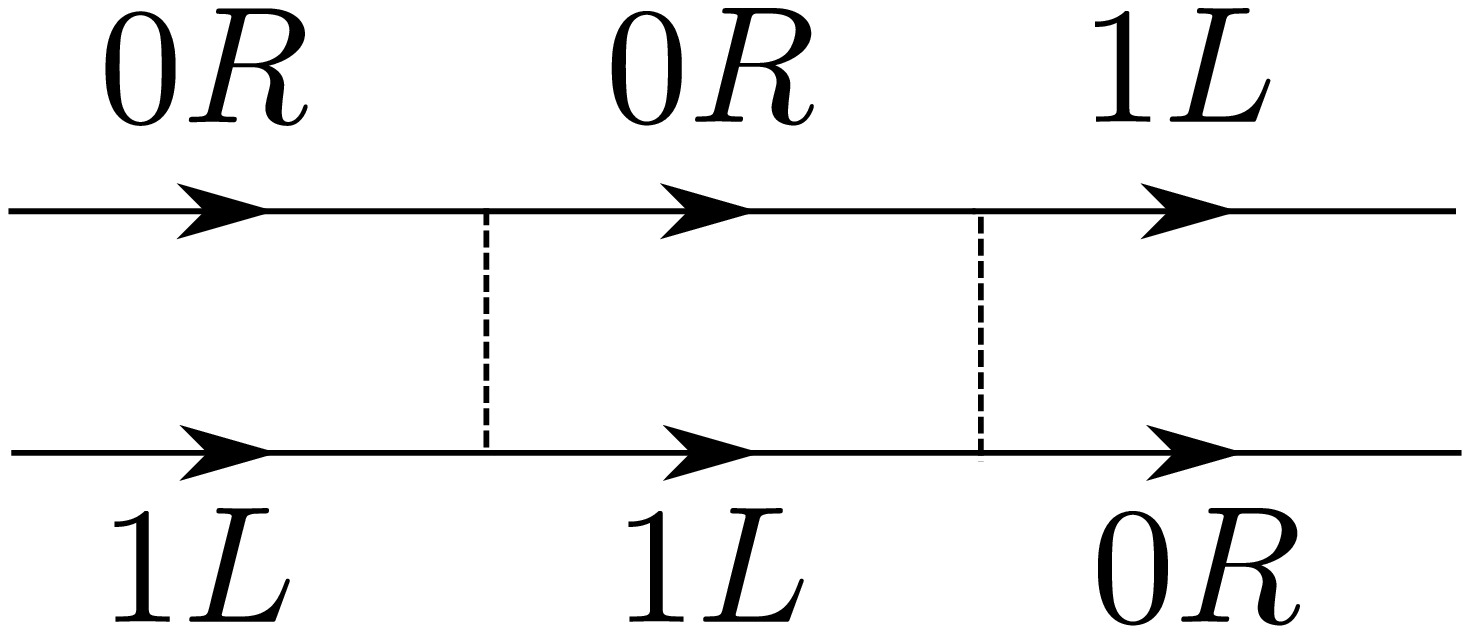}} &
		\fbox{\includegraphics[width = 0.3\textwidth]{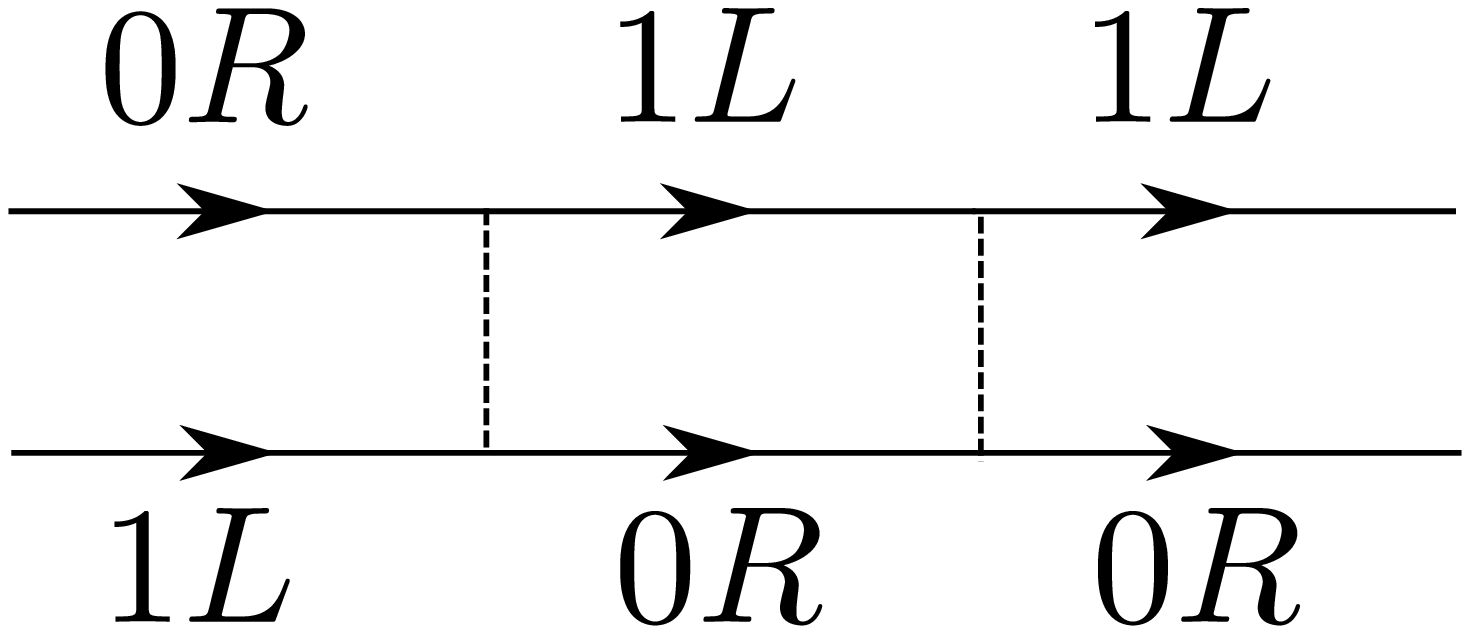}} \\ ~ \\
		\fbox{\includegraphics[width = 0.3\textwidth]{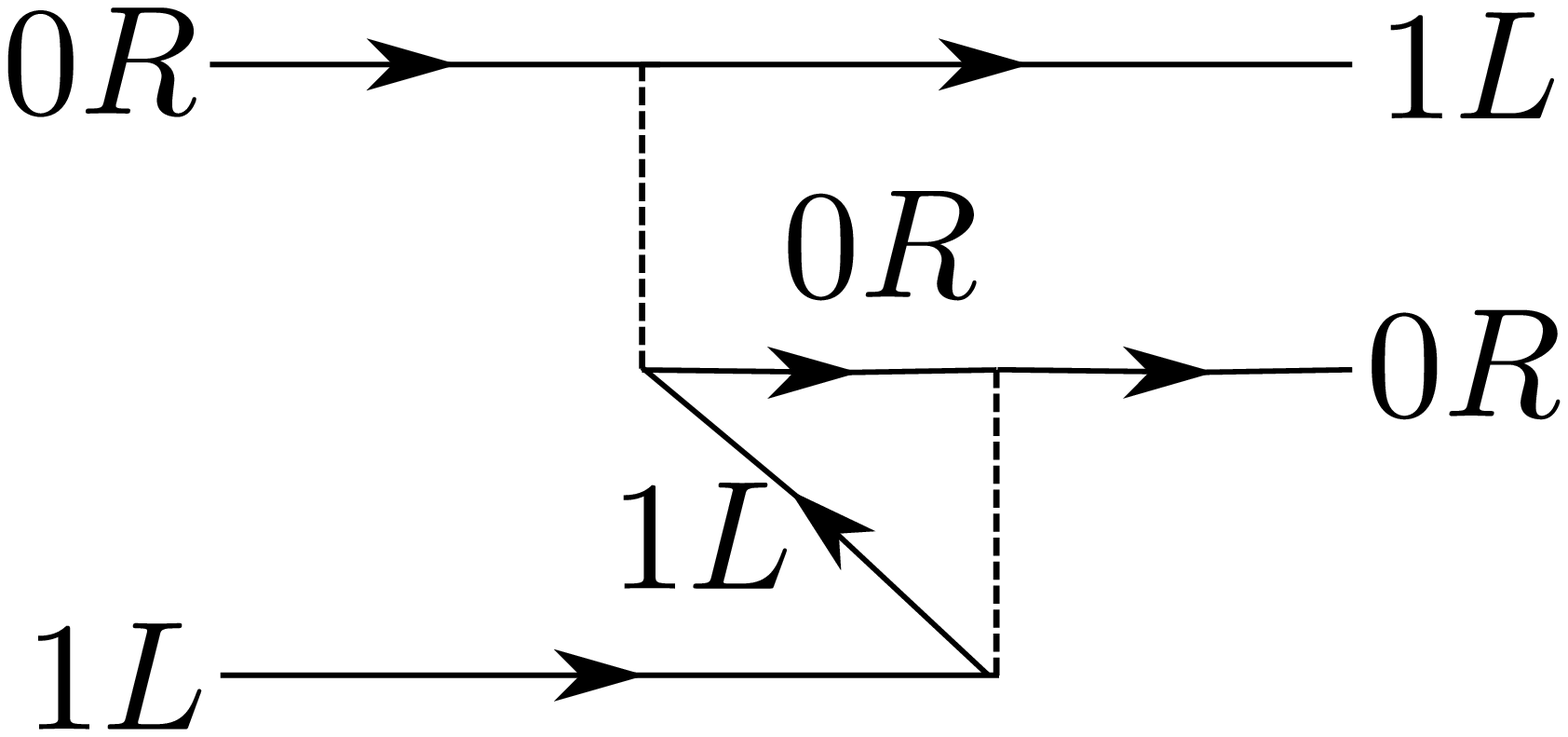}} &
		\fbox{\includegraphics[width = 0.3\textwidth]{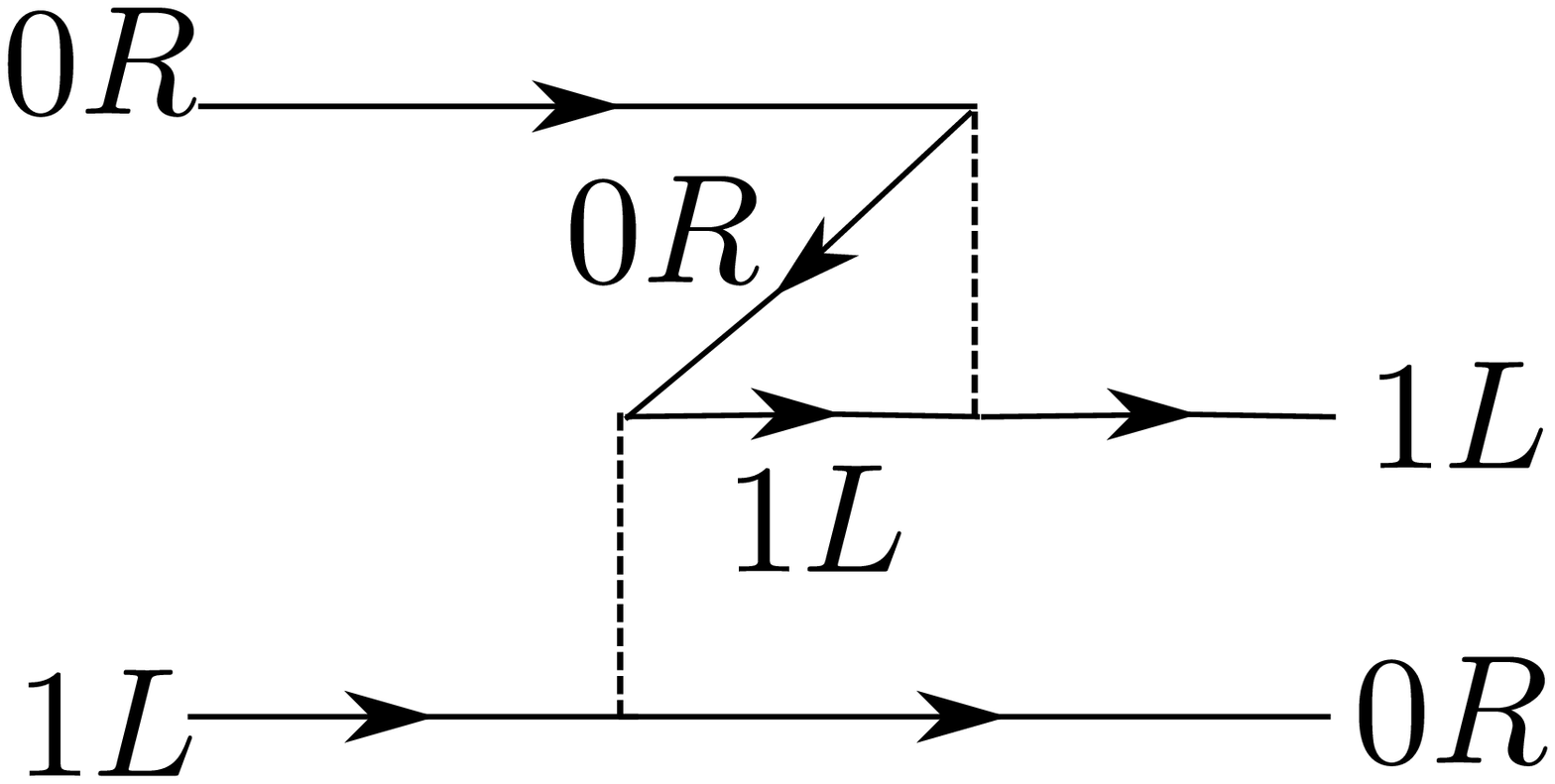}} \\ ~ \\
		\fbox{\includegraphics[width = 0.3\textwidth]{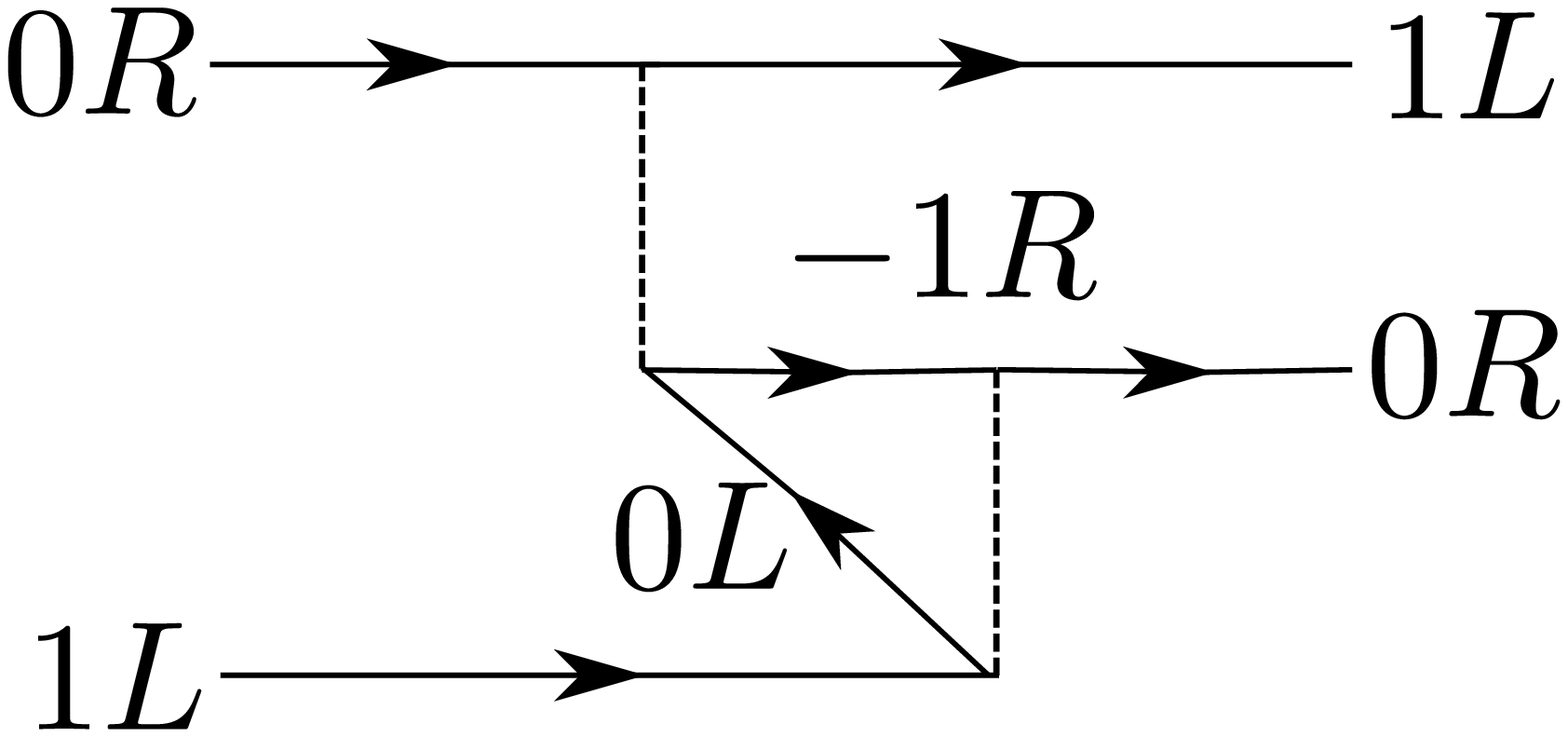}} &
		\fbox{\includegraphics[width = 0.3\textwidth]{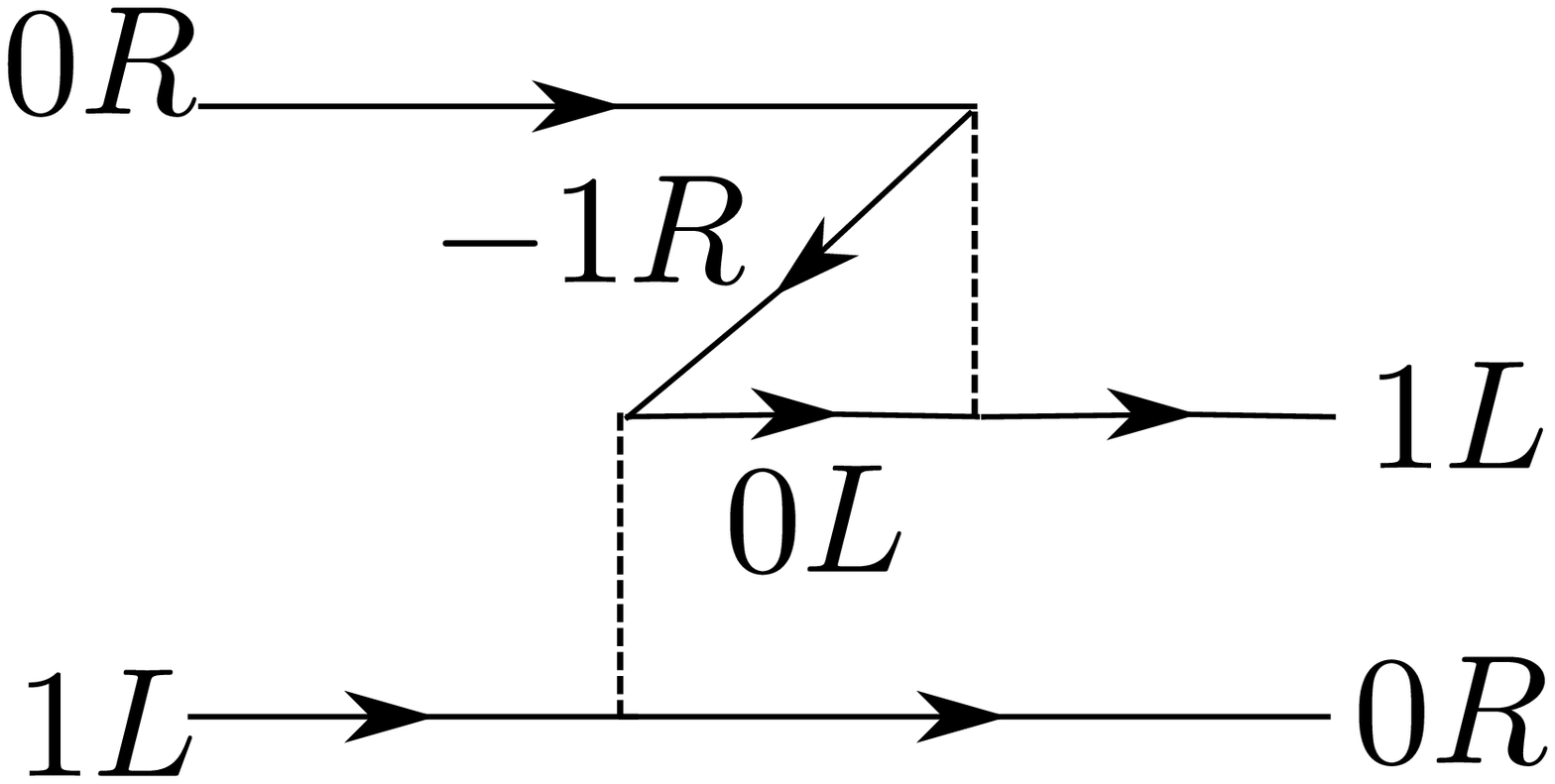}} \\ ~ \\
		\fbox{\includegraphics[width = 0.3\textwidth]{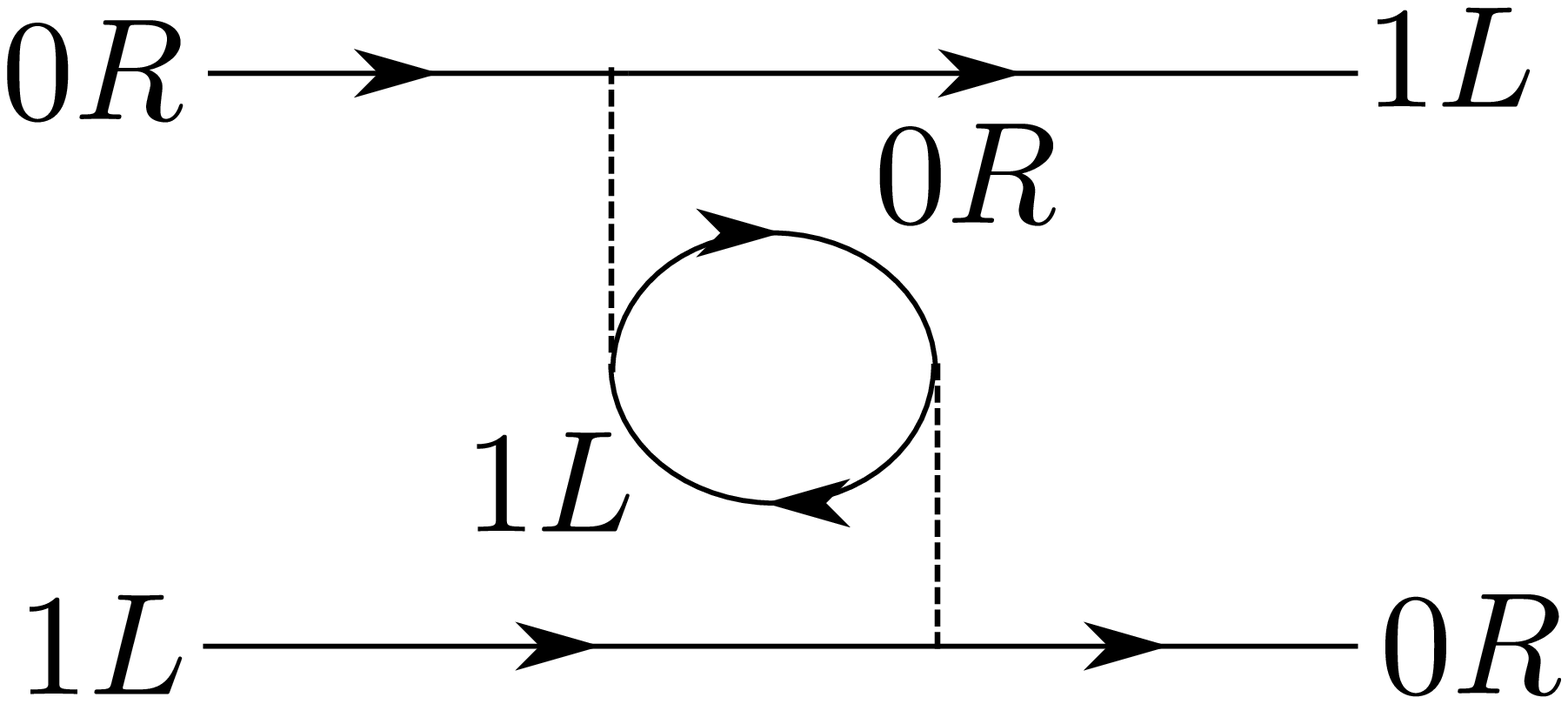}} &
		\fbox{\includegraphics[width = 0.3\textwidth]{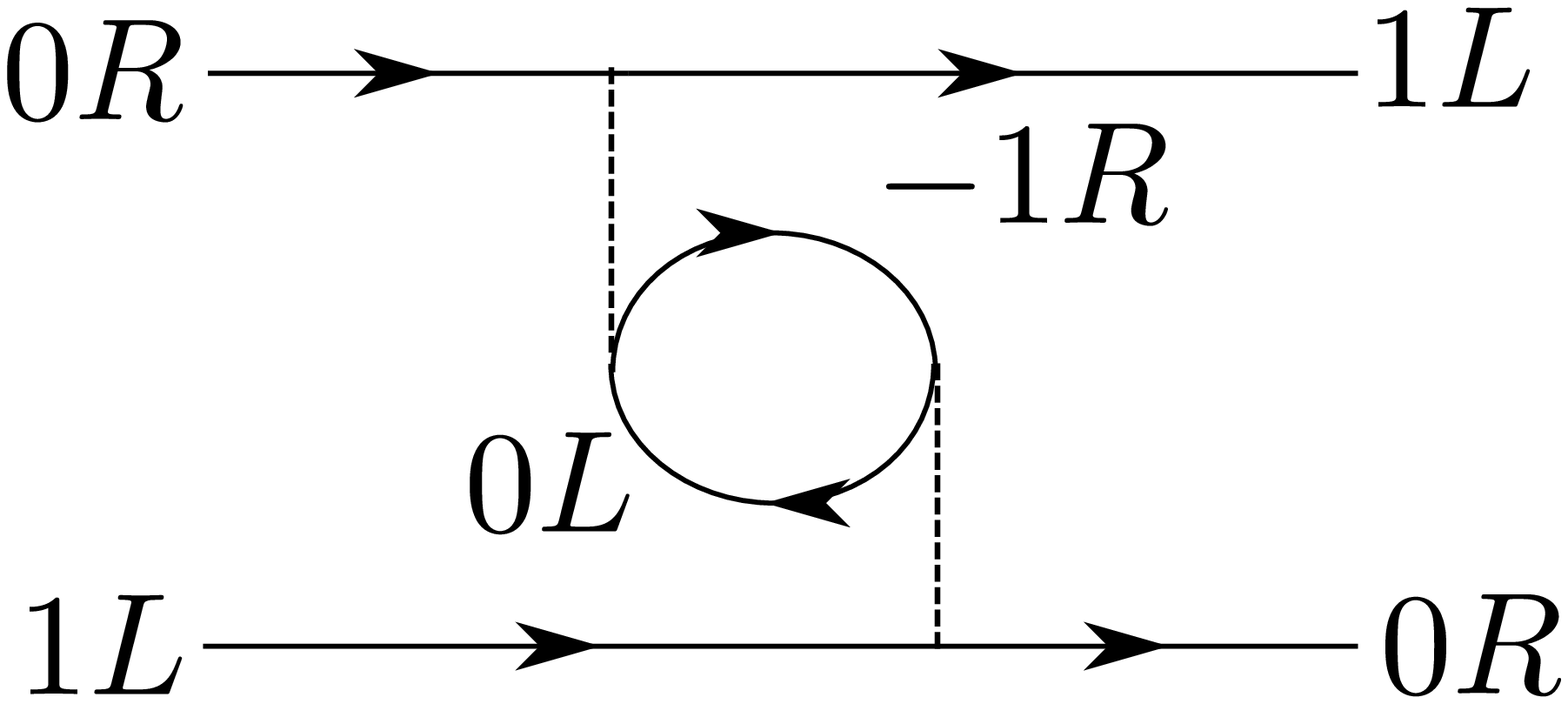}}
	\end{tabular}
	\caption{Diagrams contributing to the flow of the interaction $\widetilde{U}_{01}$. In the Cooper channel the intermediate states have $L = 1$, while in the electron-hole channels $L = -1$.}
	\label{U11}
\end{figure}

\begin{figure}
	\begin{tabular}{ccc}
		\fbox{\includegraphics[width = 0.3\textwidth]{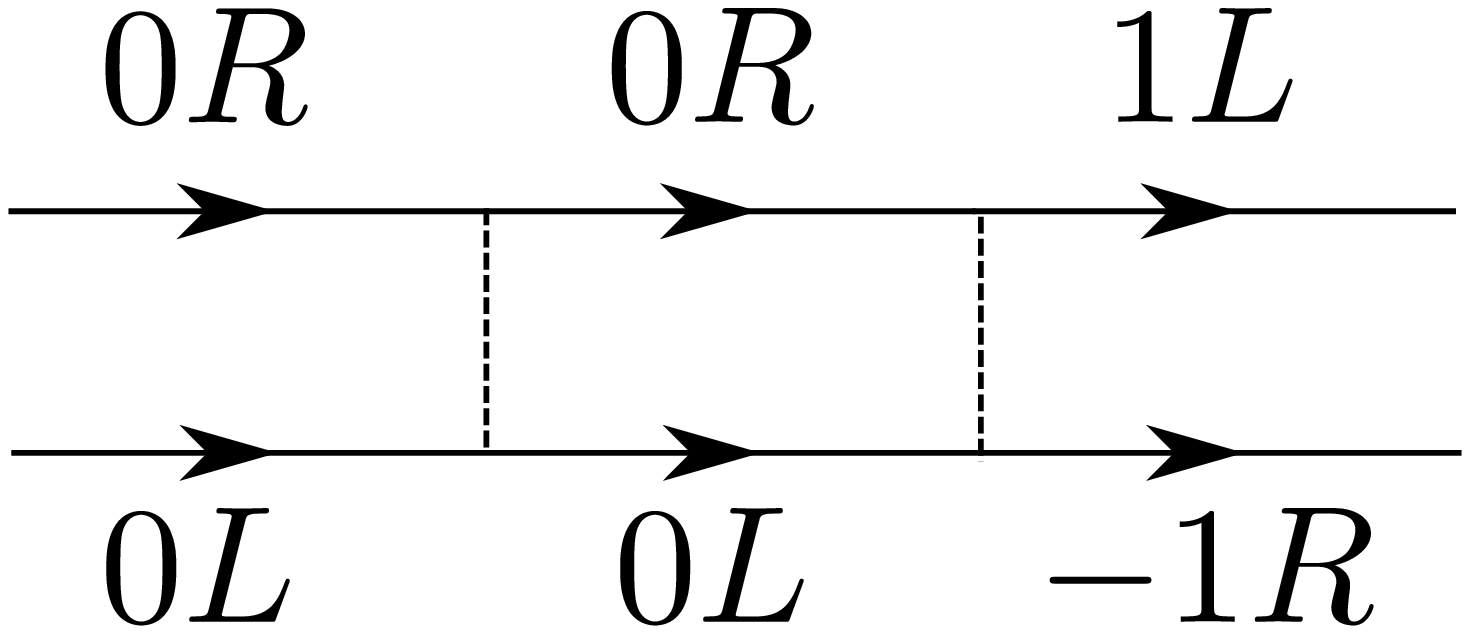}} &
		\fbox{\includegraphics[width = 0.3\textwidth]{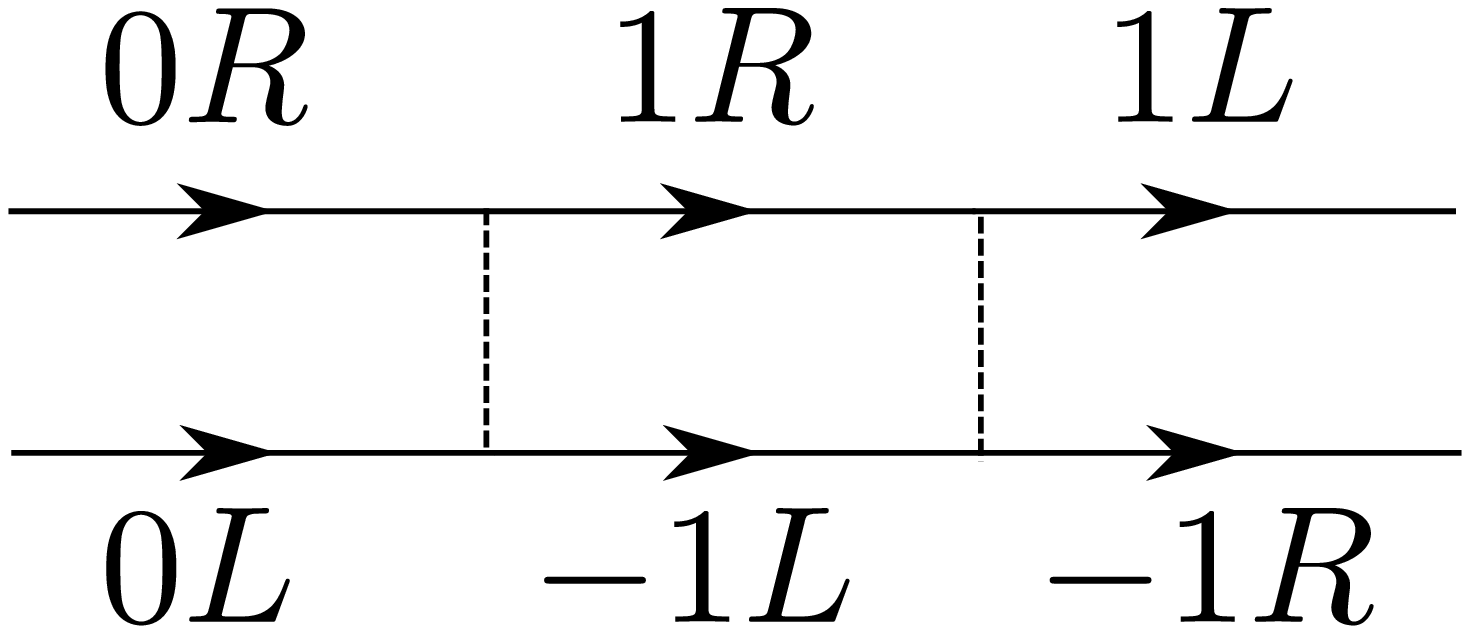}} &
		\fbox{\includegraphics[width = 0.3\textwidth]{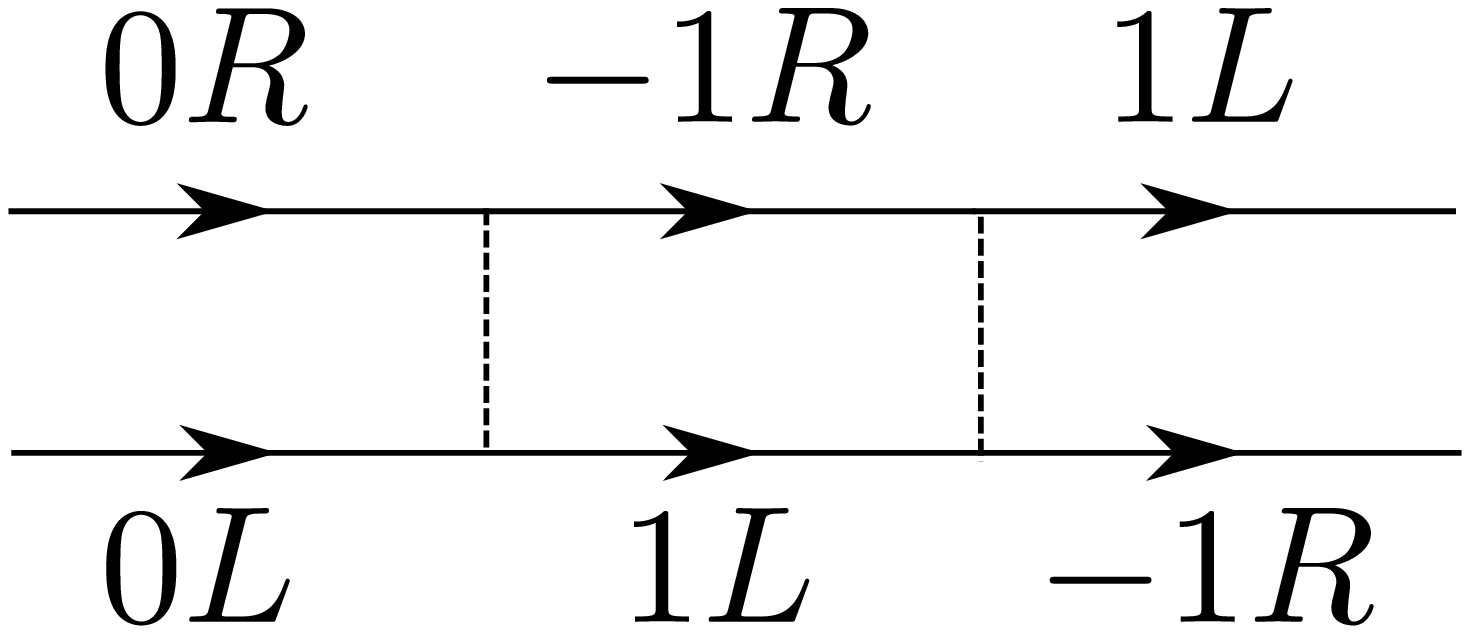}} \\ ~ \\
		\fbox{\includegraphics[width = 0.3\textwidth]{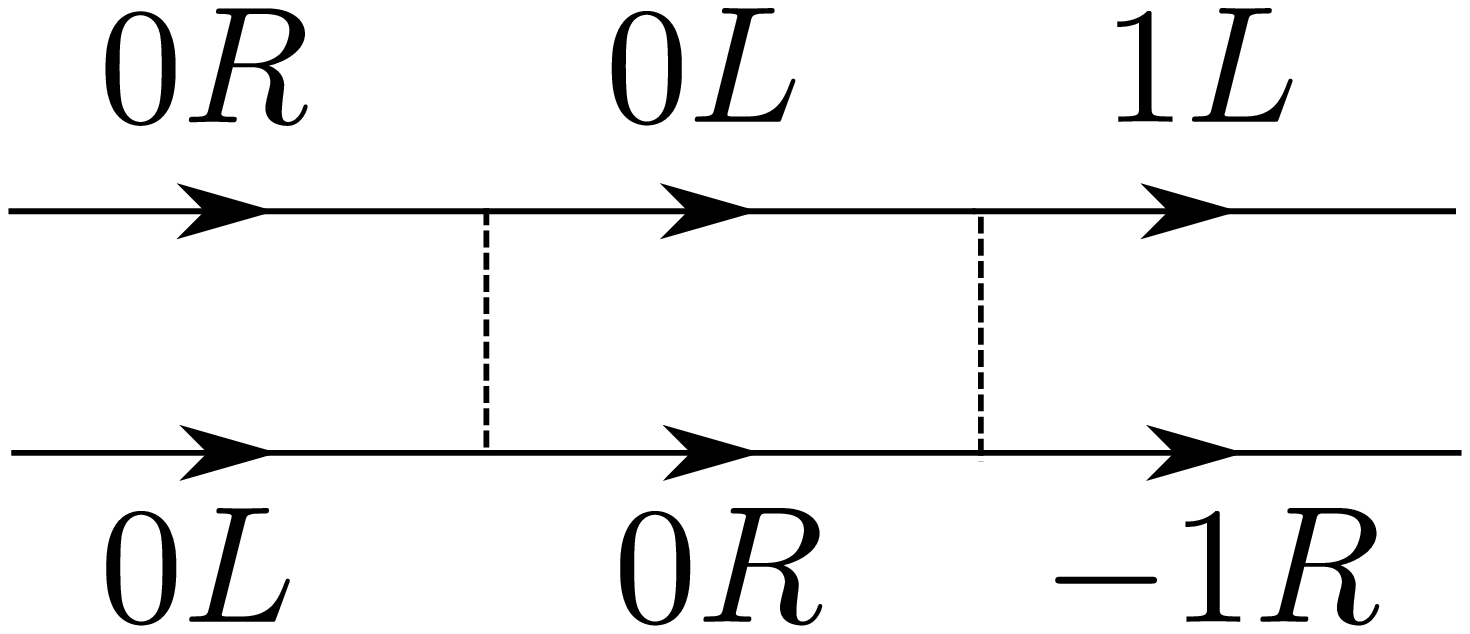}} &
		\fbox{\includegraphics[width = 0.3\textwidth]{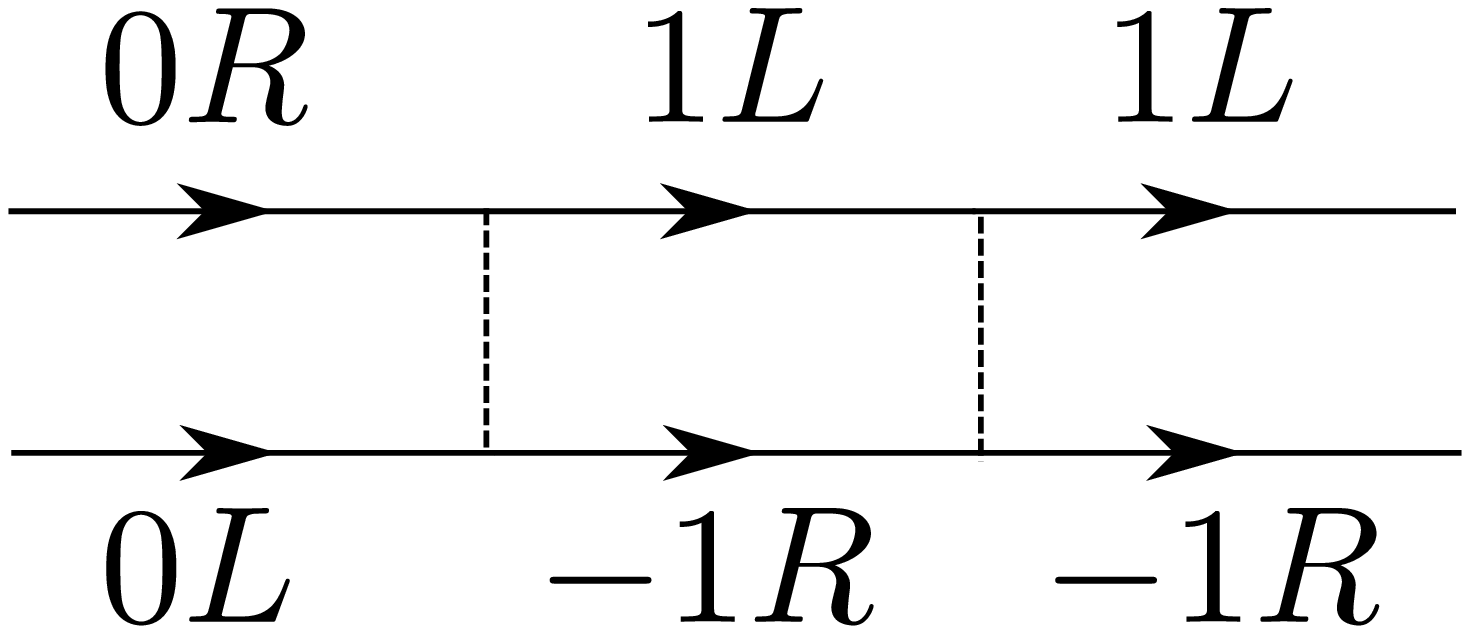}} &
		\fbox{\includegraphics[width = 0.3\textwidth]{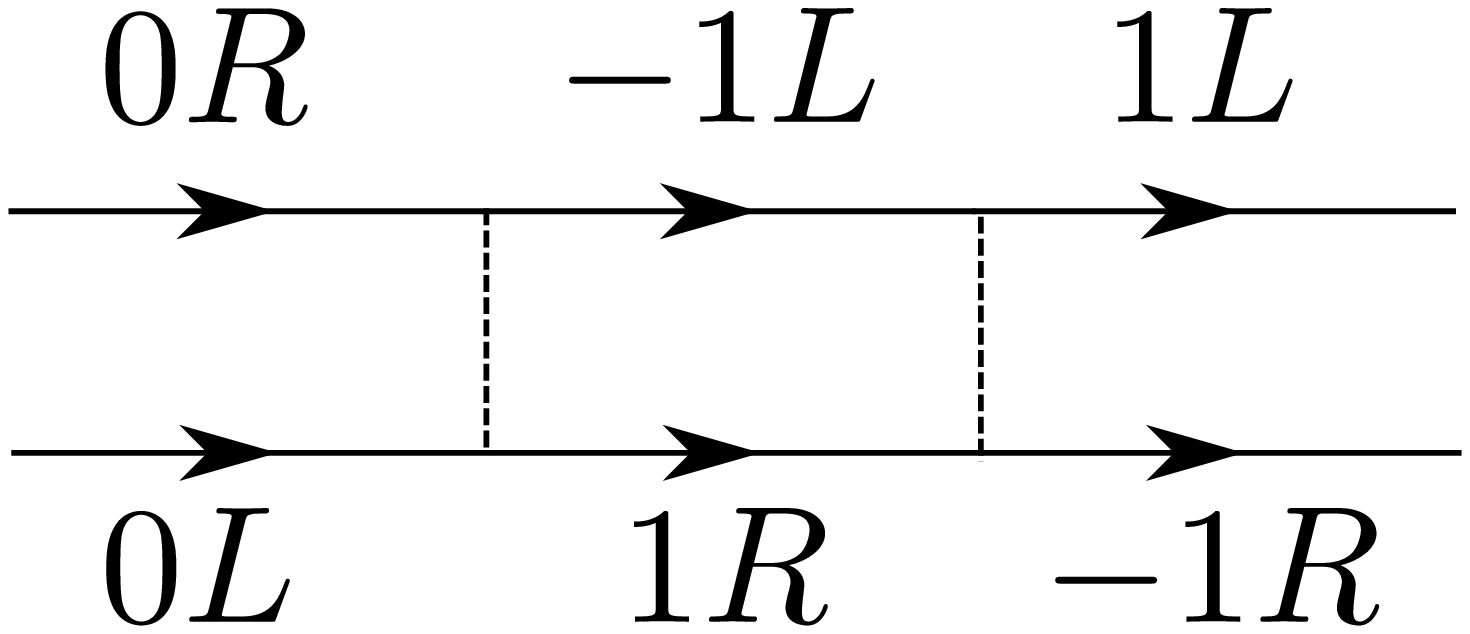}} \\ ~ \\
		\fbox{\includegraphics[width = 0.3\textwidth]{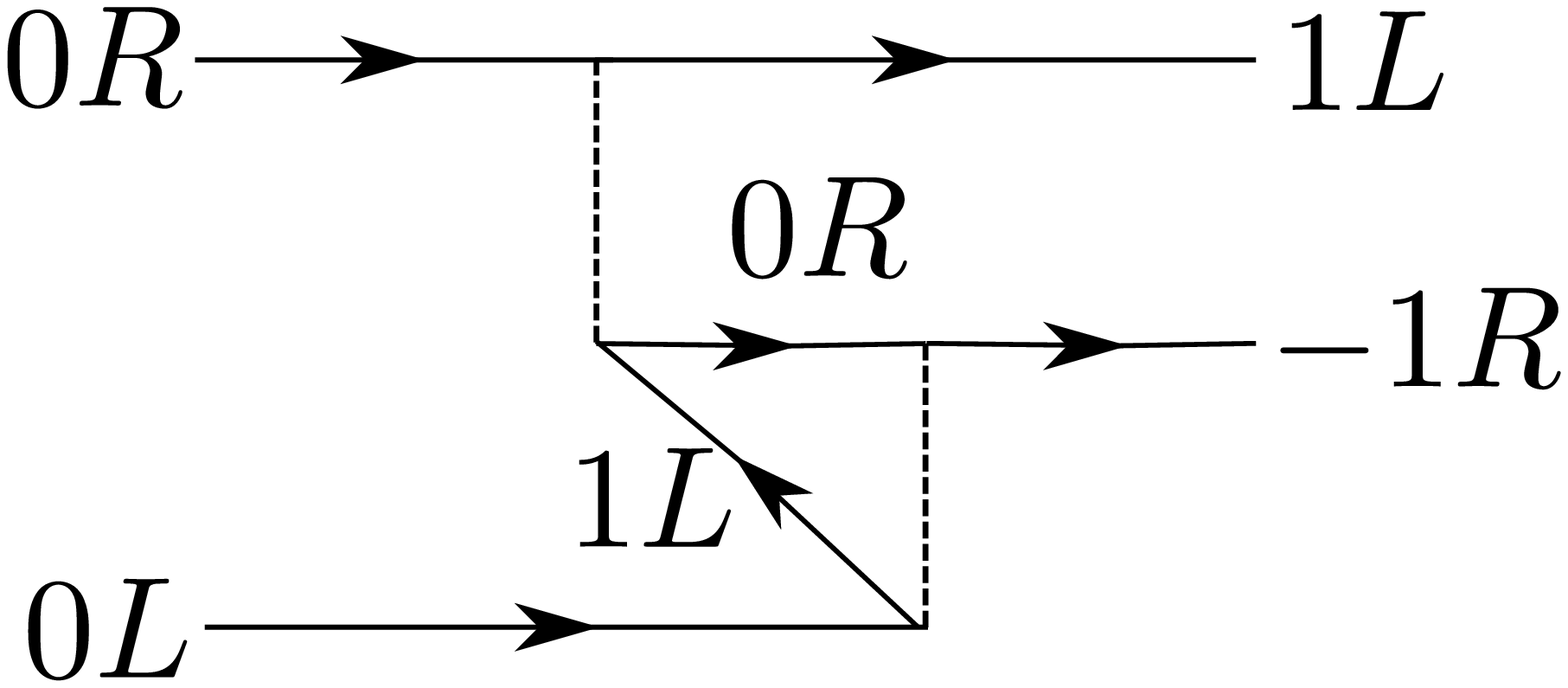}} &
		\fbox{\includegraphics[width = 0.3\textwidth]{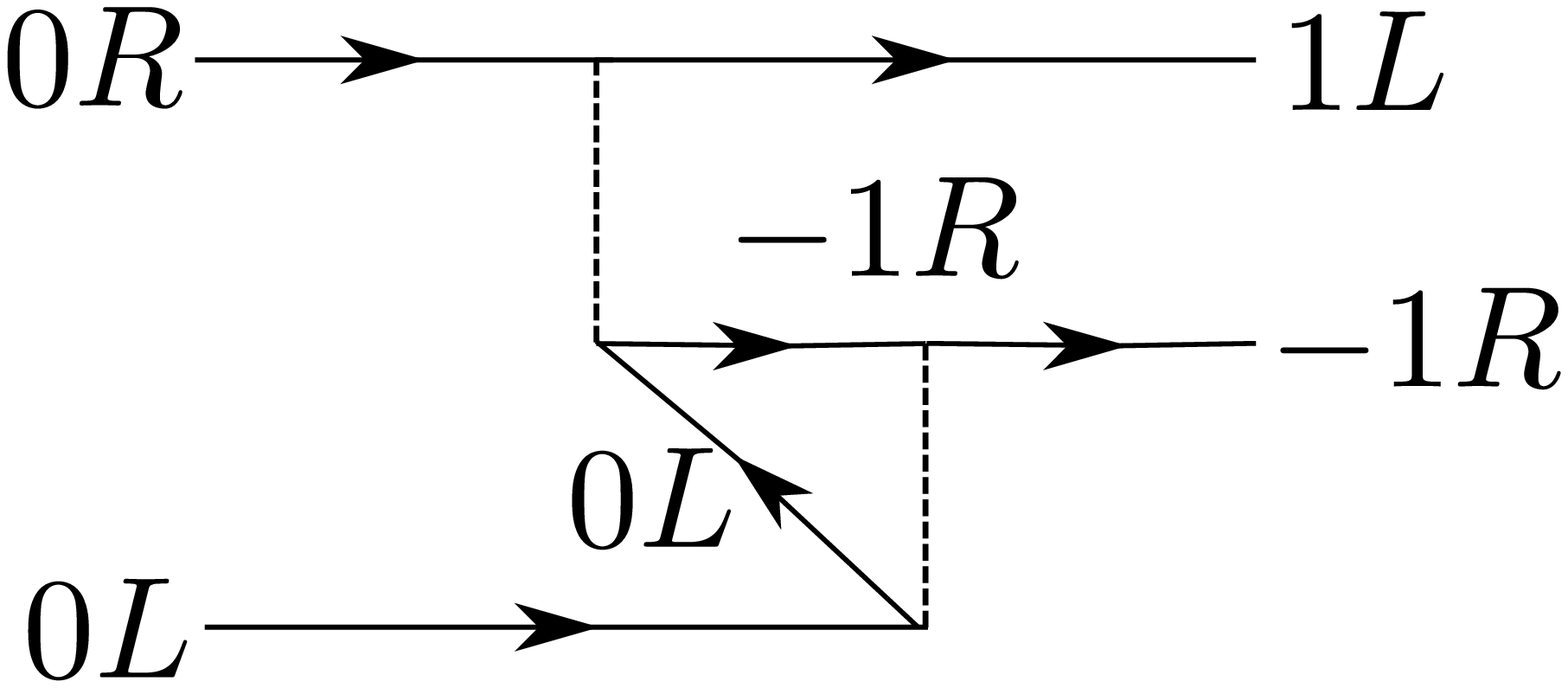}} &
		\\ ~ \\
		\fbox{\includegraphics[width = 0.3\textwidth]{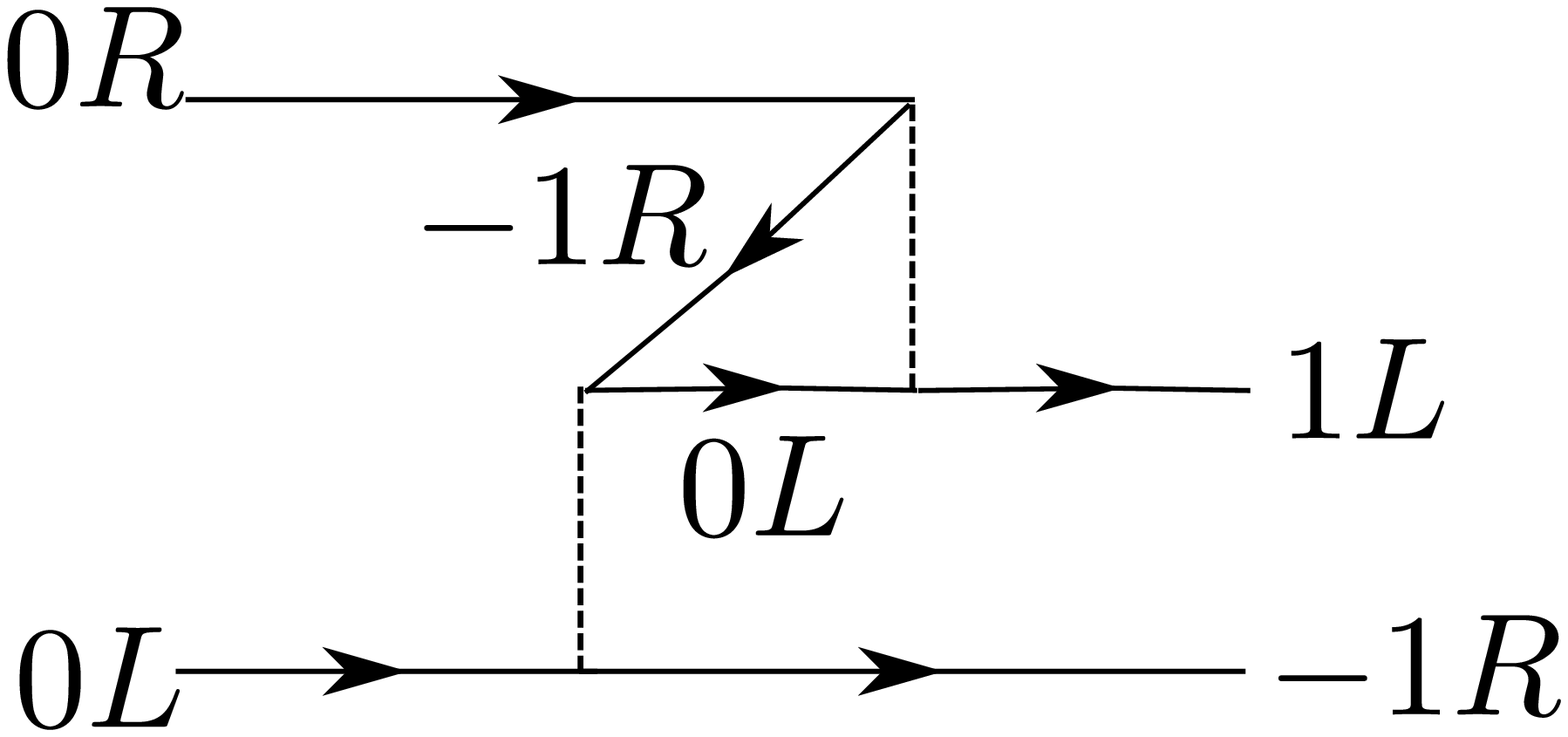}} &
		\fbox{\includegraphics[width = 0.3\textwidth]{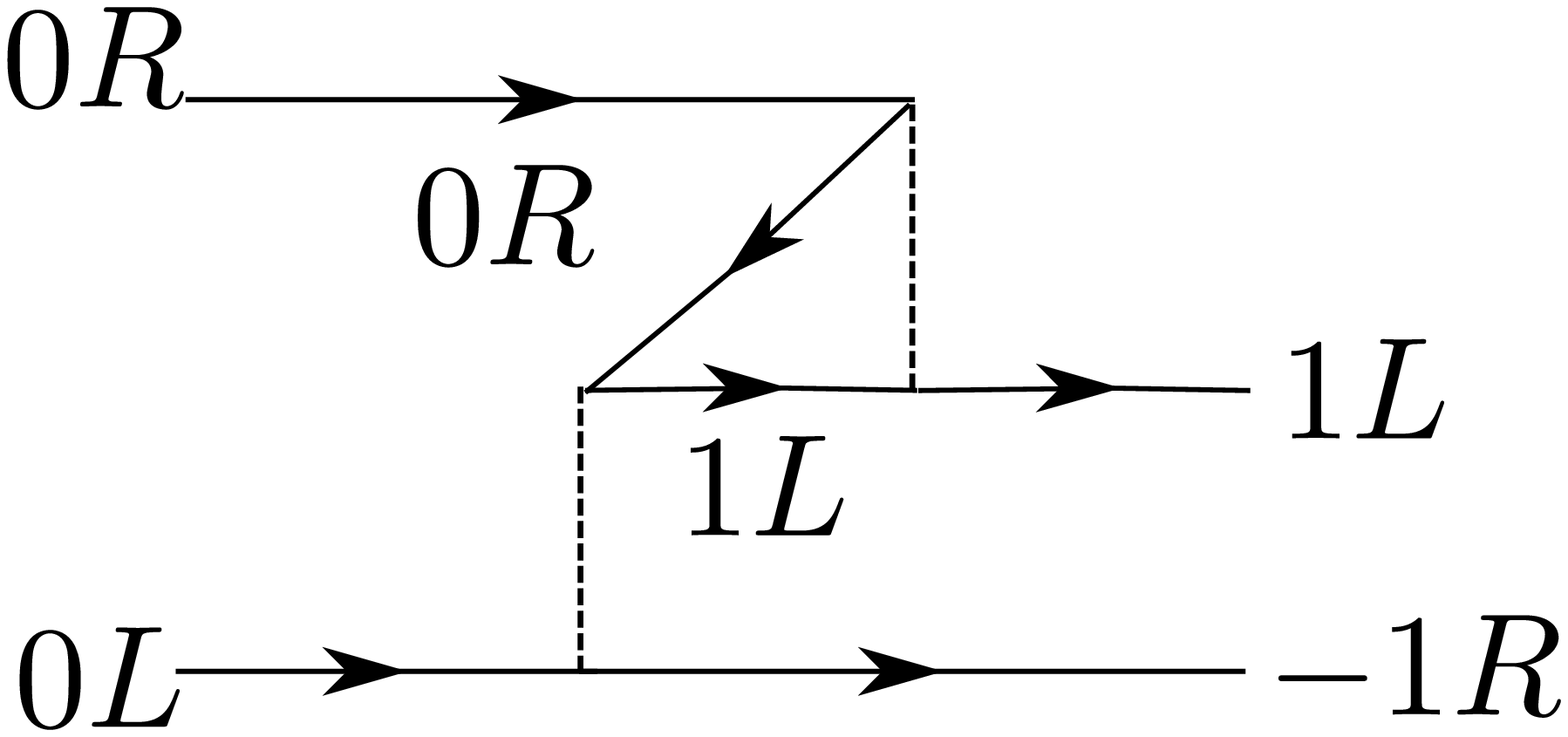}} &
		\\ ~ \\
		\fbox{\includegraphics[width = 0.3\textwidth]{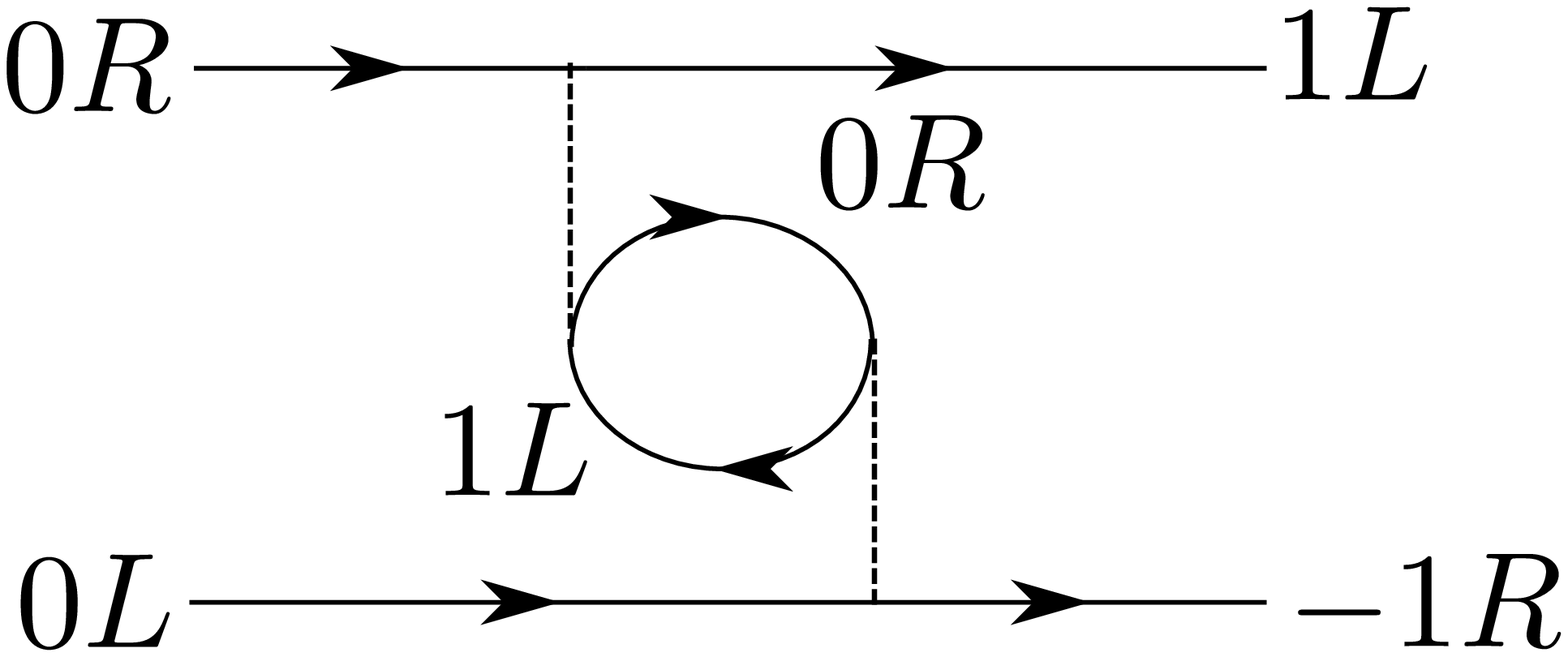}} &
		\fbox{\includegraphics[width = 0.3\textwidth]{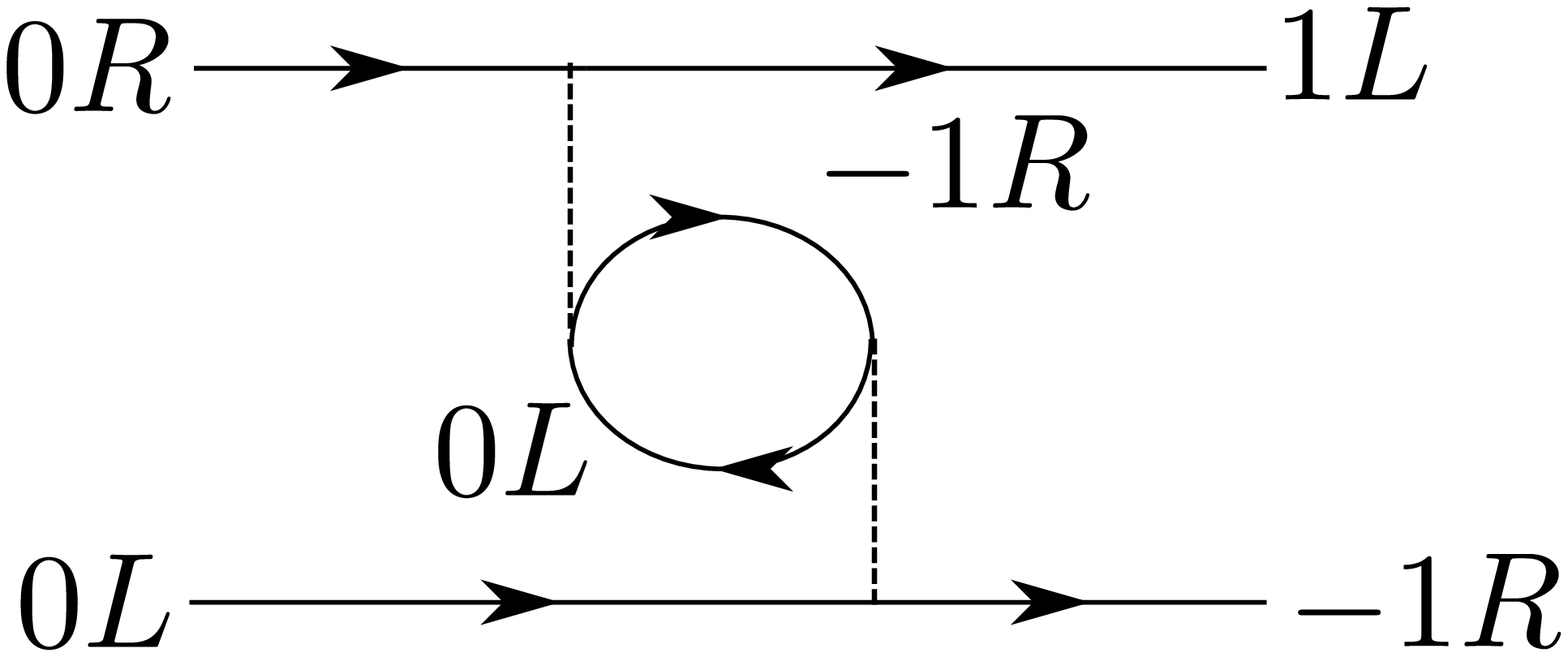}} &
	\end{tabular}
	\caption{Diagrams contributing to the flow of the interaction $\widetilde{V}_1$. In the electron-electron channels $L = 0$ in the intermediate states, while in the electron-hole channels $L = -1$.}
	\label{U12}
\end{figure}

The RG flow of the couplings to second order may be represented by the Feynman diagrams shown in Figs. ~\ref{U1}-~\ref{U12}. Each diagram shows a virtual scattering process involving one intermediate two-particle state, which consists of either two electrons propagating in opposite directions (which we refer to as the \emph{Cooper channels} following standard terminology), or an electron and hole propagating in the same direction. Integration over the intermediate states lead to a logarithmic divergence in the scattering amplitudes in second order perturbation theory, which in the perturbative RG treatment corresponds to a contribution $\propto \ln (\Lambda_0/\Lambda)$ to the running couplings, where $\Lambda_0 \approx \epsilon_F$ is the UV cutoff and $\Lambda$ is the running energy scale.

In many cases, pairs of diagrams exist which consist only of forward-scattering vertices in which a Cooper channel process may be related to a particle-hole channel process by reversal of all the lines corresponding to either left or right movers. This situation occurs, for example, in the pair of diagrams shown in the first line of Fig. ~\ref{U1}. In such a case the pair of diagrams exactly compensate each other. Similar pairs of canceling diagrams are not shown for the other interactions.

\onecolumngrid

~\\

~\\

~\\

~\\

~ \\

~ \\

~ \\

~ \\

~ \\

~ \\

